%% file: ec23.multiagent.contracts.tex
\documentclass{article}
\usepackage{arxiv}
\usepackage{booktabs} 
\usepackage{algorithm}
\usepackage[noend]{algpseudocode}
\usepackage{tikz}
\usetikzlibrary{calc}
\usepackage{thm-restate}
\usepackage{mathtools}
\usepackage{amsmath}
\usepackage{amssymb}
\usepackage{mathtools}
\usepackage{amsthm}
\usepackage{comment}
\usepackage{url}

\newtheorem{assumption}{Assumption}
\newtheorem{remark}{Remark}
\newtheorem{definition}{Definition}
\newtheorem{example}{Example}
\newtheorem{theorem}{Theorem}
\newtheorem{corollary}{Corollary}

\input{defs}
\usepackage{natbib}

\title{Multi-Agent Contract Design: How to Commission Multiple Agents with Individual Outcomes}

\author{
	Matteo Castiglioni\\
	Politecnico di Milano\\
	\texttt{matteo.castiglioni@polimi.it}
	\And
	Alberto Marchesi\\
	Politecnico di Milano\\
	\texttt{alberto.marchesi@polimi.it}
	\And
	Nicola Gatti\\
	Politecnico di Milano\\
	\texttt{nicola.gatti@polimi.it}
}

\begin{document}


\maketitle	

\input{abstract}

\input{intro}

\input{related}

\input{prelim}

\input{matroids}

\input{ir_super}

\input{dr_sub}

\input{unknown}

\newpage
\bibliographystyle{ACM-Reference-Format}
\bibliography{biblio}

\newpage
\appendix
\input{appendix_matroid}

\input{appendix_hard_ir_super}

\input{appendix_ir}

\input{appendix_hard_dr_sub}
\input{appendix_dr}
\input{appendix_bayesian}

\end{document}

%% file: defs.tex
\newcommand{\vomega}{{\boldsymbol{\omega}}}
\newcommand{\va}{{\boldsymbol{a}}}

\newcommand{\NPHard}{$\mathsf{NP}$-hard}

\newcommand{\cG}{\mathcal{G}}
\newcommand{\M}{\mathcal{M}}
\newcommand{\B}{\mathcal{B}}
\newcommand{\I}{\mathcal{I}}
\newcommand{\avec}{\boldsymbol{a}}

\newcommand{\vtheta}{\boldsymbol{\theta}}
\newcommand{\mP}{\mathcal{P}}

\newcommand{\Ocal}{\mathcal{O}}

\newcommand*\circled[1]{\tikz[baseline=(char.base)]{
		\node[shape=circle,draw,inner sep=1pt] (char) {\scriptsize #1};}}

\newcommand{\SUP}{\textnormal{\textsc{sup}}}
\newcommand{\LP}{\textnormal{\textsc{lp}}}
\newcommand{\VAL}{\textnormal{\textsc{val}}}
\newcommand{\OPT}{\textnormal{\textsc{opt}}}
\newcommand{\APX}{\textnormal{\textsc{apx}}}

\newcommand{\Hcal}{\mathcal{H}}

\newcommand{\poly}{\textnormal{poly}}

\makeatletter
\newcommand{\pushright}[1]{\ifmeasuring@#1\else\omit\hfill$\displaystyle#1$\fi\ignorespaces}
\newcommand{\pushleft}[1]{\ifmeasuring@#1\else\omit$\displaystyle#1$\hfill\fi\ignorespaces}
\newcommand{\specialcell}[1]{\ifmeasuring@#1\else\omit$\displaystyle#1$\ignorespaces\fi}
\makeatother

%% file: abstract.tex
\begin{abstract}
  We study \emph{hidden-action principal-agent problems} with \emph{multiple} agents. These are problems in which a principal commits to an outcome-dependent payment scheme (called \emph{contract}) in order to incentivize some agents to take costly, unobservable actions that lead to favorable outcomes. Previous works on multi-agent problems study models where the principal observes a single outcome determined by the actions of \emph{all} the agents. Such models considerably limit the contracting power of the principal, since payments can only depend on the joint result of all the agents' actions, and there is no way of paying each agent for their individual result. In this paper, we consider a model in which each agent determines their own \emph{individual} outcome as an effect of their action only, the principal observes all the individual outcomes separately, and they perceive a reward that jointly depends on all these outcomes. This considerably enhances the principal's contracting capabilities, by allowing them to pay each agent on the basis of their individual result. We analyze the computational complexity of finding principal-optimal contracts, revolving around two newly-introduced properties of principal's rewards, which we call \emph{IR-supermodularity} and \emph{DR-submodularity}. Intuitively, the former captures settings with \emph{increasing returns}, where the rewards grow faster as the agents' effort increases, while the latter models the case of \emph{diminishing returns}, in which rewards grow slower instead. These two properties naturally model two common real-world phenomena, namely diseconomies and economies of scale. In this paper, we first address basic instances in which the principal knows everything about the agents, and, then, more general \emph{Bayesian} instances where each agent has their own \emph{private type} determining their features, such as action costs and how actions stochastically determine individual outcomes. As a preliminary result, we show that finding an optimal contract in a non-Bayesian instance can be reduced in polynomial time to a suitably-defined maximization problem over a matroid having a particular structure. Such a reduction is needed to prove our main positive results in the rest of the paper. We start by analyzing non-Bayesian instances with IR-supermodular rewards, where we prove that the problem of computing a principal-optimal contract is inapproximable in general, but it becomes polynomial-time solvable under some mild regularity assumptions. Then, we study non-Bayesian instances with DR-submodular rewards, showing that the problem is inapproximable also in this setting, but it admits a polynomial-time approximation algorithm which outputs contracts providing a multiplicative approximation $(1-1/e)$ of the principal's reward in an optimal contract, up to a small additive loss. In conclusion, we extend our positive results to Bayesian instances. First, we provide a characterization of the principal's optimization problem, by showing that it can be approximately solved by means of a linear formulation. This is non-trivial, since in general the problem may \emph{not} admit a maximum, but only a supremum. Then, based on such a linear formulation, we provide a polynomial-time approximation algorithm that employs an \emph{ad hoc} implementation of the ellipsoid method using an \emph{approximate separation oracle}. We prove that such an oracle can be implemented in polynomial time by exploiting our positive results on non-Bayesian instances. Surprisingly, this allows us to (almost) match the guarantees obtained for non-Bayesian instances.
\end{abstract}

%% file: intro.tex
\section{Introduction}\label{sec:intro}

Over the last few years, \emph{principal-agent problems} have received a growing attention from the economics and computation community.
These problems model scenarios in which a \emph{principal} interacts with one or more \emph{agents}, with the latter playing actions that induce externalities on the former.
We focus on \emph{hidden-action} problems, where the principal only observes some stochastically-determined outcome of the actions selected by the agents, but \emph{not} the actions themselves.
The principal gets a reward associated with the realized outcome, while an agent incurs in a cost when performing an action.
Thus, the principal's goal is to incentivize agents to undertake actions which result in profitable outcomes.
This is accomplished by committing to a \emph{contract}, which is a payment scheme defining how much the principal pays each agent depending on the realized outcome.

The classical textbook example motivating the study of hidden-action principal-agent problems is that of a firm (principal) hiring a salesperson (agent) in order to sell some products.
The salesperson has to decide on the level of effort (action) to put in selling products, while the firm only observes the number of products that are actually sold (outcome).
In such a scenario, it is natural that the firm commits to pay a commission to the salesperson by stipulating a contract with them, and that such a commission only depends on the number of products being sold.

Nowadays, the study of principal-agent problems is also motivated by the fact that they are ubiquitous in several real-world settings, such as, \emph{e.g.}, crowdsourcing platforms~\citep{ho2016adaptive}, blockchain-based smart contracts~\citep{cong2019blockchain}, and healthcare~\citep{bastani2016analysis}.

The computational aspects of principal-agent problems with a single agent have been widely investigated in the literature.
Instead, only few works study problems with multiple agents.
Some notable examples are the papers by~\citet{babaioff2006combinatorial}~and~\citet{emek2012computing}, and the very recent preprint by~\citet{duetting2022multi}.
These works address models where the principal observes a single outcome determined by the actions of \emph{all} the agents.
Such models considerably limit the contracting power of the principal, since payments can only depend on the joint result of all the agents' actions, and there is no way of paying each agent for their individual result.
%

In this paper, we introduce and study principal-agent problems with multiple agents---compactly referred to as \emph{principal-multi-agent problems}---in which each agent determines their own \emph{individual} outcome as an effect of their action only, the principal observes all the individual outcomes separately, and they perceive a reward that jointly depends on all these outcomes.
%
%
%
Our model fits many real-world applications.
For instance, in settings where a firm wants to hire multiple salespersons, it is natural that the firm can observe the number of products being sold by each of them individually.
Additionally, as we show in this paper, our model also allows to circumvent the equilibrium-selection issues raised by the problems studied in~\citep{babaioff2006combinatorial,emek2012computing,duetting2022multi}.
Indeed, as we discuss later in Section~\ref{sec:related}, such issues originate from the appearance of externalities among the agents, which are instead \emph{not} present in our setting.

\subsection{Original Contributions}

We investigate the computational complexity of finding optimal contracts in our principal-multi-agent problems with agents' individual outcomes.
%
%
Our analysis revolves around two properties of principal's rewards, which we call \emph{IR-supermodularity} and \emph{DR-submodularity}.
Intuitively, the former captures settings with \emph{increasing returns}, where the rewards grow faster as the agents' effort increases, while the latter models the case of \emph{diminishing returns}, in which rewards grow slower as the effort increases. 
These two properties naturally model two common real-world phenomena, namely diseconomies and economies of scale, respectively.
%
%

In the first sections of the paper (namely Sections~\ref{sec:preliminaries_problem},~\ref{sec:matroids},~\ref{sec:ir_super},~and~\ref{sec:dr_sub}), we study basic principal-multi-agent problems in which the principal knows everything about agents, \emph{i.e.}, their action costs and the probability distributions that their actions induce over (individual) outcomes.
Then, in Section~\ref{sec:bayesian}, we switch the attention to the far more general \emph{Bayesian} settings in which each agent's action costs and probability distributions depend on a \emph{private agent's type}, which is unknown to the principal, but randomly drawn according to a commonly-known probability distribution.

After introducing, in Section~\ref{sec:preliminaries_problem}, all the preliminary concepts related to the non-Bayesian version of our principal-multi-agent problems, in Section~\ref{sec:matroids} we provide a useful preliminary result.
We show that the problem of computing an optimal contract in a non-Bayesian instance can be reduced in polynomial time to the maximization of a suitably-defined set function over a matroid having a particular structure.
Specifically, we call the matroids introduced by our reduction $1$-partition matroids, since their ground sets are partitioned into some classes and their independent sets are all the subsets which contain at most one element for each class.
At the end of the section (more precisely in Section~\ref{sec:matroids_additional}), we also provide an additional preliminary result, by showing that there exists a polynomial-time algorithm for maximizing particular set functions, which we call \emph{ordered-supermodular} functions, over $1$-partition matroids.
This will be useful to derive our positive result in the following Section~\ref{sec:ir_super}, and it may also be of independent interest.

In Section~\ref{sec:ir_super}, we provide our main results on non-Bayesian instances with IR-supermodular principal's rewards.
We start with a negative result: for any $\rho > 0$, it is \NPHard~to design a contract providing a multiplicative approximation $\rho$ of the principal's expected utility in an optimal contract, even when both the number of agents' actions and the number of outcomes are fixed.
Then, we show how to circumvent such a negative result by introducing a mild regularity assumption.
Specifically, we prove that, in instances with IR-supermodular principal's rewards that additionally satisfy a particular \emph{first-order stochastic dominance} (FOSD) condition, an (exact) optimal contract can be found in polynomial time.
This is accomplished by exploiting the reduction introduced in Section~\ref{sec:matroids}, and by proving that, for such instances, the resulting set function is ordered-supermodular.

In Section~\ref{sec:dr_sub}, we switch our attention to non-Bayesian instances with DR-submodular principal's rewards.
Similarly to the preceding section, we start with a negative result: for any $\alpha > 0$, it is \NPHard~to design a contract providing a multiplicative approximation $n^{1-\alpha}$---with $n$ being the number of agents---of the principal's expected utility in an optimal contract, even when both the number of agents' actions and the dimensionality of the outcomes are fixed.
Next, we complement such a negative result by providing a polynomial-time approximation algorithm for the problem.
In particular, we exploit the reduction to matroid optimization introduced in Section~\ref{sec:matroids} and a result by~\citet{sviridenko2017optimal} in order to design an algorithm that, with high probability, outputs a contract providing a multiplicative approximation $(1-1/e)$ of the principal's reward in an optimal contract, up to a small additive loss $\epsilon > 0$, in time polynomial in the instance size and $1/\epsilon$.

Finally, we conclude the paper, in Section~\ref{sec:bayesian}, by providing our results on Bayesian principal-multi-agent problems.
First, we extend the model recently introduced by~\citet{Castiglioni2022Designing} to our multi-agent setting.
The key feature of such a model is that, by taking inspiration from classical mechanism design, it adds a \emph{type-reporting} stage in which each agent is asked to report their type to the principal.
In such a setting, the principal is better off committing to a \emph{menu of randomized contracts} rather than a single contract.
This specifies a collection of probability distributions over (non-randomized) contracts, where each distribution is employed to draw a contract upon a different combination of types reported by the agents.
Surprisingly, we show that it is possible to implement a polynomial-time approximation algorithm for the problem of computing an optimal menu of randomized contracts, whose guarantees (almost) match those obtained for non-Bayesian instances.
In order to obtain the result, we first provide a characterization of the principal's optimization problem, by showing that it can be approximately solved by means of a  \emph{linear program} (LP) with polynomially-many variables and exponentially-many constraints.
Notice that this step is non-trivial, since in general the principal's optimization problem may \emph{not} admit a maximum, but only a supremum.
Our algorithm is based on an \emph{ad hoc} implementation of the ellipsoid method, which approximately solves such an LP, provided that it has access to a suitably-defined, polynomial-time \emph{approximate separation oracle}.
Such an oracle can be implemented by using the algorithms developed in Sections~\ref{sec:ir_super}~and~\ref{sec:dr_sub} for non-Bayesian instances.
%

%% file: related.tex
\subsection{Related Works}\label{sec:related}

Next, we survey the most-related computational works on hidden-action principal-agent problems.
%

\paragraph{Works on Principal-Agent Problems with a Single Agent.}
Most of these works focus on non-Bayesian settings.
Among the most related to ours, \citet{dutting2021complexity}~and~\citet{dutting2022combinatorial} study models whose underlying structure is combinatorial.
In particular, the latter analyze the case in which the outcome space is defined implicitly through a succinct representation, while the former address settings in which the agent selects a subset of actions (rather than a single one).
Moreover, \citet{babaioff2014contract} study the complexity of contracts in terms of the number of different payments that they specify, while~\citet{dutting2019simple} use the computational lens to analyze the efficiency (in terms of principal's expected utility) of \emph{linear contracts} with respect to optimal ones.
Recently, some works also considered the more realistic Bayesian settings~\citep{guruganesh2021contracts,alon2021contracts,castiglioni2022bayesian,DBLP:conf/sigecom/CastiglioniM022}.
In particular,~\citet{DBLP:conf/sigecom/CastiglioniM022} introduce the idea of menus of randomized contracts, showing that in Bayesian settings they enjoy much nicer computational properties than menus of deterministic (\emph{i.e.}, non-randomized) contracts, which were previously studied in~\citep{guruganesh2021contracts,alon2021contracts}.

\paragraph{Works on Principal-Agent Problems with Multiple Agents.}
All the previous works on multi-agent settings are limited to non-Bayesian instances. 
\citet{babaioff2006combinatorial} are the first to study a model with multiple agents (see also its extended version~\citep{babaioff2012combinatorial} and its follow-ups~\citep{babaioff2009free,babaioff2010mixed}).
They study a setting in which agents have binary actions, called \emph{effort} and \emph{no effort}, and the outcome is determined according to a probability distribution that depends on the set of agents that decide to undertake \emph{effort}.
This model induces \emph{externalities} among the agents, since the realized outcome (and, in turn, the agents' payments) depends on the actions taken by all the agents.
\citet{babaioff2006combinatorial} show that finding an optimal contract is \#$\mathsf{P}$-complete	even when the outcome-determining function is represented as a ``simple'' read-once network.
\citet{emek2012computing} extend the work by~\citet{babaioff2006combinatorial} by showing that the problem is \NPHard~even for a special class of submodular functions, while admitting an FPTAS.
Finally, a very recent preprint by~\citet{duetting2022multi} considerably extends previous works by providing constant-factor approximation algorithms for problems with submodular and XOS rewards.

%% file: prelim.tex
\section{The Principal--Multi--Agent Problem}\label{sec:preliminaries_problem}

An instance of the \emph{principal-multi-agent problem} is characterized by a tuple $(N,\Omega,A)$,\footnote{For ease of notation, in this paper we assume that all the numerical quantities that define a principal-multi-agent problem instance, such as costs, rewards, and probabilities, are attached to their corresponding elements in the sets $N$, $\Omega$, and $A$, so that we can simply write $I \coloneqq (N,\Omega,A)$ to identify an instance of the problem.} where: $N$ is a finite set of $n \coloneqq |N|$ \emph{agents}; $\Omega$ is a finite set of $m \coloneqq |\Omega|$ possible \emph{(individual) outcomes} of an agent's action, and $A$ is a finite set of $\ell \coloneqq |A|$ \emph{actions} available to each agent.\footnote{For ease of presentation, we assume that all the agents share the same action set and outcome set. Our results continue to hold even if each agent $i \in N$ has their own action set $A_i$ and their actions induce outcomes in an agent-specific set $\Omega_i$.}

For each agent $i \in N$, we introduce $F_{i, a} \in \Delta_\Omega$ to denote the probability distribution over outcomes $\Omega$ induced by action $a \in A$ of agent $i$,\footnote{In this paper, given a finite set $X$, we denote by $\Delta_X$ the set of all the probability distributions defined over elements of $X$.} while $c_{i, a} \in [0,1]$ denotes the agent's cost for playing such action.\footnote{For ease of presentation, costs and rewards are in $[0,1]$. All the results can be easily generalized to an arbitrary range.}
For ease of presentation, we let $F_{i, a, \omega} $ be the probability that $F_{i, a}$ assigns to $\omega \in \Omega$, so that it holds $\sum_{\omega \in \Omega} F_{i, a, \omega}  =1$.
We define $\va \in A^n \coloneqq \bigtimes_{i \in N} A$ as a tuple of agents' actions, whose $i$-th component is denoted by $a_i$ and represents the action played by agent $i$.
Moreover, we let $\vomega \in \Omega^n \coloneqq \bigtimes_{i \in N} \Omega$ be a tuple of outcomes, whose $i$-th component $\omega_i$ is the individual outcome achieved by agent $i$.
Each tuple $\vomega \in \Omega^n$ has an associated reward to the principal, which we denote by $r_\vomega \in [0,1]$.
As a result, whenever the agents play the actions defined by a tuple $\va \in A^n$, the principal achieves an expected reward equal to $R_{\va} \coloneqq \sum_{\vomega \in \Omega^n} r_\vomega \prod_{i \in N} F_{i,a_i, \omega_i}  $.

%
Notice that, in our model, the principal observes all the elements in the tuple of outcomes $\vomega \in \Omega^n$ reached by the agents, which consists in an individual outcome $\omega_i$ for each agent $i \in N$.
This is in contrast with previous works on principal-multi-agent problems (see, \emph{e.g.},~\citep{babaioff2006combinatorial,emek2012computing,duetting2022multi}), which assume that the principal can only observe a single outcome that is jointly determined by the tuple of all the agents' actions.

\subsection{Contracts and Principal's Optimization Problem}\label{sec:preliminaries_contracts}

In a principal-multi-agent problem, the goal of the principal is to maximize their expected utility by committing to a \emph{contract}, which specifies payments from the principal to each agent contingently on the actual individual outcome achieved by the agent.
Formally, a contract is defined by a matrix $p \in {\mathbb{R}_+^{n \times m}}$, whose entries $p_{i,\omega} \ge 0$ define a payment for each agent $i \in N$ and outcome $\omega \in\Omega$.\footnote{W.l.o.g., we can restrict the attention to contracts that define payments independently for each agent, rather than dealing with contracts which specify payments based on the tuple of outcomes resulting from the actions of all agents. This is because each agent $i \in N$ induces a specific outcome $\omega_i$ with their action (independently of what the others do), and such outcome is observed by the principal. As a consequence, we also have that, in our setting, there are \emph{no externalities} among the agents, since an agent's expected utility does \emph{not} depend on the actions played by other agents.}
%
Notice that the assumption that payments are non-negative (\emph{i.e.}, they can only be from the principal to agents) is common in contract theory, where it is known as \emph{limited liability}~\citep{carroll2015robustness}.
When agent $i \in N$ selects an action $a \in A$ under a contract $p \in {\mathbb{R}_+^{n \times m}}$, the expected payment from the principal to agent $i$ is $P_{i,a} \coloneqq \sum_{\omega \in \Omega} F_{i, a, \omega} \, p_{i,\omega}$, while the agent's expected utility is $P_{i,a} - c_{i, a}$.
%

Given a contract $p \in {\mathbb{R}_+^{n \times m}}$, each agent $i \in N$ selects an action such that:
\begin{enumerate}
	\item it is \emph{incentive compatible} (IC), \emph{i.e.}, it maximizes their expected utility among actions in $A$;
	\item it is \emph{individually rational} (IR), \emph{i.e.}, it has non-negative expected utility (if there is no IR action, then agent $i$ abstains from playing so as to maintain the \emph{status quo}).
\end{enumerate}

For ease of presentation, we make the following w.l.o.g. assumption:
%

\begin{assumption}[Null action]\label{ass:null_action}
	There exists an action $a_\varnothing \in A$ such that $c_{i,a_\varnothing} = 0$ for all $i \in N$.
\end{assumption}
Such an assumption implies that each agent has an action providing them with a non-negative utility, thus ensuring that any IC action is also IR and allowing us to focus w.l.o.g. on incentive compatibility only.
In the following, given a contract $p \in {\mathbb{R}_+^{n \times m}}$, we denote by $A^\ast_i(p) \subseteq A$ the set of actions that are IC for agent $i \in N$ under that contract.
Formally, it holds $A^\ast_i(p) \coloneqq \arg\max_{a \in A} \left\{ P_{i,a}- c_{i,a}  \right\}$.
Furthermore, given an action $a \in A$ of agent $i \in N$, we let $\mP^{i,a} \subseteq \mathbb{R}^{n \times m}_+$ be the set of contracts such that action $a$ is IC for agent $i$ under them; formally, $\mP^{i,a} \coloneqq \left\{ p \in \mathbb{R}_+^{n \times m} \mid  a \in A_i^*(p) \right\}$.

Given a contract $p \in {\mathbb{R}_+^{n \times m}}$, the resulting set $A^\ast_i(p)$ of IC actions for an agent $i\in N$ may contain more than one action.
Thus, it is necessary to adopt a suitable tie-breaking-rule assumption.
\begin{remark}[On classical tie-breaking rules]\label{rem:tie_break}
Most of the works on principal-agent problems usually assume that, whenever an agent is indifferent among multiple IC actions, they break ties in favor of the principal (see, \emph{e.g.},~\citep{dutting2021complexity}).
Such an assumption is unreasonable in our setting.
Indeed, as we show in Corollary~\ref{cor:hard_label}, the problem of computing a utility-maximizing tuple of agents' actions that are IC under a given contract $p \in \mathbb{R}_+^{n \times m}$ is \NPHard.
%
%
%
%
%
%
\end{remark}
We circumvent the issue of classical tie-breaking rules highlighted in Remark~\ref{rem:tie_break} by slightly abusing terminology and extending the notion of contract to also include action recommendations for the agents.
Formally, we identify a contract with a pair $(p,\avec^\ast)$, where $p \in \mathbb{R}_+^{n \times m}$ defines the payments and $\avec^\ast \in \bigtimes_{i \in N} A_i^\ast(p)$ specifies a tuple of agents' actions, which should be interpreted as action recommendations suggested by the principal to the agents.
Given that the actions in $\avec^\ast$ are IC under $p$, we assume w.l.o.g. that the agents stick to such recommendations.

In conclusion, the principal's optimization problem reads as follows:
\begin{definition}[Principal's Optimization Problem]\label{def:princ_problem}
	Given an instance $(N,\Omega,A)$ of principal-multi-agent problem, compute an \emph{optimal} contract $(p,\avec^\ast)$---with $p \in \mathbb{R}^{n \times m}$ and $\avec^\ast \in \bigtimes_{i \in N}  A_i^\ast(p)$---, which is defined as a pair $(p,\avec^\ast)$ maximizing the principal's expected utility:
	\[
		R_{\avec^\ast} - \sum_{i \in N} P_{i,a_i^\ast} = \sum_{\vomega \in \Omega^n} r_\vomega \prod_{i \in N} F_{i,a_i, \omega_i}  -\sum_{i \in N}\sum_{\omega \in \Omega} F_{i, a, \omega} \, p_{i,\omega}.
	\]
\end{definition}

\subsection{On the Representation of Principal's Rewards}

Representing principal's rewards in a principal-multi-agent problem becomes unfeasible when there are many agents, since the number of possible tuples of outcomes grows as $m^n$.
%
Thus, we work with a succinct representation of principal's rewards, which we formally introduce in the following.
We remark that, with arbitrary rewards, an optimal contract can be found in time polynomial in the instance size (\emph{i.e.}, in time depending polynomially on $m^n$), as we show in Section~\ref{sec:matroids_reduction}.
%
%

We say that a principal-multi-agent problem instance $(N,A,\Omega)$ has \emph{succinct rewards} if:
\begin{enumerate}
	\item outcomes can be represented as non-negative $q$-dimensional vectors with $q \in \mathbb{N}_{>0}$ representing the dimensionality of the outcome space, namely $\Omega$ is a finite subset of $ \mathbb{R}_+^q$;
	\item the principal's rewards can be expressed by means of a reward function $g: \mathbb{R}_+^{nq} \to \mathbb{R}$ such that $r_\vomega = g(\vomega)$ holds for every tuple of outcomes $\vomega \in \Omega^n$, and, thus, we can also write $R_{\va} = \sum_{\vomega \in \Omega^n} g(\vomega) \prod_{i \in N} F_{i,a_i, \omega_i}  $ for every $\va \in A^n$.\footnote{Notice that, since we assume that $r_\vomega \in [0,1]$ for all $\vomega \in \Omega^n$, while the function $g$ is allowed to take any real value over its domain $\mathbb{R}_+^{nq}$, it has to hold that $g(\vomega) \in [0,1]$ for all $\vomega \in \Omega^n$.}
\end{enumerate}
%
%
%
Let us remark that, for ease of presentation and overloading notation, we denote tuples of outcomes as vectors, namely $\vomega \in  \Omega^n \subseteq \mathbb{R}_+^{nq}$, where we let $\omega_{i,j}$ be the $j$-th component of the vector that identifies the outcome achieved by agent $i$, for all $i \in N$ and $j \in [q]$.\footnote{In this paper, given a positive natural number $x \in \mathbb{N}_{>0}$, we let $[x] \coloneqq \{ 1,\ldots,x \}$ be the set of the first $x$ natural numbers.}
Moreover, in the following, we assume that $g: \mathbb{R}_+^{nq} \to \mathbb{R}$ can be accessed through an oracle that, given $\vomega \in  \Omega^n$, outputs $g(\vomega)$.\footnote{In this paper, for ease of exposition, we assume that the value of $R_{\va}$ for any given $\va \in A^n$ can be computed in polynomial time, without enumerating tuples of outcomes. The value of $R_{\va}$ can be approximated up to any arbitrarily small error with high probability by sampling each $\omega_i$ independently from $F_{i,a}$, and evaluating $g(\vomega)$. All the results in the paper can be easily extended to also account for this additional (arbitrarily small) approximation.}

In this work, we make the following common assumption on principal's rewards:
\begin{assumption}[Increasing rewards]\label{ass:inc_rew}
	The principal's reward function $g: \mathbb{R}_+^{nq} \to \mathbb{R}$ is increasing; formally, it holds that $g(\vomega) \geq g(\vomega')$ for all $\vomega, \vomega' \in \mathbb{R}_+^{nq} : \vomega \geq \vomega'$.
\end{assumption} 
Moreover, we will focus on two particular classes of reward functions, which, as we show next, enjoy some useful properties and are met in many real-world settings.
%
%
%
%
%
%
\begin{definition}[DR-submodularity and IR-supermodularity]\label{def:dr_ir}
	A reward function $g: \mathbb{R}_+^{nq} \to \mathbb{R}$ is \emph{diminishing-return submodular (DR-submodular)} if, for all $\vomega, \vomega', \vomega'' \in \mathbb{R}_+^{nq}: \vomega\le \vomega'$, it holds
	\[
	g(\vomega +\vomega'')- g(\vomega) \geq g(\vomega' +\vomega'')- g(\vomega') .
	\]
	Moreover, a reward function $g: \mathbb{R}_+^{nq} \to \mathbb{R}$ is \emph{increasing-return supermodular (IR-supermodular)} if its opposite function $-g$ is DR-submodular.
	%
	%
\end{definition}

Let us remark that, when the reward function $g$ is continuously differentiable, then the property that characterizes DR-submodular functions has a more intuitive interpretation.
Indeed, as shown by~\citet{Bian2017Guaranteed}, when $g$ is continuously differentiable, $g$ is DR-submodular if and only if: 
\[
	\nabla g(\vomega) \geq \nabla g(\vomega') \quad \forall \vomega, \vomega' \in \mathbb{R}_+^{nq}: \vomega \geq \vomega'.
\] 
Intuitively, this means that, if a tuple of outcomes $\vomega'$ dominates component-wise another tuple $\vomega$, then in $\vomega'$ the reward function grows slower than in $\vomega$ along all of its components.
This property is satisfied in many real-world scenarios, as we show in the following specific example.

\begin{example}[Selling multiple products]\label{ex:sales_multi}
	Consider a principal-agent problem modeling the interaction between a firm and a salesperson (the example can be easily generalized to the case of multiple salespersons).
	The firm wants to sell $q \in \mathbb{N}_{>0}$ different products, and the salesperson can sell a variable quantity of each product, depending on the level of effort put in selling each of them.
	Thus, the outcome achieved by the salesperson can be encoded by a vector $\vomega \in \mathbb{R}_+^{1q}$ whose $j$-th component $\omega_{1,j}$ represents the quantity of product $j$ being sold.
	In such a setting, a DR-submodular reward function $g$ models scenarios in which the firm is subject to diseconomies of scale, and, thus, the marginal return of each unit of product sold decreases as the quantity sold increases.
	This may be due to the fact that, \emph{e.g.}, the firm has to sustain much higher operational costs in order increase its selling capacity.
	On the other hand, an IR-supermodular reward function $g$ models cases in which there are economies of scale, and, thus, the marginal return of each unit of product sold increases with quantity (since, \emph{e.g.}, the fixed costs are more efficiently covered).
	%
	%
\end{example}

%% file: matroids.tex
\section{Reducing Principal--Multi--Agent Problems to Matroids}\label{sec:matroids}

In this section, we show that computing an optimal contract in principal-multi-agent problems can be reduced in polynomial time to a maximization problem defined over a special class of matroids.
First, in Section~\ref{sec:matroids_prelim}, we introduce some preliminary definitions on matroids and optimization problems over matroids.
Then, in Section~\ref{sec:matroids_reduction}, we provide the reduction.

We conclude the section with Section~\ref{sec:matroids_additional}, in which we provide a preliminary technical result for the problem of maximizing functions defined over $1$-partition matroids and satisfying a particular (stronger) notion of supermodularity.
This result will be useful in the following Section~\ref{sec:ir_super}.

\subsection{Preliminaries on Matroids}\label{sec:matroids_prelim}

A matroid $\M \coloneqq (\cG,\I)$ is defined by a finite ground set $\cG$ and a collection $\I$ of independent sets, which are subsets of $\cG$ satisfying some characteristic properties, namely:
\begin{enumerate}
	\item the empty set is independent, \emph{i.e.}, $\varnothing \in \I$;
	\item every subset of an independent set is independent, \emph{i.e.}, for $S' \subseteq S \subseteq \cG$, if $S \in \I$ then $S' \in \I$;
	\item if $S \in \I$ and $S' \in \I$ are two independent sets such that $S$ has more elements than $S'$, \emph{i.e.}, $|S| > |S'|$, then there exists an element $x \in S \setminus S'$ such that $S' \cup\{ x \} \in \I$.  
\end{enumerate}
Any subset $S \subseteq \cG$ such that $S \notin \I$ is said to be \emph{dependent}.
The \emph{bases} of the matroid $\M$ are all the maximal independent sets of $\M$, where an independent set is said to be maximal if it becomes dependent by adding any element of $\cG$ to it.   
We denote by $\B(\M) \subset 2^{\cG}$ the set of the bases of $\M$.
We refer the reader to~\citep{schrijver2003combinatorial} for a detailed treatment of matroids.

In the following, we will also consider optimization problems defined over matroids.
In particular, given a set function $f: 2^{\cG} \to \mathbb{R}$ assigning a value to each subset of the ground set, the associated maximization problem over a matroid $\M \coloneqq (\cG,\I)$ is defined as $\max_{S \in \I} f(S)$.

\subsection{Reduction to Matroid Optimization}\label{sec:matroids_reduction}

In order to provide our reduction, we need to introduce the following class of matroids:
\begin{definition}[$1$-Partition Matroid]\label{def:unit_part_matr}
	A matroid $\M \coloneqq (\cG, \I)$ is a \emph{$1$-partition matroid} if there exists $d \in \mathbb{N}_+$ subsets $\cG_i \subseteq \cG$ of ground elements such that:
	\begin{enumerate}
		\item $\cG = \bigcup_{i \in [d]} \cG_i$ and $\cG_i \cap \cG_j =\varnothing$ for all $i, j \in [d]: i \neq j$;
		\item $\I = \left\{ S \subseteq \cG \, : \, | S \cap \cG_i | \leq 1 \,\, \forall i \in [d] \right\}$.
	\end{enumerate} 
	%
	%
\end{definition}
Intuitively, in a $1$-partition matroid, the ground set $\cG$ is partitioned into $d$ disjoint subsets $\cG_i$, and the independent sets are all and only the subsets of $\cG$ that contain \emph{at most} one element of each subset $\cG_i$.
In the following, we denote by $\M \coloneqq ( \left\{ \cG_i \right\}_{i \in [d]}, \I )$ a $1$-partition matroid with $\cG \coloneqq \bigcup_{i \in [d]} \cG_i$, and, for ease of notation, we let $k_i \coloneqq |\cG_i|$ for every $i \in [d]$.
Notice that, as it is immediate to check, the set $\B(\M)$ of the bases of a $1$-partition matroid $\M \coloneqq ( \left\{ \cG_i \right\}_{i \in [d]}, \I )$ is made by all the subsets of $\cG$ containing exactly one element for each subset $\cG_i$.

Next, we show how the problem of computing an optimal contract in principal-multi-agent problems can be reduced in polynomial time to a maximization problem defined over a suitably-constructed $1$-partition matroid, which is formally defined as follows:

\begin{definition}[Mapping from principal-multi-agent problems to $1$-partition matroids] \label{def:mapping}
	Given an instance of principal-multi-agent problem, say $I \coloneqq (N, \Omega, A)$, we define its corresponding $1$-partition matroid $\M^I \coloneqq ( \left\{ \cG_i^I \right\}_{i \in N}, \I^I )$ as follows:
	\begin{enumerate}
		\item $\cG_i^I \coloneqq \left\{ (i,a) : a \in A \right\}$ for all $i \in N$, with $\cG^I \coloneqq \bigcup_{i \in [d]} \cG_i^I$;
		\item $\I^I \coloneqq \left\{ S \subseteq \cG^I \, : \, |S \cap \cG_i^I| \leq 1 \,\, \forall i \in [d] \right\}$.
	\end{enumerate}
	%
\end{definition}

It is immediate to check that $\M^I$ is indeed a $1$-partition matroid.
Moreover, its \emph{bases} correspond one-to-one to agents' action profiles $\avec \in A^n$.
In particular, an independent set $S \in \I^I$ of $\M^I$ assigns an action to each agent in $N_S \coloneqq \left\{ i \in N : |S \cap \cG^I_i| = 1  \right\}$; we denote by $a_{S,i} \in A$ the action associated to agent $i \in N_S$. 
Since a base of a $1$-partition matroid is any independent set $S \in \I^I$ containing one element for each $\cG_i$, it completely specifies an agents' action profile, which we denote by $\avec_S = (a_{S,i})_{i \in N}$.
For ease of presentation, in the following we overload notation and write $\avec_S = (a_{S,i})_{i \in N}$ also for independent sets $S \in \I^I$ that are \emph{not} bases, by letting all the unspecified actions be equal to the null one; formally, $a_{S,i} = a_\varnothing$ for all $i \in N \setminus N_S$.

The following theorem formalizes our reduction:
\begin{restatable}{theorem}{theoremMat}\label{thm:eq_matr}
	Given an instance $I \coloneqq (N, \Omega, A)$ of principal-multi-agent problem, the problem of computing a contract maximizing the principal's expected utility can be reduced in polynomial time to solving $\max_{S \in \I^I} f^I(S)$ over the $1$-partition matroid $\M^I = ( \left\{ \cG_i^I \right\}_{i \in N}, \I^I )$, where $f^I: 2^{\cG^I} \to \mathbb{R}$ is a set function such that, for every independent set $S \in \I^I$, it holds:
	\[
		f^I(S)\coloneqq R_{\avec_S} - \sum_{i\in N} \widehat P_{i,a_{S,i}}, \quad \text{where} \quad \widehat P_{i,a_{S,i}} = \min_{p \in \mP^{i, a_{S,i}} } \sum_{\omega \in \Omega} F_{i,a_{S,i},\omega} \, p_{i,\omega}.
	\]
\end{restatable}

Intuitively, $f^I(S)$ is equal to the maximum possible expected utility that the principal can get by means of contracts under which the actions in $\avec_S$ are IC and are those recommended by the principal to the agents.
The proof of Theorem~\ref{thm:eq_matr} relies on the following useful lemma, which shows that the optimal value of $f^I$ is always attained at a base of $\M^I$.
\begin{restatable}{lemma}{lemmaOptBase}\label{lem:opt_base}
	There always exists a base $S^* \in \B(\M^I)$ of $\M^I$ such that $f^I(S^*) = \max_{S \in \I^I} f^I(S)$.
\end{restatable}

Let us also remark that Lemma~\ref{lem:opt_base} and Theorem~\ref{thm:eq_matr} immediately provide a polynomial-time algorithm for finding an optimal contract in principal-multi-agent instances without succinct rewards.
Indeed, since the optimal value of $f^I$ is always attained at least one base of the matroid $\M^I$ (Lemma~\ref{lem:opt_base}), in order to find an optimal contract it is sufficient to enumerate all the bases of $\M^I$, which are $m^n$.
Without a succinct reward representation, the size of an instance of principal-multi-agent problem grows as $m^n$ (there is a reward value for each tuple of agents' outcomes), and, thus, the enumerative algorithm runs in time polynomial in the instance size.

\subsection{Preliminary Technical Results on $1$-Partition Matroids}\label{sec:matroids_additional}

%

We first introduce a particular class of set functions defined over $1$-partition matroids, which we call \emph{ordered-supermodular} functions.
In order to do this, we first need some additional notation.
Given a $1$-partition matroid $\M \coloneqq ( \left\{ \cG_i \right\}_{i \in [d]}, \I )$, for each $i \in [d]$ we introduce a bijective function $\pi_i : [k_i] \to \mathcal{G}_i$ to denote an ordering of the subset $\cG_i$ in which the elements are ordered from $\pi_i(1)$ to $\pi_i(k_i)$.
Given two independent sets $S, S' \in \I$ of the matroid, we denote by $S \wedge S'$ the partition-wise ``maximum'' of the two sets, \emph{i.e.}, the set made by an element $x \in \left( S \cup S' \right) \cap \cG_i$ with maximal value of $\pi_i^{-1}(x)$ for each partition $i \in [d]$ (notice that $\left( S \cup S' \right) \cap \cG_i$ contains at most one element of $\cG_i$ for each of the two sets $S$ and $S'$).
Analogously, we define $S \vee S'$ as the partition-wise ``minimum'' of the two sets.
Then, a set function is said ordered-supermodular if there exist some orderings of the sets $\cG_i$ such that the function satisfies the classical condition of supermodularity over the independent sets of the matroid, with the usual union and intersection operators replaced by the partition-wise ``maximum'' $\wedge$ and ``minimum'' $\vee$, respectively.
Formally:
\begin{definition}[Ordered-supermodular function]\label{def:ord_sup}
	A set function $f: 2^{\mathcal{G}} \to\mathbb{R}$ defined over a $1$-partition matroid $\mathcal{M} \coloneqq (\left\{ \mathcal{G}_i \right\}_{i \in [d]}, \mathcal{I})$ is said to be \emph{ordered-supermodular} if there exist bijective functions $\pi_i : [k_i] \to \mathcal{G}_i$ for $i \in [d]$ such that, for every pair of independent sets $S', S \in \I$:
	\[
		f(S \land S') + f(S \lor S') \ge f(S)+ f(S') .
	\] 
	%
	%
\end{definition}
Notice that, if one restricts the attention to independent sets $S', S \in \I$ such that $S \cup S' \in \I$, then the condition for ordered-supermodularity coincides with that for supermodularity.
Thus, intuitively, the former can be seen as a way of tightening the latter in order to also account for cases in which the union of independent sets is \emph{not} independent.

%

Finally, we show that the characteristic feature of ordered-supermodular functions allows us to reduce their optimization to solving maximization problems of supermodular functions defined over \emph{rings of sets}, which can be done in polynomial time~\citep{schrijver2000combinatorial,bach2019submodular}.\footnote{We recall that a ring of sets is a family of sets $\mathcal{R}$ that is closed under both union and intersection. Formally, given any two sets $S, S' \in \mathcal{R}$, it holds $S \cup S' \in \mathcal{R}$ and $S \cap S' \in \mathcal{R}$~\citep{birkhoff1937rings}.}
%
%
%
\begin{restatable}{theorem}{redToRIND}\label{thm:redToRInd}
	The problem of maximizing an ordered-supermodular function over a $1$-partition matroid can be reduced in polynomial time to maximizing a supermodular function over a ring of sets.
	%
\end{restatable}

\begin{restatable}{corollary}{polyOrderSup}\label{cor:polyOrderSup}
	The problem of maximizing an ordered-supermodular function over a $1$-partition matroid admits a polynomial-time algorithm.
\end{restatable}

%% file: ir_super.tex
\section{Principal-Multi-Agent Problems with IR-supermodular Rewards}\label{sec:ir_super}

In this section, we study principal-multi-agent problems with succinct rewards specified by IR-supermodular functions.
First, in Section~\ref{sec:hard_ir_super}, we prove that in such setting the problem of computing an optimal contract is inapproximable in polynomial time.
Then, in Section~\ref{sec:ir_super_regular} we show that, under mild assumptions, the problem can be solved in polynomial time.
%

\subsection{Inapproximability Result}\label{sec:hard_ir_super}

In order to prove the negative result, we provide a reduction from the \textsf{LABEL-COVER} problem, which consists in assigning labels to the vertexes of a bipartite graph in order to satisfy some given constraints that define which pairs of labels can be assigned to vertexes connected by an edge.
In particular, we consider the \emph{promise version} of the problem, in which, given an instance such that either there exists an assignment of labels satisfying at least a fraction $c$ of the constraints or all the possible assignments satisfy less than a fraction $s$ of them (with $s \leq c$), one has to establish which one of the two cases indeed holds.
Such a problem is known to be \NPHard~\citep{raz1998parallel,arora1998proof}.
We refer the reader to Appendix~\ref{app:hard_ir_super} for a formal definition of the problem.

Our inapproximability result formally reads as follows:
\begin{theorem}\label{thm:hard_label}
	For any constant $\rho > 0$, in principal-multi-agent problems with succinct rewards specified by an IR-supermodular function, it is \textnormal{\textsf{NP}}-hard to design a contract providing a $\rho$-approximation of the principal's expected utility in an optimal contract, even when both the number of outcomes $m$ and the number of agents' actions $\ell$ are fixed. 
	%
\end{theorem}

Indeed, the proof of Theorem~\ref{thm:hard_label} provides an even stronger hardness result. 
It also shows that it is \NPHard\ to find a tuple of agents' actions $\avec \in A^n$ that is ``approximately'' optimal for the principal under a given contract $p\in \mathbb{R}^{n \times m}_+$.
Formally, the following corollary holds:
\begin{corollary} \label{cor:hard_label}
	For any constant $\rho > 0$, in principal-multi-agent problems with succinct rewards specified by an IR-supermodular function, it is \textnormal{\textsf{NP}}-hard to compute a $\rho$-approximate solution to the problem of finding the best (for the principal) tuple of IC agents' actions $\avec \in \bigtimes_{i \in N} A_i^*(p)$ for a given contract $p\in \mathbb{R}^{n \times m}_+$, even when both the number of outcomes $m$ and that of agents' actions $\ell$ are fixed.
	%
\end{corollary}

Corollary~\ref{cor:hard_label} is readily proved by noticing that the proof of Theorem~\ref{thm:hard_label} continues to hold even if we restrict it to the null contract in which all the payments are zero.
%

\subsection{A Polynomial-time Algorithm for Instances Satisfying the FOSD Condition}\label{sec:ir_super_regular}

In the following, we show how to circumvent the negative result established by Theorem~\ref{thm:hard_label}.
In particular, we prove that, in principal-multi-agent problems with succinct rewards specified by an IR-supermodular function, under some mild additional assumptions the problem of computing an optimal contract can indeed be solved in polynomial time.

We consider instances satisfying a particular \emph{first-order stochastic dominance} (FOSD) condition, which is similar to several properties that are commonly studied in the contract theory literature (see, \emph{e.g.},~\citep{tadelis2005lectures}).
Moreover, such a condition is reasonably satisfied in many real-world settings.
Intuitively, it states that the higher the cost of an agent's action, the bigger the probability with which such an action induces ``good'' outcomes.
%
For instance, in salesperson problems with multiple products (Example~\ref{ex:sales_multi}), such a condition is always satisfied, since outcome vectors represent the quantity of each product being sold and action costs encode the effort levels undertaken by the agents.
Naturally, the salesperson undertaking an higher level of effort in selling products will result in a bigger probability of generating large volumes of sales.

In order to formally define the FOSD condition, we first need to introduce some additional notation.
Given a subset of outcomes $\Omega' \subseteq \Omega$, we say that $\Omega'$ is \emph{comprehensive} whenever, for every $\omega \in \Omega'$ and $\omega' \in \Omega$, if $\omega' \leq \omega$ then $\omega' \in \Omega'$.
Moreover, for ease of presentation, with a slight abuse of notation and w.l.o.g. we assume that the actions of each agent $i \in N$ are re-labeled so that $A = \left\{ a_1,\ldots,a_{\ell} \right\}$ with $c_{i, a_j} \leq c_{i, a_{j+1}}$ for every $j \in [\ell-1]$.
Then, we have the following definition:
%
%
%
%
\begin{definition}[First order stochastic dominance]\label{def:fosd}
	An instance of principal-multi-agent problem is said to satisfy the \emph{first-order stochastic dominance} (FOSD) condition if, for every agent $i \in N$ and action index $j \in [\ell-1]$, the following holds for all the comprehensive sets $\Omega' \subseteq \Omega$:
	\[   
		\sum_{\omega \in \Omega'} F_{i,a_{j+1},\omega} \leq  \sum_{\omega \in \Omega'} F_{i,a_{j},\omega}.
	\]
	%
\end{definition}

\begin{remark}
	A condition similar to Definition~\ref{def:fosd}, called \emph{monotone likelihood ratio property (MLRP)}, has been considered by~\citet{dutting2019simple} limited to the case in which outcomes are identified by scalar values.
	In such settings, the MLRP is strictly stronger than the FOSD condition.
	Our definition of FOSD generalizes the classical FOSD condition (see, \emph{e.g.},~\citep{tadelis2005lectures}) from settings in which the outcomes are scalar values to those where they are vector values.
\end{remark}
%
%
%
%

Next, we prove how to design a polynomial-time algorithm for the problem of finding an optimal contract by exploiting the FOSD condition.
Intuitively, the idea of the proof is the following.
First, thanks to Theorem~\ref{thm:eq_matr}, we can reduce in polynomial time an instance $I \coloneqq (N, A, \Omega)$ of the principal-multi-agent problem to the optimization of a suitably-defined set function $f^I$ over a $1$-partition matroid $\M^I$ (see Theorem~\ref{thm:eq_matr} and Definition~\ref{def:mapping} for the definition of $f^I$ and $\M^I$, respectively).
Moreover, by Corollary~\ref{cor:polyOrderSup}, if $f^I$ is ordered-supermodular we can solve the optimization problem in polynomial time.
Hence, in order to prove the result, we simply need to show that, whenever the FOSD condition is satisfied, the function $f^I$ is indeed ordered-supermodular.

First, we prove the following preliminary result which follows from~\citep{OSTERDAL20101222}.
\begin{restatable}{lemma}{dominated}\label{lm:dominated}
	In principal-multi-agent problems with succinct rewards that satisfy the FOSD condition, for every agent $i \in N$ and pair $a_j, a_k \in A$ of agent $i$'s actions such that $j <k$, there exists a collection of probability distributions $\mu^{\omega} \in \Delta_{\Omega^-}$, one per outcome $\omega \in \Omega$, which are supported on the finite subset of the positive orthant $\Omega^- \coloneqq \mathbb{R}_+^m \cap \left\{ \omega - \omega' \mid \omega,\omega' \in \Omega \right\}$ and satisfy the following equations:
	\[ 
		F_{i, a_k,\omega}  = \sum_{\omega' \in \Omega} F_{i, a_j,\omega'} \, \mu^{\omega'}_{\omega-\omega'} \quad \forall \omega \in \Omega.
	\]
	%
\end{restatable}

Given Lemma~\ref{lm:dominated}, we are ready to show that, if the instance $I$ meets the FOSD condition, then its corresponding set function $f^I$ is indeed ordered-supermodular over the $1$-partition matroid $\M^I$.
\begin{restatable}{lemma}{orderSup} \label{lm:orderSup}
	Given an instance $I \coloneqq (N,\Omega,A)$ of principal-multi-agent problem that (i) has succinct rewards specified by an IR-supermodular function and (ii) satisfies the FOSD condition, the set function $f^I$ defined over the $1$-partition matroid $\M^I = ( \left\{ \cG_i^I \right\}_{i \in N}, \I^I )$ is ordered-supermodular.
	%
\end{restatable}

Finally, Lemma~\ref{lm:orderSup} allows us to prove the main positive result of this section:
\begin{restatable}{theorem}{IR}\label{thm:IR}
	For principal-multi-agent problem instances that (i) have succinct rewards specified by an IR-supermodular function and (ii) satisfy the FOSD condition, the problem of computing an optimal contract admits a polynomial-time algorithm.
	%
\end{restatable}

%% file: dr_sub.tex
\section{Principal-Multi-Agent Problems with DR-submodular Rewards}\label{sec:dr_sub}

In this section, we switch the attention to principal-multi-agent problems with succinct rewards specified by DR-submodular functions.
First, similarly to the case of IR-supermodular reward functions, we provide a strong negative result.
In particular, we show that the problem of computing an optimal contract cannot be approximated up to within any constant factor, even when either the number of actions or the dimensionality of outcome vectors is fixed.

In order to prove the negative result, we provide a reduction from the \emph{promise version} of the well-known $\textsf{INDEPENDENT-SET}$ problem.
In such a version of the problem, one is given an undirected graph $G \coloneqq (V,E)$ such that either there exists an independent set of size at least $|V|^{1-\alpha}$---for some $\alpha > 0$---or all the independent sets have size at most $|V|^\alpha$, and is asked to decide which one of the two cases holds.   
Such a problem is known to be \NPHard~for any $\alpha  > 0$~\citep{hastad1999clique,Zuckerman2007linear}.
This is exploited by our reduction in order to prove Theorem~\ref{thm:hard_is}.
The reader can find more details on the definition of the promise version of $\textsf{INDEPENDENT-SET}$ in Appendix~\ref{app:hard_dr_sub}.
\begin{theorem}\label{thm:hard_is}
	For any constant $\alpha > 0$, in principal-multi-agent problems with succinct rewards specified by a DR-submodular function, it is \NPHard~to design a contract providing an $n^{1-\alpha}$--approximation of the principal's expected utility in an optimal contract, even when both the number of agents' actions $\ell$ and the dimensionality $q$ of outcome vectors are fixed.
	%
\end{theorem}

Next, we complement the inapproximability result in Theorem~\ref{thm:hard_is} by providing a polynomial-time approximation algorithm for the problem.
In order to do so, we exploit the fact that, in settings with succinct rewards specified by DR-submodular functions, the set function $f^I$ constructed in Theorem~\ref{thm:eq_matr} is always a submodular function over the $1$-partition matroid $\M^I$.
However, this is \emph{not} sufficient, since such a function is \emph{non}-monotone and \emph{non}-positive, and, thus, we need to deploy some non-standard tools in order to come up with a polynomial-time approximation algorithm.

As a first step, given an instance $I \coloneqq (N,\Omega,A)$ of principal-multi-agent problem, we extend the definition of the function $f^I$ to all the subsets of $\cG^I$ (notice that Theorem~\ref{thm:eq_matr} provides a value of $f^I$ only for the independent sets $\I$).
To do this, we first need to introduce some additional notation.

For ease of presentation, in the rest of this section we will make the following w.l.o.g. assumption:
\begin{assumption}[Null outcome]\label{ass:omega_null}
	There exists an outcome $\omega_{\varnothing} \in \Omega$ such that $\omega_{\varnothing}  = \mathbf{0} \in \mathbb{R}^q$ and, for every agent $i \in N$, it holds that $F_{i,a_{\varnothing},\omega_{\varnothing}} = 1$ and $F_{i,a,\omega_{\varnothing}} = 0$ for all $a \in A \setminus \{ a_\varnothing \}$.
\end{assumption}
Then, by slightly abusing notation, given any $S\subseteq \cG^I$ we let $F_{i,S} \coloneqq \sum_{(i,a)\in S \cup\{(i,a_\varnothing) \mid i\in N\}} F_{i,a}$ be the probability distribution of the sum of independent random variables distributed as $F_{i,a}$, one for each pair $(i,a)$ in $S \cup\{(i,a_\varnothing) \mid i\in N\}$.
Notice that the probability distributions defined above are no longer supported on the set of outcomes $\Omega$, but rather on the set of all the possible vectors in $\mathbb{R}_+^q$ that can be obtained as the sum of at most $n \ell$ (possibly repeated) vectors in $\Omega$.
We denote such a set by $\tilde \Omega \subseteq \mathbb{R}_+^q$, and let $F_{i,S,\omega}$ be the probability that $F_{i,S}$ assigns to $\omega \in \tilde \Omega$.
Moreover, we let $\tilde \Omega^n \coloneqq \bigtimes_{i \in N} \tilde \Omega$.
Finally, we overload notation and let $R_{S} \coloneqq \sum_{\vomega \in \tilde \Omega^n} r_{\vomega} \prod_{i \in N} F_{i,S,\omega_i}$ for any $S\subseteq \cG^I$.
Notice that, since any independent set $S \in \I^I$ includes at most one pair $(i,a)$ for each agent $i \in N$, it is easy to check that $R_{S} = R_{\va_S}$ (see Section~\ref{sec:ir_super} for the definition of $R_{\va_S}$).
%

We are now ready to provide the formal definition of the extension of $f^I$:
\begin{definition}[Extension of $f^I$]\label{def:ext_f}
	Given an instance $I\coloneqq (N,\Omega,A)$ of principal-multi-agent problem, the extension of $f^I$ to all the subsets of $\cG^I$ is such that, for every $S\subseteq \cG^I$:
	\[
		f^I(S)\coloneqq R_{S} - \sum_{(i,a)\in S} \widehat P_{i,a}, \quad \text{where} \quad \widehat P_{i,a} \coloneqq \min_{p \in \mP^{i, a }} \sum_{\omega \in \Omega} F_{i,a,\omega} \, p_{i,\omega}.
	\]
	%
	%
\end{definition}

The crucial result that we need in order to design a polynomial-time approximation algorithm is the following Lemma~\ref{lem:sub_dr}, which shows that the extended function $f^I$ can be decomposed as the sum of a monotone-increasing submodular function and a linear one.
Formally:
\begin{restatable}{lemma}{subdr}\label{lem:sub_dr}
	Given an instance $I \coloneqq (N, \Omega, A)$ of principal-multi-agent problem with succinct rewards specified by a DR-submodular function, the extended set function $f^I$ (see Definition~\ref{def:ext_f}) can be defined as $f^I(S) \coloneqq \mathsf{f}^I(S)+\mathsf{l}^I(S)$ for every $S \subseteq \cG^I$, where $\mathsf{f}^I:2^{\cG^I}\rightarrow \mathbb{R}_+$ is a monotone-increasing submodular function and $\mathsf{l}^I:2^{\cG^I}\rightarrow \mathbb{R}$ is a linear function, both defined over the $1$-partition matroid $\M^I$.
	%
\end{restatable}

Lemma~\ref{lem:sub_dr} allows us to apply a result by~\citet{sviridenko2017optimal}, who provide a polynomial-time approximation algorithm for the problem of optimizing the sum of a monotone-increasing submodular function and a linear one over a matroid.
%
%
%
%
%
%
This immediately gives the following result:
%
%

\begin{restatable}{theorem}{drdue}\label{thm:DR}
	In principal-multi-agent problems with succinct rewards specified by a DR-submodular function, the problem of computing an optimal contract admits a polynomial-time approximation algorithm that, for any $\epsilon>0$ given as input, outputs a contract with principal's expected utility at least $ (1-1/e) R_{(p,\avec^*)} - P_{(p,\avec^*)}-\epsilon$ for any contract $(p,\avec^*)$ with high probability, where $R_{(p,\avec^*)} \in [0,1]$, respectively $P_{(p,\avec^*)} \in \mathbb{R}_+$, denotes the expected reward, respectively payment, under $(p,\avec^*)$.
	%
	%
\end{restatable}
%

%% file: unknown.tex
\section{Bayesian Principal-multi-agent Problems}\label{sec:bayesian}

In this last section, we study \emph{Bayesian} principal-multi-agent problems in which each agent has a \emph{private type} determining their action costs and distributions over outcomes.
In particular, we extend the Bayesian model recently introduced by~\citet{Castiglioni2022Designing} to multi-agent settings.
%

First, in Section~\ref{sec:bayesian_model} we formally introduce {Bayesian} principal-multi-agent problems and all their related concepts.
Then, Section~\ref{sec:bayesian_formulation} provides a formulation of the computational problem that the principal has to solve in Bayesian settings.
Next, in Section~\ref{sec:bayesian_lp} we show how such a problem can be ``approximately formulated'' as an LP with exponentially-many variables and polynomially-many constraints.
Finally, in Section~\ref{sec:bayesian_algo} we exploit such a formulation to design a polynomial-time approximation algorithm for the problem, based on an \emph{ad hoc} implementation of the \emph{ellipsoid method} that uses an \emph{approximate separation oracle} that can be implemented in polynomial time in settings having the same properties as those in which we derived our positive results in Sections~\ref{sec:ir_super}~and~\ref{sec:dr_sub}.
%

\subsection{The Model}\label{sec:bayesian_model}

An instance of the \emph{Bayesian principal-multi-agent problem} is characterized by a tuple $(N,\Theta,\Omega,A)$, where $N$, $\Omega$, and $A$ are defined as in non-Bayesian instances, while $\Theta$ is a finite set of agents' types.\footnote{For ease of exposition, all agents share the same set $\Theta$. Our results can be easily extended to the case of agent-specific sets.}
We denote by $\vtheta \in \Theta^n \coloneqq \bigtimes_{i \in N} \Theta$ a tuple of agents' types, whose $i$-th component $\theta_i$ represents the type of agent $i$.
We assume that agents' types are jointly determined according to a probability distribution $\lambda \in \Delta_{\Theta^n}$ supported on a subset $\text{supp}(\lambda) \subseteq \Theta^n$ of tuples of agents' types---with $\lambda_{\vtheta}$ being the probability assigned to $\vtheta \in \text{supp}(\lambda)$---, and that such a distribution is commonly known to the principal and all the agents.\footnote{Let us remark that, as it is the case for action costs and distributions over outcomes, as well as rewards, the probabilities defining the distribution $\lambda$ are part of the representation of a Bayesian principal-multi-agent problem instance, and, thus, they are part of the input to the principal's optimization problem. Hence, the running time of any polynomial-time algorithm for such a problem must depend polynomially on the size of $\text{supp}(\lambda)$. It is crucial that only probabilities $\lambda_{\vtheta}$ corresponding to tuples of agents' types $\vtheta \in \text{supp}(\lambda)$ in the support of $\lambda$ are specified as input, otherwise the size of the input representation would always be exponential in $n$, rendering the task of designing polynomial-time algorithms straightforward.}
Action costs and distributions over outcomes are extended so that they also depend on the agent's type; formally, they are denoted as $F_{i,\theta,a}$ and $c_{i,\theta,a}$, where $\theta \in \Theta$ is the type of agent $i \in N$.
Similarly, we extend the definition of expected reward, denoted as $R_{\vtheta,\va}$.
Moreover, w.l.o.g., we modify Assumption~\ref{ass:null_action} so that the null action $a_\varnothing$ now satisfies $c_{i,\theta, a_\varnothing} = 0$ for all $i \in N$ and $\theta \in \Theta$.
Finally, for an agent $i \in N$ of type $\theta \in \Theta$, we define $A^\ast_{i,\theta}(p) \subseteq A$ as the set of actions that are IC under a given contract $p \in \mathbb{R}_+^{n \times m}$, while $\mP^{i,\theta, a} \subseteq \mathbb{R}_+^{n \times m}$ denotes the set of contracts under which a given action $a \in A$ is IC.

Following the line of~\citet{Castiglioni2022Designing}, we consider the case in which the principal commits to a \emph{menu of randomized contracts}.
In our multi-agent setting, a randomized contract is defined as a probability distribution $\gamma$ supported on $\mathbb{R}_+^{n \times m}$.
Then, a menu consists in a collection $\Gamma = (\gamma^{\vtheta})_{\vtheta \in \Theta^n}$ containing a randomized contract $\gamma^{\vtheta}$ for each possible tuple of agents' types $\vtheta \in \Theta^n$.
%
%

The interaction between the principal and agents having types specified by $\vtheta \sim \lambda$ goes as follows:
\begin{enumerate}
	\item the principal commits to a menu of randomized contracts $\Gamma = (\gamma^{\vtheta})_{\vtheta \in \Theta^n}$;
	%
	%
	\item each agent $i \in N$ reports a type $\hat \theta_i \in \Theta$ to the principal (possibly different from their type $\theta_i$);
	\item the principal draws a contract $p \sim \gamma^{\widehat{\vtheta}}$, where $\widehat \vtheta \in \Theta^n$ denotes the tuple of agents' types whose $i$-th component is the type $\hat \theta_i$ reported by agent $i$;
	\item each agent $i \in N$ plays an IC action $a_i \in A_{i,\theta_i}^\ast(p)$ according to their true type $\theta_i$, resulting in a tuple of agents' actions $\avec \in \bigtimes_{i \in N} A_{i,\theta_i}^\ast(p)$.  
\end{enumerate}

As discussed in Section~\ref{sec:preliminaries_problem} (see Remark~\ref{rem:tie_break}), in our multi-agent setting a contract does \emph{not} only need to specify payments, but also action recommendations for the agents.
Thus, in the rest of this section, whenever we refer to a contract $p \in \mathbb{R}_+^{n \times m}$ belonging to the support of a randomized contract $\gamma^{\vtheta}$ for $\vtheta \in \Theta^n$, we always assume that it is paired with a tuple $\avec^\ast \in \bigtimes_{i \in N} A_{i,\theta_i}^\ast(p)$ of IC (for the types specified by $\vtheta$) action recommendations for the agents.
%

In a Bayesian setting, the goal of the principal is to commit to an \emph{optimal} menu of randomized contracts, which is one maximizing their expected utility, which is obtained by extending the non-Bayesian expression in Definition~\ref{def:princ_problem} to also account for the expectation with respect to the distribution $\lambda$ of agents' types and the distributions $\gamma^{\vtheta}$ defining the randomized contracts in the menu (see Objective~\eqref{eq:obj} below for a formal mathematical formula).   
%

As in single-agent settings~\citep{Castiglioni2022Designing}, it is possible to focus w.l.o.g. on menus of randomized contracts that are \emph{dominant-strategy incentive compatible} (DSIC).\footnote{It is easy to show that focusing on DSIC menus of randomized contracts is w.l.o.g. by using a revelation-principle-style argument. See the book by~\citet{shoham2008multiagent} for some examples of these kinds of arguments.}
These are menus such that the agents are always incentivized to truthfully report their type to the principal, no matter the types reported by others (see Constraints~\eqref{eq:linearize} for a formalization of the DSIC conditions).

\subsection{Formulating the Principal's Optimization Problem}\label{sec:bayesian_formulation}

Next, we show how to formulate the problem of computing an optimal DSIC menu of randomized contacts in Bayesian principal-multi-agent problems.
The formulation that we propose in the following is specifically tailored so as to ease the design of our approximation algorithm.

As a first step, we show that we can focus w.l.o.g. on randomized contracts $\gamma^{\vtheta}$ having a finite support $\text{supp}(\gamma^{\vtheta}) \subseteq \mathbb{R}_+^{n \times m}$. 
Such a result is already known for single-agent settings (see Lemma~1 in~\citep{Castiglioni2022Designing} and Theorem~1 in~\citep{gan2022optimal}), but it can be easily generalized to our multi-agent problems.
In particular, since in our model there are no externalities among the agents, it is immediate to adapt the results of~\citet{Castiglioni2022Designing}~and~\citet{gan2022optimal} in order to show that there always exists an optimal DSIC menu of randomized contracts such that, for every agent $i \in N$ and tuple of agents' types $\vtheta \in \Theta^n$, the contracts in the support $\text{supp}(\gamma^{\vtheta})$~of~$\gamma^{\vtheta}$ specify at most one different agent $i$'s payment scheme for each action $a \in A$.
Moreover, such agent $i$'s payment scheme is such that action $a$ is IC when the type of agent $i$ is $\theta_i$.
Formally:
\begin{restatable}{lemma}{smallSup}\label{lem:small_supp}
	Given an instance $I \coloneqq (N,\Theta,\Omega,A)$ of Bayesian principal-multi-agent problem and a DSIC menu of randomized contracts, there always exists another DSIC menu of randomized contracts $\Gamma = (\gamma^{\vtheta})_{\vtheta \in \Theta^n}$ with at least the same principal's expected utility such that, for every $i \in N$ and $\vtheta \in \Theta^n$, it holds that $\left| \left\{  p_i \, \mid \, p \in \textnormal{supp}(\gamma^{\vtheta}) \wedge p \in \mP^{i,\theta_i,a}  \right\} \right| \leq 1$ for all $a \in A$, where $p_i \in \mathbb{R}_+^m$ denotes the $i$-th row of matrix $p$ (i.e., the agent $i$'s payment scheme under contract $p$).
\end{restatable}

Lemma~\ref{lem:small_supp} allows us to identify the contracts in the support $\text{supp}(\gamma^{\vtheta})$ of $\gamma^{\vtheta}$ with their corresponding tuples of action recommendations for the agents, since there could be at most one different contract for each one of such tuples.
%
%
%
%
%
%
Thus, in order to characterize the elements defining a menu of randomized contracts which are needed for our purposes, it is sufficient to specify:
\begin{itemize}
	\item for every tuple of agents' types $\vtheta \in \Theta^n$ and tuple of agents' actions $\avec \in A^n$, the probability $t_{\vtheta,\avec} \in [0,1]$ that the randomized contract $\gamma^{\vtheta}$ places on the contract whose corresponding action recommendations for the agents are specified by $\avec$;
	\item for every agent $i \in N$, tuple of agents' types $\vtheta \in \Theta^n$, and action $a \in A$, the probability $\xi_{i,\vtheta,a} \in [0,1]$ with which agent $i$ is recommended to play action $a$ after the agents collectively reported the types specified by $\vtheta$ to the principal;
	\item for every agent $i \in N$, tuple of agents' types $\vtheta \in \Theta^n$, action $a \in A$, and outcome $\omega \in \Omega$, the payment $p_{i, \vtheta, a, \omega} \geq 0$ from the principal to agent $i$ when the agents reported the types in $\vtheta$ to the principal, agent $i$ is recommended action $a$, and the realized outcome is $\omega$.
\end{itemize}

We are now ready to provide our formulation of the problem of computing an optimal DSIC menu of randomized contracts in Bayesian principal-multi-agent problems.
Before doing that, for ease of presentation, we introduce some additional notation.
In particular, we let $\tilde \Theta^n \coloneqq \text{supp}(\lambda)$ be the set of tuples of agents' types that could be possibly reported to the principal if the agents truthfully reveal their types. 
Moreover, given any $\vtheta \in \Theta^n$, we let $\vtheta_{-i}$ be the tuple obtained by dropping agent $i$'s type $\theta_i$ from $\vtheta$.
Then, given a type $\theta \in \Theta$, we write $(\theta, \vtheta_{-i})$ to denote the tuple obtained by adding $\theta$ to $\vtheta_{-i}$ as agent $i$'s type, so that $\vtheta = (\theta_i, \vtheta_{-i})$. 
Finally, for every agent $i \in N$, we denote by $\tilde \Theta_{-i} \coloneqq \left\{ \vtheta_{-i} \, : \, \vtheta \in \tilde \Theta^n \right\}$ the set of all tuples of types that could be possibly reported to the principal by agents other than $i$, assuming that they truthfully reveal their types.\footnote{Using the sets $\tilde \Theta^n$ and $\tilde \Theta^n_{-i}$ to index the variables appearing in Problem~\eqref{pr:types} is crucial in order to guarantee that the number of variables and that of constraints defining the problem is polynomial in the size of $\text{supp}(\lambda)$. Indeed, indexing the variables over all the tuples of agents' types in $\Theta^n$ would lead to a number of variables and constraints exponential in $n$.}

We can now formulate the principal's optimization problem as follows:
\begin{subequations}\label{pr:types}
	\begin{align}
		\sup & \,\, \sum_{\vtheta \in \tilde \Theta^n} \lambda_{\vtheta} \sum_{\va\in A^n} t_{\vtheta,\va} \, R_{\vtheta, \va}  - \sum_{i \in N} \sum_{\vtheta \in \tilde \Theta^n} \lambda_{\vtheta}  \sum_{a \in A} \xi_{i,\vtheta,a} \sum_{\omega \in \Omega} F_{i,\theta_i,a,\omega}  \, p_{i,\vtheta,a,\omega} \quad\quad\quad\quad\quad\,\,\, \text{s.t.}\quad \label{eq:obj}  \\
		&  \sum_{a \in A} \xi_{i,\vtheta,a} \left( \sum_{\omega \in \Omega}  F_{i,\theta_i,a,\omega} \, p_{i,\vtheta,a,\omega} - c_{i,\theta_i,a} \right) \geq \nonumber \\
		& \specialcell{\,\,\,\, \sum_{a \in A} \xi_{i,(\theta,\vtheta_{-i}),a} \max_{a' \in A} \left\{ \sum_{\omega \in \Omega}  F_{i,\theta_i,a',\omega} \, p_{i,(\theta,\vtheta_{-i}),a,\omega} -  c_{i,\theta_i,a'} \right\} \hfill \forall  i\in N, \forall \vtheta \in \tilde \Theta^n, \forall \theta \in \Theta}  \label{eq:linearize} \\
		&  \specialcell{\sum_{a \in A} \xi_{i,(\theta, \vtheta_{-i}),a}=1 \hfill \forall i \in N, \forall \theta \in \Theta, \forall \vtheta_{-i} \in \tilde \Theta^n_{-i}} \label{eq:sum_one_1}\\
		&  \specialcell{ \sum_{\va \in A^n : a_i=a} t_{\vtheta,\va} = \xi_{i,\vtheta,a}\hfill \forall i \in N,	\forall \vtheta \in \tilde \Theta^n, \forall a \in A} \label{eq:sum_one_2} \\
		& \specialcell{t_{\vtheta,\avec} \ge 0 \hfill \forall \vtheta \in \tilde \Theta^n_{-i}, \forall \avec \in A^n}  \\
		&  \specialcell{\xi_{i, (\theta, \vtheta_{-i}), a} \ge 0 \hfill \forall i \in N, \forall \theta \in \Theta, \forall \vtheta_{-i} \in \tilde \Theta^n_{-i}, \forall a \in A} \\
		& \specialcell{p_{i,(\theta, \vtheta_{-i}),a,\omega} \ge 0 \hfill \forall i \in N, \forall \theta \in \Theta, \forall \vtheta_{-i} \in \tilde \Theta^n_{-i}, \forall a \in A , \forall \omega \in \Omega,}  
	\end{align}
\end{subequations}
where Objective~\eqref{eq:obj} is the principal's expected utility for the menu of randomized contracts encoded by the variables in the problem, Constraints~\eqref{eq:linearize} specify the conditions ensuring that the menu is DSIC (for $\theta \neq \theta_i$), as well as the conditions guaranteeing that any action $a$ such that $\xi_{i,\vtheta,a}>0$ is IC for an agent $i$ of type $\theta_i$ under the payments defined by variables $p_{i,\vtheta,a,\omega}$, while Constraints~\eqref{eq:sum_one_1}~and~\eqref{eq:sum_one_2} ensure that the menu of randomized contracts is well defined.
%
%
%
%

Notice that Problem~\eqref{pr:types} is defined in terms of $\sup$ rather than $\max$.
This is because, as shown in~\citep{Castiglioni2022Designing}, even in single-agent settings the problem of computing an optimal DSIC menu of randomized contracts may \emph{not} admit a maximum.
In the following, for ease of presentation, we let $\SUP$ be the optimal value of Problem~\eqref{pr:types} (\emph{i.e.}, the value of the supremum).

\subsection{An ``Approximately-optimal'' LP Formulation}\label{sec:bayesian_lp}

As a preliminary step towards the design of our approximation algorithm (see Section~\ref{sec:bayesian_algo}), we show how to find an ``approximately-optimal'' DSIC menu of randomized contracts by solving an LP which features exponentially-many variables and polynomially-many constraints.

In the following, we will make extensive use of the set $A_{i,\theta} \subseteq A$ of actions which are \emph{inducible} for an agent $i \in N$ of type $\theta \in \Theta$.
This is the set of all actions that are IC for an agent $i$ of type $\theta$ under at least one contract; formally, $A_{i,\theta} \coloneqq \left\{ a \in A \mid \exists p  \in \mathbb{R}^{n \times m}_+ : a \in A^*_{i,\theta}(p) \right\}$.

First, we can prove the following useful result:
\begin{restatable}{lemma}{bounded}\label{lem:bounded}
	There exists a function $\tau : \mathbb{N} \to \mathbb{R}$ such that $\tau(x)$ is $O(2^{\poly(x)})$---with $\poly(x)$ being a polynomial in $x$---and, for every instance $I \coloneqq (N,\Theta,\Omega,A)$ of Bayesian principal-multi-agent problem, agent $i \in N$, type $\theta\in \Theta$, and inducible action $a \in A_{i,\theta}$, there exists a contract $p \in \mathbb{R}^{n \times m}_+$ such that $a\in A^*_{i,\theta}(p)$ and $p_{i, \omega} \le \tau(|I|)$ for all $\omega \in \Omega$, where $|I|$ is the size of instance $I$.\footnote{In the rest of the section, we always assume that the size of a problem instance is expressed in terms of number of bits.}
	%
	%
\end{restatable}

Intuitively, Lemma~\ref{lem:bounded} states that, if an action $a$ is inducible for an agent $i$ of type $\theta$, then there exists a contract under which such an action is IC and whose payments are ``small'', in the sense that they can be represented with a number of bits that is upper bounded by a quantity depending polynomially on the size of the problem instance.
As we show next, such a result is crucial for proving Theorem~\ref{thm:regularize}, as it allows to satisfactorily bound the principal's expected utility loss due to solving an LP rather than Problem~\eqref{pr:types}.

Next, we formally introduce LP~\eqref{lp:types}, which is obtained from Problem~\eqref{pr:types} by (i) replacing each product of two variables $\xi_{i,\vtheta,a} \, p_{i,\vtheta,a,\omega}$ with a single variable $y_{i,\vtheta,a,\omega}$, (ii) considering the inducible actions in $A_{i,\theta}$ as the only actions available to an agent $i $ of type $\theta $, and (iii) linearizing the $\max$ operator in Constraints~\eqref{eq:linearize}.
By letting $A^{n,\vtheta} \coloneqq \bigtimes_{i \in N} A_{i,\theta_i}$ for every $\vtheta \in \tilde \Theta^n$, we can write:
\begin{subequations}\label{lp:types}
	\begin{align}
		\max & \,\, \sum_{\vtheta \in \tilde \Theta^n} \lambda_{\vtheta} \sum_{\avec \in A^{n,\vtheta}} t_{\vtheta,\avec} \, R_{\vtheta, \va} - \sum_{i \in N} \sum_{\vtheta \in \tilde \Theta^n} \lambda_{\vtheta} \sum_{a\in A_{i,\theta_i}} \sum_{\omega \in \Omega} F_{i,\theta_i,a,\omega} \, y_{i,\vtheta,a,\omega}  \quad\quad\quad\quad\quad\quad  \text{s.t.}\quad \\
		& \specialcell{\sum_{a \in A_{i,\theta_i}} \left( \sum_{\omega \in \Omega} y_{i,\vtheta,a,\omega} \, F_{i,\theta_i,a,\omega}-  \xi_{i,\vtheta,a} \, c_{i,\theta_i,a} \right) \ge  \sum_{a \in A_{i,\theta}} \gamma_{i,\vtheta,\theta,a} \hfill \forall i \in N, \forall \vtheta \in \tilde \Theta^n,\forall \theta \in \Theta} \label{eq:lp_cons_1}\\
		& \gamma_{i,\vtheta,\theta,a} \ge \sum_{\omega \in \Omega} y_{i,(\theta,\vtheta_{-i}),a,\omega} \, F_{i,\theta_i,a',\omega} -  \xi_{i,(\theta,\vtheta_{-i}),a} \, c_{i,\theta_i,a'} \nonumber\\
		& \specialcell{\hfill \forall  i \in N, \forall \vtheta \in \tilde \Theta^n, \forall \theta \in \Theta, \forall a \in A_{i,\theta_i}, \forall a' \in A_{i,\theta_i}} \label{eq:lp_cons_2}\\
		& \specialcell{\sum_{a \in A_{i,\theta}} \xi_{i,(\theta, \vtheta_{-i}),a}=1 \hfill \forall i \in N, \forall \theta \in \Theta, \forall \vtheta_{-i} \in \tilde \Theta^n_{-i}} \label{eq:lp_cons_3} \\
		& \specialcell{\sum_{\va \in A^{n,\vtheta}:a_i= a} t_{\vtheta,\va}  = \xi_{i,\vtheta, a}\hfill \forall i \in N, \forall \vtheta \in \tilde \Theta^n, \forall a \in A_{i,\theta_i}} \label{eq:lp_cons_4}\\
		& \specialcell{t_{\vtheta,\avec} \ge 0 \hfill \forall \vtheta \in \tilde \Theta^n, \forall \avec \in A^{n,\vtheta}}  \label{eq:lp_cons_5}\\
		&  \specialcell{\xi_{i, (\theta, \vtheta_{-i}), a} \ge 0 \hfill \forall i \in N, \forall \theta \in \Theta, \forall \vtheta_{-i} \in \tilde \Theta^n_{-i}, \forall a \in A_{i,\theta}} \label{eq:lp_cons_6}\\
		& \specialcell{y_{i,(\theta, \vtheta_{-i}),a,\omega} \ge 0 \hfill \forall i \in N, \forall \theta \in \Theta, \forall \vtheta_{-i} \in \tilde \Theta^n_{-i}, \forall a \in A_{i,\theta} , \forall \omega \in \Omega} \label{eq:lp_cons_7} \\
		& \specialcell{\gamma_{i,\vtheta, \theta, ,a}  \hfill \forall i \in N, \forall \vtheta \in \tilde \Theta^n, \forall \theta \in \Theta, \forall a \in A_{i,\theta}.} \label{eq:lp_cons_8}
	\end{align}
\end{subequations}
%
%

By letting $\LP$ be the optimal value of LP~\eqref{lp:types}, the following lemma shows that such a value is always at least as large as the value of the supremum defined in Problem~\eqref{pr:types}.
\begin{restatable}{lemma}{lpGeq}\label{lem:lp_geq_sup}
	For every instance of Bayesian principal-multi-agent problem, it holds $\LP\ge \SUP$.
\end{restatable}

Lemma~\ref{lem:lp_geq_sup} is proved by showing that, given any feasible solution to Problem~\eqref{pr:types}, it is possible to recover a feasible solution to LP~\eqref{lp:types} having the same objective function value.
However, the converse is \emph{not} true in general, \emph{i.e.}, given a feasible solution to LP~\eqref{lp:types}, it is \emph{not} always possible to build a feasible solution to Problem~\eqref{pr:types} having at least the same value.
Thus, it might be the case that $\SUP < \LP$.
%
%
%
This is caused by the existence of what we call \emph{irregular} feasible solutions to LP~\eqref{lp:types}:
%
\begin{definition}
	A feasible solution to LP~\eqref{lp:types} is said to be \emph{irregular} if there exists an agent $i \in N$, a tuple of agents' types $\vtheta \in \Theta$, an inducible action $a \in A_{i,\theta_i}$, and an outcome $\omega \in \Omega$ such that $y_{i,\vtheta,a,\omega}>0$ and $\xi_{i,\vtheta,a}=0$.
	A feasible solution to LP~\eqref{lp:types} is said to be \emph{regular} if it is not irregular.
\end{definition}

It is easy to see that, given a regular feasible solution to LP~\eqref{lp:types}, we can recover a feasible solution to Problem~\eqref{pr:types} with the same objective function value by simply letting $p_{i,\vtheta,a,\omega}= y_{i,\vtheta,a,\omega}/\xi_{i,\vtheta,a}$ for every $i \in N$, $\vtheta \in \tilde \Theta^n$, $a \in A_{i , \theta_i}$, and $\omega \in \Omega$.
However, the same is \emph{not} true for irregular solutions, as the operation above is clearly ill defined in that case.
Nevertheless, we show that, given any irregular feasible solution to LP~\eqref{lp:types}, it is always possible to build a regular solution by only incurring in an arbitrarily small loss in objective function value.
Formally:
\begin{restatable}{lemma}{fromIRR}\label{lem:from_irr_reg}
	Given an instance $I \coloneqq (N,\Theta,\Omega,A)$ of Bayesian principal-multi-agent problem and an irregular solution to LP~\eqref{lp:types} with value $\VAL$, for any $\epsilon>0$, it is possible to recover a regular solution to LP~\eqref{lp:types} with value at least $\VAL-\epsilon (n \, \tau(|I|)+1)$ in time polynomial in $|I|$ and $\frac{1}{\epsilon}$, where $\tau$ is a function defined as per Lemma~\ref{lem:bounded} and $|I|$ denotes the size of instance $I$.
\end{restatable}

Finally, we are ready to prove that solving LP~\eqref{lp:types} in place of Problem~\eqref{pr:types} allows us to recover in polynomial time a DSIC menu of randomized contracts that only incurs in an arbitrarily small loss with respect to the value $\SUP$ of the supremum of Problem~\eqref{pr:types}.
Formally:
\begin{restatable}{theorem}{regularize}\label{thm:regularize}
	Given an instance $I \coloneqq (N,\Theta,\Omega,A)$ of Bayesian principal-multi-agent problem and an optimal solution to LP~\eqref{lp:types}, for any $\epsilon > 0$, it is possible to recover a feasible solution to Problem~\eqref{pr:types} with value at least $\SUP -\epsilon (n \, \tau(|I|)+1)$ in time polynomial in $|I|$ and $\frac{1}{\epsilon}$, where $\tau$ is a function defined as per Lemma~\ref{lem:bounded} and $|I|$ denotes the size of instance $I$.
	%
\end{restatable}

\subsection{Approximation Algorithm}\label{sec:bayesian_algo}

LP~\eqref{lp:types} features exponentially-many variables and polynomially-many constraints, and, thus, it can be solved in polynomial time by applying the ellipsoid method ot its dual, provided access to a suitable polynomial-time \emph{separation oracle} for the constraints of the dual~\citep{grotschel2012geometric}.
%

In this last section, we show that, despite an (exact) polynomial-time separation oracle may \emph{not} be available in our setting, it is always possible to design a polynomial-time \emph{approximate separation oracle}.
This, together with some \emph{ad hoc} modifications to the ellipsoid method, allows us to design the desired approximation algorithm for the problem of interest.
Indeed, in instances that satisfy the FOSD condition and have IR-supermodular succinct rewards, it is possible to design an (exact) polynomial-time separation oracle.
Instead, in instances having DR-submodular succinct rewards, this is \emph{not} possible, and, thus, we need an approximate separation oracle.\footnote{Notice that  the existence of an exact oracle for instances with DR-submodular rewards would contradict Theorem~\ref{thm:hard_is}.}
%
%

We start by introducing a relaxation of LP~\eqref{lp:types} (see LP~\eqref{lp:typesRelaxed} below) and by showing that the two LPs are indeed equivalent (see Lemma~\ref{lem:lps_equiv} below).
Such a preliminary step allows us to obtain a dual LP which has additional constraints on its variables, which will be crucial in order to design a polynomial-time approximate separation oracle.
The relaxation of LP~\eqref{lp:types}, which is obtained by replacing the `$=$' in Constraints~\eqref{eq:lp_cons_4} with a `$\leq$', reads as follows:
\begin{subequations}\label{lp:typesRelaxed}
	\begin{align}
	\max & \,\, \sum_{\vtheta \in \tilde \Theta^n} \lambda_{\vtheta} \sum_{\avec \in A^{n,\vtheta}} t_{\vtheta,\avec} \, R_{\vtheta, \va} - \sum_{i \in N} \sum_{\vtheta \in \tilde \Theta^n} \lambda_{\vtheta} \sum_{a\in A_{i,\theta_i}} \sum_{\omega \in \Omega} F_{i,\theta_i,a,\omega} \, y_{i,\vtheta,a,\omega}  \quad \text{s.t.} \quad\\
	& \specialcell{\sum_{\va \in A^{n,\vtheta}:a_i= a} t_{\vtheta,\va} \leq \xi_{i,\vtheta, a}    \hfill \forall i \in N, \forall \vtheta \in \tilde \Theta^n, \forall a \in A_{i,\theta_i}} \\
	& \textnormal{Constraints~\eqref{eq:lp_cons_1}---\eqref{eq:lp_cons_3}~and~\eqref{eq:lp_cons_5}---\eqref{eq:lp_cons_8}.} \nonumber
	\end{align}
\end{subequations}

\begin{restatable}{lemma}{lpsEquiv}\label{lem:lps_equiv}
	For every instance of Bayesian principal-multi-agent problem, LP~\eqref{lp:types} and LP~\eqref{lp:typesRelaxed} have the same optimal value. Moreover, given a feasible solution to LP~\eqref{lp:typesRelaxed}, it is always possible to recover in polynomial time a feasible solution to LP~\eqref{lp:types} having at least the same value.
\end{restatable}

By Lemma~\ref{lem:lps_equiv}, we can solve LP~\eqref{lp:typesRelaxed} instead of LP~\eqref{lp:types}.
%
%
%
%
The dual problem of LP~\eqref{lp:typesRelaxed} reads as follows:\footnote{Notice that, in LP~\eqref{lp:typesDual}, we used $\mathbf{1} \left\{ \cdot \right\}$ to the denote the indicator function for the event written within curly braces}
%
%
\begin{subequations}\label{lp:typesDual}
	\begin{align}
		\min & \,\, \sum_{i \in N} \sum_{ \theta \in \Theta} \sum_{\vtheta_{-i} \in \tilde \Theta^n_{-i}}  x_{i,\theta, \vtheta_{-i}} \quad\quad\quad\quad\quad\quad\quad\quad\quad\quad\quad\quad\quad\quad\quad\quad\quad\quad\quad\quad\quad\quad\quad \text{s.t.} \quad \\
		&\specialcell{- \mathbf{1} \left\{ (\theta,\vtheta_{-i}) \in \Tilde \Theta^n \right\}  \sum_{\theta' \in \Theta} y_{i,(\theta,\vtheta_{-i}),\theta'} \, c_{i,\theta,a}+ \sum_{\substack{\theta' \in \Theta: \\ (\theta',\vtheta_{-i})\in \Tilde \Theta^n}} \sum_{a' \in A_{i,\theta'}} c_{i,\theta',a'} \, z_{i,(\theta',\vtheta_{-i}),\theta,a,a'}\hfill}   \nonumber\\
		& \specialcell{\quad + d_{i, \theta,\vtheta_{-i}} - \mathbf{1} \left\{ (\theta,\vtheta_{-i})\in \tilde \Theta^n \right\} \, x_{i, \theta,\vtheta_{-i}} \ge 0 \hfill  \forall i \in N, \forall \theta \in \Theta, \forall \vtheta_{-i} \in \tilde \Theta^n_{-i}, \forall a \in A_{i, \theta}} \label{eq:conFirst}\\ 
		& \specialcell{-y_{i,\vtheta,\theta} + \sum_{a' \in A_{i,\theta_i}}  z_{i,\vtheta,\theta,a,a'}  \ge 0 \hfill  \forall i \in N, \forall \vtheta \in \tilde \Theta^n, \forall \theta \in \Theta, \forall a \in A_{i, \theta}} \\ 
		&  \mathbf{1}\left\{ (\theta,\vtheta_{-i}) \in \tilde \Theta^n \right\}  \sum_{\theta' \in \Theta} y_{i,(\theta,\vtheta_{-i}),\theta'} \, F_{i,\theta,a,\omega} -\sum_{\substack{\theta' \in \Theta: \\ (\theta',\vtheta_{-i})\in \Tilde \Theta^n}} \sum_{a' \in A_{i,\theta'}} F_{i,\theta',a',\omega} \, z_{i,(\theta',\vtheta_{-i}),\theta,a,a'}  \geq \nonumber\\
		& \specialcell{\quad- \mathbf{1} \left\{ (\theta,\vtheta_{-i}) \in \tilde \Theta^n \right\} \lambda_{(\theta,\vtheta_{-i})} \, F_{i, \theta,a,\omega}  \hfill \forall  i \in N, \forall \theta \in \Theta, \forall \vtheta_{-i} \in \tilde \Theta^n_{-i}, \forall a \in A_{i, \theta}, \forall \omega \in \Omega} \label{eq:conMid1}\\
		& \specialcell{\sum_{i \in N} y_{i,\vtheta,a_i} \ge \lambda_{\vtheta} \, R_{\vtheta, \va} \hfill \forall  \vtheta \in \tilde \Theta^n, \forall \va \in A^{n,\vtheta}} \label{eq:useOracle}\\
		& \specialcell{x_{i,\theta,\vtheta_{-i}} \hfill \forall i \in N, \forall \theta \in \theta, \forall \vtheta_{-i} \in \tilde \Theta^n_{-i}} \label{eq:conMid2}\\
		& \specialcell{y_{i,\vtheta,a}  \ge 0 \hfill \forall i \in N, \forall \vtheta \in \tilde \Theta^n, \forall a \in A_{i,\theta_i}} \\
		& \specialcell{z_{i, \vtheta,\theta,a, a'} \le 0 \hfill \forall i \in N, \forall \vtheta \in \tilde \Theta^n, \forall \theta \in \Theta, \forall a \in A_{i,\theta_i} , \forall a' \in A_{i,\theta_i}}  \\
		& \specialcell{d_{i,\theta,\vtheta_{-i}}  \hfill \forall i \in N, \forall \theta \in \Theta, \forall \vtheta_{-i} \in \tilde \Theta^n_{-i},} \label{eq:conLast}
	\end{align}
\end{subequations}
where $x_{i,\theta,\vtheta_{-i}}$ are dual variables that correspond to Constraints~\eqref{eq:lp_cons_3}, $y_{i,\vtheta,a}$ to Constraints~\eqref{eq:lp_cons_4}, $z_{i, \vtheta,\theta,a, a'}$ to Constraints~\eqref{eq:lp_cons_2}, while $d_{i,\theta,\vtheta_{-i}}$ to Constraints~\eqref{eq:lp_cons_1}.
%

The dual LP~\eqref{lp:typesDual} features polynomially-many variables and exponentially-many constraints.\footnote{We recall that the distribution $\lambda$ is part of the problem instance given as input to our algorithm, and, thus, both $|\tilde \Theta^n|$ and $|\tilde \Theta^n_{-i}|$ are polynomial quantities in the size of such instance.}
Moreover, Constraints~\eqref{eq:useOracle} are the only ones which are exponential in the size of the problem instance, since there is a group of such constraints for every tuple of agents' actions $\avec \in A^{n,\vtheta}$.
Thus, in order to have the ellipsoid method running in polynomial time on LP~\eqref{lp:typesDual}, it is sufficient to design a polynomial-time separation oracle for Constraints~\eqref{eq:useOracle}, as the others can be checked one by one in polynomial time.
As we show next, only an ``approximate version'' of such a separation oracle can be implemented in polynomial time, according to the following definition.
%
%
\begin{definition}[Approximate separation oracle] \label{def:oracle}
	Given any $\alpha \in (0,1]$, an \emph{approximate separation oracle} for Constraints~\eqref{eq:useOracle} is a procedure $\Ocal_{\alpha}(\cdot,\cdot,\cdot,\cdot)$ which, given in input an instance $I \coloneqq (N,\Theta,\Omega,A)$ of Bayesian principal-multi-agent problem, a tuple of agents' types $\vtheta \in \tilde \Theta^n$, a vector $w \in \mathbb{R}^{n \ell}$ of weights---with $w_{i,a}$ denoting the vector component corresponding to agent $i \in N$ and action $a \in A_{i,\theta_i}$---, and an additive error $\epsilon > 0$, returns a tuple of agents' actions $\va \in A^{n,\vtheta}$ such that:
	\begin{align*}
		\lambda_{\vtheta} \, R_{\vtheta, \va} - \sum_{i \in N} w_{i,a_i} \ge  \alpha \, \lambda_{\vtheta} \,  R_{\vtheta, \va'} - \sum_{i \in N} w_{i, a'_i} - \epsilon \quad   \forall \va' \in A^{n,\vtheta},
	\end{align*}
	in time polynomial in $|I|$, $\max_{i \in N,a \in A} |w_{i,a}| $, and $\frac{1}{\epsilon}$, where $|I|$ denotes the size of instance $I$.
	%
\end{definition}
Notice that, by letting each weight $w_{i,a}$ be equal to $y_{i,\vtheta,a}$ for some feasible solution to LP~\eqref{lp:typesDual}, the problem solved in a call $\Ocal_{\alpha}(I,w,\vtheta,\epsilon)$ to the approximate separation oracle intuitively consists in finding the most violated constraint among Constraints~\eqref{eq:useOracle}, up to a reward-multiplying approximation factor $\alpha$ and an additive error $\epsilon$ given as input.

Next, we show that it is possible to apply an \emph{ad hoc} implementation of the ellipsoid method to LP~\eqref{lp:typesDual}, which, given access to an approximate separation oracle for Constraints~\eqref{eq:useOracle} as in Definition~\ref{def:oracle}, returns a feasible solution to LP~\eqref{lp:typesDual} that provides a desirable approximation of the optimal value of LP~\eqref{lp:typesDual}.
Such a procedure can be embedded in a suitable binary search scheme, resulting in a polynomial-time approximation algorithm for the principal's optimization problem.
%
%
\begin{restatable}{theorem}{bayesian}\label{th:bayesian}
	Given access to an approximate separation oracle $\Ocal_{\alpha}(\cdot,\cdot,\cdot,\cdot)$ with $\alpha \in (0,1]$, there exists an algorithm that, given any $\rho > 0$ and instance of Bayesian principal-multi-agent problem as input, returns a DSIC menu of randomized contracts with principal's expected utility at least $\alpha \, R_\Gamma-P_\Gamma-\rho$ for every menu of randomized contracts $\Gamma=	\{\gamma_{\vtheta}\}_{\theta \in \Theta^n}$, where $R_\Gamma \in [0,1]$, respectively $P_\Gamma \in \mathbb{R}_+$, denotes the expected reward, respectively the expected overall payment, of $\Gamma$.
	Moreover, such an algorithm runs in time polynomial in the instance size and $\frac{1}{\rho}$.   
	%
\end{restatable}

We conclude the section by showing that the approximate separation oracle $\Ocal_{\alpha}(\cdot,\cdot,\cdot,\cdot)$ can be implemented in polynomial time for two classes of Bayesian principal-multi-agent problems.
This is the last step needed to fully specify the approximation algorithm introduced in Theorem~\ref{th:bayesian}. 

In Bayesian principal-multi-agent problem instances that satisfy the FOSD condition and have IR-supermodular succinct rewards, we are able to design a polynomial-time  approximate separation oracle $\Ocal_{\alpha}(\cdot,\cdot,\cdot,\cdot)$ with $\alpha = 1$.\footnote{Notice that, in Bayesian principal-multi-agent problem instances that satisfy the FOSD condition and have IR-supermodular succinct rewards, it is easy to adapt our results so as to show that there exists an exact separation oracle (thus getting rid of the additive error $\epsilon > 0$). We decided to use an approximate separation oracle anyway, for ease of exposition. This choice does \emph{not} detriment the final approximation guarantees of the algorithm (see Corollary~\ref{cor:dr_sub}), since we cannot get rid of the additive approximation $\rho > 0$ given that Problem~\eqref{pr:types} may \emph{not} admit a maximum.}
Instead, in instances with DR-submodular succinct rewards, we sow to implement an oracle $\Ocal_{\alpha}(\cdot,\cdot,\cdot,\cdot)$ with $\alpha = 1 - 1/e$.
These implementations work by solving suitably-defined problems that resemble non-Bayesian principal-multi-agent instances.
In particular, they have their same structure, while the rewards are scaled by a factor $\lambda_{\vtheta}$ and the values $\widehat P_{i,a}$ are substituted by the weights $w_{i,a}$. 
%
%
%
%
Formally, we get the following two results:
\begin{restatable}{corollary}{corIR}\label{cor:ir_super}
	In Bayesian principal-multi-agent problem instances that (i) have succinct rewards specified by an IR-supermodular function and (ii) satisfy the FOSD condition, for any $\rho>0$, the problem of computing an optimal menu of randomized contracts admits an algorithm returning a menu with principal's expected utility at least $\OPT-\rho$ in time polynomial in the instance size and $\frac{1}{\rho}$, where $\OPT$ is the value of the optimal principal's expected utility.
\end{restatable}
\begin{restatable}{corollary}{corDR}\label{cor:dr_sub}
In Bayesian principal-multi-agent problem instances with succinct rewards specified by a DR-submodular function, the problem of computing an optimal menu of randomized contracts admits a polynomial-time approximation algorithm which, for any $\epsilon>0$ given as input, outputs a menu providing the principal with an expected utility at least of $ (1-1/e) R_\Gamma-P_\Gamma-\epsilon$ for each menu of randomized contracts $\Gamma=	\{\gamma_{\vtheta}\}_{\theta \in \Theta^n}$ with high probability,  where $R_\Gamma \in [0,1]$, respectively $P_\Gamma \in \mathbb{R}_+$, denotes the expected reward, respectively the expected payment, in contract p.
\end{restatable}

Notice that Corollary~\ref{cor:dr_sub} provides the same approximation guarantees of its corresponding result for non-Bayesian instances (see Theorem~\ref{thm:DR}), while Corollary~\ref{cor:ir_super} matches those of its corresponding non-Bayesian result up to an additive error $\rho > 0$ (see Theorem~\ref{thm:IR}).

%
%
%
%

%% file: appendix_matroid.tex
\section{Proofs Omitted from Section~\ref{sec:matroids} }

\lemmaOptBase*

\begin{proof}
	Let $S \in \I^I$ be any independent set of $\M^I$.
	Clearly, by adding to $S$ all the ground elements $(i,a_\varnothing) \in \cG^I_i$ for $i \in N \setminus N_S$, we obtain a base $S' \in \I^I$.
	By definition of the null action $a_\varnothing$, it holds $\widehat P_{i,a_{S,i}} = \widehat P_{i,a_\varnothing}=0$ for every $i \in N \setminus N_S$, which implies $f^I(S) = f^I(S')$, since $R_{\avec_S} = R_{\avec_{S'}}$ given that $\avec_{S}=\avec_{S'}$.
	This concludes the proof.
\end{proof}

\theoremMat*

\begin{proof}
	We prove the result by showing that, given any pair $(p,\avec)$---where $p \in \mathbb{R}_+^{n \times m}$ is a contract and $\avec = (a_i)_{i \in N}\in \bigtimes_{i \in N} A_i^*(p)$ is a tuple of IC agents' actions recommended to the agents---, there exists a base $S \in \I^I$ of $\M^I$ such that $f^I(S)$ is greater than or equal to the principal's expected utility under $(p,\avec)$, and, conversely, given any base $S\in \I^I$ there exists a pair $(p,\avec)$ with principal's expected utility $f^I(S)$.
	This, together with Lemma~\ref{lem:opt_base}, proves the result.
	
	\paragraph{From $(p,\avec)$ to a base.}
	Let the base $S \in \I^I$ be defined so that $S \coloneqq \left\{  (i,a_i) : i \in N \right\}$.
	Then, given that $a_i \in A_i^*(p)$ for all $i \in N$ and by the definition of $\widehat P_{i,a_{S,i}}$, it holds:
	\[
	f^I(S) = R_{\avec_S} - \sum_{i\in N} \widehat P_{i,a_{S,i}}  \geq  R_{\avec} - \sum_{i\in N}  \sum_{\omega \in \Omega} F_{i,a_i,\omega} \, p_{i,\omega} = R_{\avec} - \sum_{i\in N}  P_{i,a_{i}},
	\]
	where the inequality holds since $\avec_S = \avec$ and the fact that, for every $i \in N$, the value $\widehat P_{i,a_{S,i}}$ is defined as a minimum taken over the set $\mP^{i, a_{S,i}}$, which contains the contract $p$ given that $a_i \in A_i^*(p)$.
	
	\paragraph{From a base to $(p,\avec)$.}
	Given a base $S \in \I^I$ of the matroid $\M^I$, let $(p,\avec)$ be such that $\avec = (a_i)_{i \in N}$ satisfies $(i,a_i) \in S$ for all $i \in N$ and $p \in \arg \min_{p' \in \mP^{i, a_{S,i}} } \sum_{\omega \in \Omega} F_{i,a_{S,i},\omega} \, p'_{i,\omega}$ for every $i \in N$ (notice that such a contract can be built by defining the components $p_{i,\omega}$ for $\omega \in \Omega$ independently for each $i \in N$).
	Then, it immediately follows from the definition of the function $f^I$ that the principal's expected utility under $(p,\avec)$ is equal to $f^I(S)$.
\end{proof}

\redToRIND*

\begin{proof}
	Given a $1$-partition matroid $\mathcal{M} \coloneqq (\left\{ \mathcal{G}_i \right\}_{i \in [d]}, \mathcal{I})$ and a function $f: 2^{\mathcal{G}} \to\mathbb{R}$ that is ordered-supermodular, we show that maximizing $f$ over $\M$ is equivalent to maximizing a suitably-defined supermodular function $\tilde f: \mathcal{R} \to \mathbb{R}$ over a particular ring of sets $\mathcal{R}$.
	The latter is defined by the family of all the sets $S \subseteq \cG$ such that, if $x \in S$ and $x=\pi_i^{-1}(j)$ for some $i \in [d]$ and $j \in [k_i]$, then $\pi_i(l)\in S$ for all $l \in [k_i]:l<j$.
	Moreover, for every $S \subseteq \cG$, we let $\tilde f( S) \coloneqq f(\wedge S)$, where $\wedge S$ denotes the set obtained by taking an element $x \in S$ with maximal value of $\pi_i^{-1}(x)$ for each partition $i \in [d]$.
	Then, it is sufficient to show that $\tilde f$ is supermodular.
	Indeed, given two sets $S, S' \subseteq\cG$, it holds:
	\[
	\tilde f(S)+ \tilde f(S') = f(\wedge S)+ f(\wedge S') \le f(\wedge (S \cup S') ) + f(\wedge (S \cap S') ) = \tilde  f(S \cup S') + \tilde f(S \cap S'),
	\]
	which concludes the proof.
\end{proof}

\polyOrderSup*

\begin{proof}
	The problem can be reduced in polynomial time to the maximization of a supermodular function defined over a ring of sets by  Theorem~\ref{thm:redToRInd}.
	Such a problem is known to be solvable in polynomial time; see, \emph{e.g.},~\citep{schrijver2000combinatorial,bach2019submodular}.
\end{proof}

%% file: appendix_hard_ir_super.tex

\section{Proof of Theorem~\ref{thm:hard_label}}\label{app:hard_ir_super}

In order to prove the theorem, we employ a reduction from a {promise problem} associated with \textsf{LABEL-COVER} instances, whose definition follows.

\begin{definition}[\textnormal{\textsf{LABEL-COVER}} instance]
	An instance of \textnormal{\textsf{LABEL-COVER}} is a tuple $( G, \Sigma, \Pi)$:
	\begin{itemize}
		\item $G \coloneqq (U,V,E)$ is a \emph{bipartite graph} defined by two disjoint sets of nodes $U$ and $V$, connected by the edges in $E \subseteq U \times V$, which are such that all the nodes in $U$ have the same degree;
		\item $\Sigma$ is a finite set of \emph{labels}; and
		\item $\Pi \coloneqq \left\{ \Pi_e : \Sigma \to \Sigma \mid e \in E \right\}$ is a finite set of \emph{edge constraints}.
	\end{itemize}
	Moreover, a \emph{labeling} of the graph $G$ is a mapping $\pi: U \cup V \to \Sigma$ that assigns a label to each vertex of $G$ such that all the edge constraints are satisfied.
	Formally, a labeling $\pi$ satisfies the constraint for an edge $e =(u,v) \in E$ if it holds that $ \pi(v) = \Pi_e(\pi(u))$.
\end{definition}

The classical \textsf{LABEL-COVER} problem is the search problem of finding a valid labeling for a \textsf{LABEL-COVER} instance given as input.
In the following, we consider a different version of the problem, which is the {promise problem} associated with \textsf{LABEL-COVER} instances.

\begin{definition}[\textnormal{\textsf{GAP-LABEL-COVER}$_{c,s}$}]
	For any pair of numbers $0 \le s \le c \le 1$, we define \textnormal{\textsf{GAP-LABEL-COVER}}$_{c,s}$ as the following promise problem.
	\begin{itemize}
		\item \textnormal{\texttt{Input:}} An instance $(G,\Sigma, \Pi)$ of \textnormal{\textsf{LABEL-COVER}} such that either one of the following is true:
		\begin{itemize}
			\item there exists a labeling $\pi$ that satisfies at least a fraction $c$ of the edge constraints in $\Pi$;
			\item any labeling $\pi$ satisfies less than a fraction $s$ of the edge constraints in $\Pi$.
		\end{itemize}
		\item \textnormal{\texttt{Output:}} Determine which of the above two cases hold.
	\end{itemize}
\end{definition}

To prove Theorem~\ref{thm:hard_label}, we use the following result due to~\citet{raz1998parallel}~and~\citet{arora1998proof}.

\begin{theorem}[\citet{raz1998parallel,arora1998proof}]\label{thm:hard_gap_label}
	For any $\epsilon > 0$, there exists a constant $k_\epsilon \in \mathbb{N}$ that depends on $\epsilon$ such that the promise problem \textnormal{\textsf{GAP-LABEL-COVER}}$_{1,\epsilon}$ restricted to inputs $(G, \Sigma, \Pi)$ with $|\Sigma| = k_\epsilon$ is \textnormal{\textsf{NP}}-hard.
\end{theorem}

Now, we are ready to prove Theorem~\ref{thm:hard_label}.

\begin{proof}[Proof of Theorem~\ref{thm:hard_label}]
	%
	%
	%
	Given an approximation factor $\rho>0$, we reduce from the problem \textnormal{\textsf{GAP-LABEL-COVER}}$_{1,\rho}$.
	Our construction is such that, if the \textsf{LABEL-COVER} instance admits a labeling that satisfies all the edge constraints, then the corresponding  principal-multi-agent problem admits a contract providing the principal with an overall expected utility of at least $1$.
	Otherwise, if at most a fraction $\rho$ of the constraints are satisfied, then  any contract provides the principal with an overall expected utility of at most $\rho$. Since $\rho>0$ can be an arbitrarily small constant, this is sufficient to prove the statement.
	
	\paragraph{Construction.}
	Given an instance of \textnormal{\textsf{GAP-LABEL-COVER}}$_{1,\rho}$ $(G,\Sigma,\Pi)$ with a bipartite graph $G = (U,V,E)$, we build a principal-multi-agent instance as follows.
	The set of agents includes an agent $n_v$ for every node $v \in U \cup V$ of $G$.
	The outcome space has $k_\rho$ dimensions, \emph{i.e.}, $\Omega =\mathbb{R}_+^{k_\rho}$.
	Each agent $n_v$, $v \in V \cup U$ has an action $a_\sigma$ for each label $\sigma \in \Sigma$.
	Given an label $\sigma$, let $\omega^\sigma \in \mathbb{R}_+^{k_\rho}$ be the outcome with $\omega^\sigma_{\sigma}=1$ and $\omega^\sigma_{\sigma'}=0$ for each $\omega'\neq \sigma$, where for ease of exposition we rename the set $\Sigma$ as $\{1,\dots,k_{\rho}\}$.
	For each agent $n_v$ and each action $a_\sigma$, with $\sigma \in \Sigma$, cost $c_{n,a_\sigma}=0$ and $a_\sigma$ induces the outcome $\omega^\sigma$ deterministically, \emph{i.e.}, $F_{n,a_\sigma,\omega^\sigma}=1$.
	Finally, the principal's reward function $g$ is defined as follows. For each vector $\vomega \in \Omega^{n}$,  
	\[ g(\vomega) = \sum_{(v,u) \in E} \sum_{\sigma \in \Sigma} \mathbf{1}\{\omega_{n_v,\sigma}=1 \land \omega_{n_u,\Pi_e(\sigma)}=1 \}/|E|.\]
	It is easy to see that the function is IR-supermodular in $[0,1]^{n|\Sigma|}$  and hence for all the inducible outcomes $\vomega$.\footnote{It is easy to construct an arbitrary good approximation of $g(\cdot)$ that is IR-supermodular on all the domain $\mathbb{R}_+^{n|\Sigma|}$. For instance, we can set  $g(\vomega) = e^{M(\omega_{n_v,\sigma}+\omega_{n_u,\Pi_e(\sigma)}-2)}/|E|$ for an arbitrary large $M$. }

	\paragraph{Completeness.}
	Suppose that the instance of \textnormal{\textsf{GAP-LABEL-COVER}}$_{1,\rho}$ $(G,\Sigma,\Pi)$ admits a labeling $\pi : U \cup V \to \Sigma$ that satisfies all the edge constraints in $\Pi$.
	Let us define a contract that recommends action $a_{\pi(v)}$ for every node $v \in U \cup V$, while all the payments are set to $0$, \emph{i.e.}, $p_{n, \omega}=0$ for each $n\in N$ and $\omega \in \Omega$.
	Notice that the agents  follow the recommendations since they are indifferent among all the actions.
	It is easy to see that the utility is $1$ since for each edge $(u,v)$,  $\omega_{n_v,\pi(v)}=1$ and $\omega_{n_u,\Pi_e(\pi(u))}=1$.
	This concludes the first part of the proof.

	\paragraph{Soundness.}
	We show that, if the \textsf{LABEL-COVER} instance is such that every labeling $\pi : U \cup V \to \Sigma$ satisfies at most a fraction $\rho$ of the edge constraints in $\Pi$, then, in the corresponding principal-agent setting, any contract provides the principal with an expected utility at most $\rho$.

	 Let $\hat \va$ be the tuple of action recommendations and recall that each action $\hat a_{n_v}$, $v \in V \cup U$ induces deterministically an outcome $\omega_{n_v} \in \{\omega^{\sigma}\}_{\sigma \in \Sigma}$.
	As a first step,  notice that for each edge $e=(u,v)$, $\sum_{\sigma \in \Sigma} 1(\omega_{n_v,\sigma}=1 \land \omega_{n_u,\Pi_e(\sigma)}=1)/|E|$ is at most $1/|E|$ since there is exactly one $\sigma$ such that $\omega_{n_v,\sigma}=1$.
	Suppose by contradiction that there exists a contract with utility strictly larger than $\rho$. Then, there are strictly more than $\rho|E|$ edges such that $\sum_{\sigma \in \Sigma} 1(\omega_{n_v,\sigma}=1 \land \omega_{n_u,\Pi_e(\sigma)}=1  )=1$. 
	Consider the assignment that assign to each variable $v\in V \cup U$ the label $\sigma$ such that $\hat a_{n_v}=a_{\sigma}$.
	 It is easy to see that this assignment satisfies strictly more than a $\rho$ fraction of the edges, reaching a contradiction.
\end{proof}

%% file: appendix_ir.tex
\section{Proofs Omitted from Section~\ref{sec:ir_super} }

To prove the results in this section it will be useful to employ the definition of supermodularity for continuous functions.
Indeed, the properties introduced in Definition~\ref{def:dr_ir} are special cases of the classical submodularity and supermodularity properties which are usually considered in the literature. 
Formally, by letting $\max\{\vomega, \vomega\}$, respectively $\min\{\vomega , \vomega' \}$, be the component-wise maximum, respectively minimum, between two given vectors $\vomega, \vomega' \in \mathbb{R}_+^{nq}$, the following definition holds:
\begin{definition} \label{def:sub}
	A reward function $g: \mathbb{R}_+^{nq} \to \mathbb{R}$ is \emph{submodular} if the following holds:
	\[
	g(\vomega)+ g(\vomega') \geq g(\max\{\vomega, \vomega\})+ g(\min\{\vomega, \vomega\}) \quad \forall \vomega, \vomega' \in \mathbb{R}_+^{nq}.
	\]
	Moreover, a reward function $g: \mathbb{R}_+^{nq} \to \mathbb{R}$ is \emph{supermodular} if its opposite function $-g$ is submodular.
	%
	%
\end{definition}
It is well known that any DR-submodular, respectively IR-supermodular, function is also submodular, respectively supermodular, but the converse is \emph{not} true~\citep{Bian2017Guaranteed}. 

\dominated*

\begin{proof}
	The proof follows from Theorem~1 in~\citep{OSTERDAL20101222}.
	In particular, for every agent $i \in N$ and action index $j \in [\ell-1]$, given that the FOSD condition ensures that $\sum_{\omega \in \Omega'} F_{i,a_{j+1},\omega} \le  \sum_{\omega \in \Omega'} F_{i,a_{j},\omega} $ for all comprehensive sets $\Omega' \subseteq \Omega$, Theorem~1 in~\citep{OSTERDAL20101222} states that $F_{i,a_{j}}$ can be derived from $F_{i, a_{j+1}}$ by means of a finite sequence of \emph{deteriorating bilateral transfers (of mass)}.
	These are operations which consist in moving probability mass from an outcome $\omega \in \Omega$ to another outcome $\omega' \in \Omega$ such that $\omega' \leq \omega$, while maintaining the probability mass on outcomes that are different from $\omega$ and $\omega'$ untouched.
	As a result, given an agent $i \in N$ and a pair $a_j, a_k \in A$ of agent $i$'s actions such that $j <k$, it is easy to check that the probability $F_{i, a_k,\omega}$ which $F_{i, a_k} $ places on an outcome $\omega \in \Omega$ can be expressed as a suitable combinations of the probabilities $F_{i, a_j,\omega'}$ which $F_{i, a_j,\omega'}$ places on outcomes $\omega' \in \Omega$ such that $\omega' \leq \omega$ (since all the bilateral transfers involved in the processes of turning $F_{i, a_{k}}$ into $F_{i, a_{j}}$ are deteriorating).
	This concludes the proof.
	%
	%
	%
	%
\end{proof}

\orderSup*

\begin{proof}
	%
	In order to show the result, we prove that, for every pair of independent sets $S, S' \in \I^I$:
	\[
	f^I(S \land S') + f^I(S \lor S') \ge f^I(S)+ f^I(S') ,
	\] 
	where the partition-wise ``maximum'' $\wedge$ and ``minimum'' $\vee$ are defined with respect to the bijective functions $\pi_i: [k_i] \to \cG_i^I$ (with $k_i = \ell$) constructed according to the (agent-dependent) ordering of the action set $A$.
	In particular, for every $i \in N$ and $j \in [\ell]$, it holds $\pi_i(j) = (i,a_j)$.

	For ease of presentation, in the rest of the proof we let $\va_1 \coloneqq \va_{S \land S'}$ and $\va_2 \coloneqq \va_{S \lor S'}$, so that $a_{1,i}$, respectively $a_{2,i}$, denotes the $i$-th component of $\avec_1$, respectively $\avec_2$.
	%
	%
	%
	
	First, let us notice that $\sum_{i \in N} \widehat P_{i,a_{1,i}} + \sum_{i \in N} \widehat P_{i,a_{2,i}} = \sum_{i \in N} \widehat P_{i,a_{S,i}} + \sum_{i \in N} \widehat P_{i,a_{S',i}}$, which holds since, by definition of partition-wise ``maximum'' $\wedge$ and ``minimum'' $\vee$, for every agent $i \in N$ the pair of actions $a_{1,i}, a_{2,i}$ exactly coincides (up to ordering) with $a_{S,i}, a_{S',i}$.
	Thus, given the definition of $f^I$ (see Theorem~\ref{thm:eq_matr}), in order to prove the result it is sufficient to prove that $R_{\va_1}+R_{\va_2}\ge R_{\va_S}+R_{\va_{S'}} $.
	%

	By definition of $\avec_1$ and $\avec_2$, we have that $\pi_i^{-1}(i, a_{1,i}) \geq \pi_i^{-1}(i, a_{2,i})$ for every $i \in N$. 
	Then, thanks to Lemma~\ref{lm:dominated} and how actions are ordered, for every agent $i \in N$, there exists a collection of probability distributions $\mu^{i,\omega}  \in \Delta_{\Omega^-}$, one per outcome $\omega \in \Omega$, such that $F_{i, a_{1,i},\omega}=\sum_{\omega' \in \Omega} F_{i, a_{2,i},\omega'} \mu^{i,\omega'}_{\omega-\omega'} $, where we recall that the $\mu^{i,\omega}$ are the probability distributions that allow to turn $F_{i, a_{2,i}}$ into $F_{i, a_{1,i}}$. 
	Moreover, notice that, whenever $a_{1,i}=a_{2,i}$, it holds $\mu^{i, \omega}_{\mathbf{0}}=1$ for every $\omega \in \Omega$.
	Then, we can write:	
	\begin{align*}
		R_{\avec_1} = \sum_{\vomega \in \Omega^n} \left(  \prod_{i \in N} F_{i,a_{1,i},\omega_i} \right) g(\vomega) = \sum_{\vomega \in \Omega}   \prod_{i \in N}  \left(   \sum_{\omega' \in \Omega} F_{i, a_{2,i},\omega'} \mu^{i,\omega'}_{\omega-\omega'} \right) g(\vomega) ,
	\end{align*}
	\begin{align*} 
		R_{\avec_S} &= \sum_{\vomega \in \Omega^n}  \left( \prod_{i \in N} F_{i,a_{S,i},\omega_i} \right) g(\vomega) \\
		&= \sum_{\vomega \in \Omega^n}  \left(  \prod_{\substack{i \in N: \\a_{S,i}=a_{1,i}}} \sum_{\omega' \in \Omega} F_{i, a_{2,i},\omega'} \mu^{\omega'}_{\omega-\omega'} \right)  \left(  \prod_{\substack{i \in N: \\ a_{S,i}\neq a_{1,i}}}  F_{i,a_{2,i},\omega_i} \right)  g(\vomega), 
	\end{align*}
	and
	\begin{align*}
		R_{\avec_{S'}} & = \sum_{\vomega \in \Omega^n}  \left(  \prod_{i \in N} F_{i,a_{S',i},\omega_i} \right)  g(\vomega) \\
		& = \sum_{\vomega \in \Omega^n}  \left(  \prod_{\substack{i \in N: \\ {a_{S',i}=a_{1,i}}}}  \sum_{\omega' \in \Omega} F_{i, a_{2,i},\omega'} \mu^{\omega'}_{\omega-\omega'} \right) \left( \prod_{\substack{i \in N: \\a_{S',i}\neq {a_{1,i}}}}  F_{i,a_{2,i},\omega_i} \right)  g(\vomega).
	\end{align*}
	
	In the following, for ease of presentation, given a pair of tuples of agents' outcomes $\vomega, \vomega' \in \Omega^n$ such that $\omega_i \geq \omega_i'$ for every $i \in N$, we denote by $ \vomega_1^{\vomega,\vomega'}, \vomega_2^{\vomega,\vomega'} \in \Omega^n$ another pair of tuples of agents' outcomes, which depend on $\vomega$, $\vomega'$ and
	are defined as follows:
	\begin{itemize}
		\item if agent $i \in N$ is such that $a_{1,i} = a_{S,i}$ and $a_{2,i} = a_{S', i}$, then it holds $ \omega_{1,i}^{\vomega,\vomega'} = \omega_i$ and $\omega_{2,i}^{\vomega,\vomega'} = \omega'_i$;
		\item if agent $i \in N$ is such that $a_{1,i} = a_{S',i}$ and $a_{2,i} = a_{S, i}$, then it holds $ \omega_{1,i}^{\vomega,\vomega'} = \omega'_i$ and $ \omega_{2,i}^{\vomega,\vomega'} = \omega_i$.
	\end{itemize}
	Notice that, as it is easy to check, it holds $\vomega = \vomega_1^{\vomega,\vomega'} \wedge \vomega_2^{\vomega,\vomega'}$ and $\vomega = \vomega_1^{\vomega,\vomega'} \vee \vomega_2^{\vomega,\vomega'}$.
	Thus, it is also the case that $g(\vomega)+g(\vomega')\le g( \vomega_1^{\vomega,\vomega'}) +g(  \vomega_2^{\vomega,\vomega'})$, sine the reward function $g$ is IR-supermodular and hence supermodular (see Definition~\ref{def:sub}). 
	%
	Let $N_1\coloneqq \{i \in N:a_{S,i}=a_{1,i}\}$, and $N_2\coloneqq \{i \in N:a_{S',i}=a_{1,i}\}$.
	In order to conclude the proof, we show that the following holds:
	\begin{subequations}
		\begin{align}
			R_{\va_1}  + R_{\va_2}&=\sum_{\vomega \in \Omega^n} \left( \prod_{i \in N} F_{i, a_{1,i},\omega_i} \right) g(\vomega) + \sum_{\vomega \in \Omega^n} \left( \prod_{i \in N} F_{i, a_{2,i},\omega_i} \right)g(\vomega)\\
			&=\sum_{\vomega \in \Omega^n}  \left( \prod_{i \in N}  \sum_{\omega' \in \Omega} F_{i, a_{2,i},\omega'} \mu^{i,\omega'}_{\omega_i-\omega'} \right) g(\vomega)  + \sum_{\vomega \in \Omega^n}  \left( \prod_{i \in N} F_{i,a_{2,i},\omega_i} \right) g(\vomega) \label{eq:fosd1}\\
			&= \sum_{\vomega \in \Omega^n} \sum_{\vomega'\in \Omega^n} \left(  \prod_{i \in N} F_{i, a_{2,i},\omega'_i}   \right) \left( \prod_{i \in N}  \mu^{i,\omega'_i}_{\omega_i-\omega'_i} \right)  g(\vomega)  + \sum_{\vomega \in \Omega^n}  \left(  \prod_{i \in N} F_{i,a_{2,i},\omega_i} \right) g(\vomega) \\
			&= \sum_{\vomega' \in \Omega^n} \left( \prod_{i \in N}   F_{i, a_{2,i},\omega'_i} \right) \left[ \sum_{\vomega\in \Omega^n} \left(   \prod_{i \in N}  \mu^{i,\omega'_i}_{\omega_i-\omega'_i} \right) g(\vomega) + g(\vomega') \right] \\
			&= \sum_{\vomega' \in \Omega^n} \left( \prod_{i \in N}   F_{i, a_{2,i},\omega'_i} \right) \left[ \sum_{\vomega\in \Omega^n} \left(   \prod_{i \in N}  \mu^{i,\omega'_i}_{\omega_i-\omega'_i} \right) g(\vomega) + \sum_{\vomega\in \Omega^n}\left(   \prod_{i \in N}  \mu^{i,\omega'_i}_{\omega_i-\omega'_i} \right) g(\vomega') \right] \label{eq:sum_1}\\
			&\ge \sum_{\vomega' \in \Omega^n}\left(  \prod_{i \in N}   F_{i, a_{2,i},\omega'_i} \right)  \sum_{\vomega \in \Omega^n}  \left( \prod_{i \in N}  \mu^{i,\omega'_i}_{\omega_i-\omega'_i} \right) \Big[ g(\vomega_1^{\vomega,\vomega'}) + g(\vomega_2^{\vomega,\vomega'})  \Big] \label{eq:fosdge}\\
			& =  \sum_{\vomega' \in \Omega^n} \left( \prod_{i \in N}  F_{i,a_{2,i},\omega'_i} \right) \left[  \sum_{\substack{\vomega \in \Omega^n:\\ \omega_i=\omega'_i \forall i \in N_2}}  \prod_{\substack{i \in N: \\a_{S,i}=a_{1,i}}}  \mu^{i,\omega'_i}_{\omega_i-\omega'_i} \, g(\vomega_1^{\vomega,\vomega'}) \right.\nonumber\\
			& \quad\quad \left. +\sum_{\substack{\vomega \in \Omega^n:\\\omega_i=\omega'_i \forall i \in N_1}}  \prod_{\substack{i \in N: \\a_{S',i}=a_{1,i}}}  \mu^{i,\omega'_i}_{\omega_i-\omega'_i} \, g(\vomega_2^{\vomega,\vomega'}) \right] \label{eq:prob}\\
			& =  \sum_{\vomega \in \Omega^n} \left( \prod_{\substack{i \in N: \\a_{S,i} \neq a_{1,i}}}  F_{i,a_{2,i},\omega_i} \right)   \sum_{\substack{\vomega' \in \Omega^n:\\\omega_i=\omega'_i \forall i\in N_2} } \prod_{\substack{i \in N: \\a_{S,i}=a_{1,i}}} F_{i,a_{2,i},\omega'_i} \, \mu^{i,\omega'_i}_{\omega_i-\omega'_i} \, g(\vomega_1^{\vomega,\vomega'}) \nonumber\\
			&\quad\quad+\sum_{\vomega' \in \Omega^n} \left( \prod_{\substack{i \in N: \\a_{S',i} \neq a_{1,i}}}  F_{i,a_{2,i},\omega'_i} \right) \sum_{\substack{\vomega \in \Omega^n:\\\omega_i=\omega'_i \forall i \in N_1}}  \prod_{\substack{i \in N: \\a_{S',i}=a_{1,i}}}  \mu^{i,\omega'_i}_{\omega_i-\omega'_i} \, g(\vomega_2^{\vomega,\vomega'}) \\
			& =  \sum_{\vomega \in \Omega^n} \left( \prod_{\substack{i \in N: \\a_{S,i} \neq a_{1,i}}}  F_{i,a_{2,i},\omega_i} \right)   \sum_{\substack{\vomega' \in \Omega^n:\\\omega_i=\omega'_i \forall i\in N_2}}  \prod_{\substack{i \in N: \\a_{S,i}=a_{1,i}}} F_{i,a_{2,i},\omega'_i} \, \mu^{i,\omega'_i}_{\omega_i-\omega'_i} \, g(\vomega) \nonumber\\
			&\quad\quad+\sum_{\vomega \in \Omega^n} \left( \prod_{\substack{i \in N: \\a_{S',i} \neq a_{1,i}}}  F_{i,a_{2,i},\omega_i} \right) \sum_{\substack{\vomega' \in \Omega^n:\\\omega_i=\omega'_i \forall i \in N_1}}  \prod_{\substack{i \in N: \\a_{S',i}=a_{1,i}}}  \mu^{i,\omega'_i}_{\omega_i-\omega'_i} \, g(\vomega) \label{eq:bo} \\
			& =  \sum_{\vomega \in \Omega^n} \left( \prod_{\substack{i \in N: \\a_{S,i} \neq a_{1,i}}}  F_{i,a_{2,i},\omega_i} \right) \left(   \prod_{\substack{i \in N: \\a_{S,i}=a_{1,i}}}  \sum_{\omega' \in \Omega} F_{i,a_{2,i},\omega'_i} \, \mu^{i,\omega'_i}_{\omega_i-\omega'}\right) \, g(\vomega) \nonumber \\
			&\quad\quad +  \sum_{\vomega \in \Omega^n} \left( \prod_{\substack{i \in N: \\a_{S',i} \neq a_{1,i}}}  F_{i,a_{2,i},\omega_i} \right) \left(   \prod_{\substack{i \in N: \\a_{S',i}=a_{1,i}}}  \sum_{\omega' \in \Omega} F_{i,a_{2,i},\omega'_i} \, \mu^{i,\omega'_i}_{\omega_i-\omega'}\right) \, g(\vomega)  \\
			&= R_{\va_S}+ R_{\va_{S'}},\label{eq:last}
		\end{align}
	\end{subequations}
	where Equation~\eqref{eq:fosd1} follows from the definition of $R_{a_1}$, Equation~\eqref{eq:sum_1} comes from the fact that $\sum_{\vomega \in \Omega^n}  \prod_{i \in N} \mu^{i,\omega'_i}_{\omega_i-\omega'_i} = 1$, Equation~\eqref{eq:fosdge} holds since $g(\vomega)+g(\vomega')\le g( \vomega_1^{\vomega,\vomega'}) +g(  \vomega_2^{\vomega,\vomega'})$ by supermodularity, Equation~\eqref{eq:prob} follows from $\vomega_{1}^{\vomega,\vomega'}=\vomega_{1}^{\vomega,\vomega''}$ whenever $\omega'_i=\omega''_i$ for all $i \in N_1$,   $\vomega_2^{\vomega,\vomega'}=\vomega_2^{\vomega,\vomega''}$ whenever $\omega'_i=\omega''_i$ for all $i \in N_2$, and $\sum_{\omega \in \Omega} \mu^{i,\omega'}_{\omega_i-\omega'}=1$, Equation~\eqref{eq:bo} comes from $\vomega_1^{\vomega,\vomega'}=\vomega_1^{\vomega,\vomega''}$ whenever $\omega'_i=\omega''_i$ for all $i \in N_1$ and  $\vomega_2^{\vomega,\vomega'}=\vomega_2^{\vomega,\vomega''}$ whenever $\omega'_i=\omega''_i$ for all $i \in N_2$, while Equation~\eqref{eq:last} follows from the definition of $R_{\va_S}$+ $R_{\va_{S'}}$.	
	%
	%
	%
\end{proof}

\IR*

\begin{proof}
	By Theorem~\ref{thm:eq_matr}, computing a utility-maximizing contract in a principal-multi-agent problem instance $I \coloneqq (N,\Omega,A)$ can be reduced in polynomial time to the problem of maximizing a suitably-defined set function $f^I$ over a particular $1$-partition matroid $\M^I$.
	Moreover, by Lemma~\ref{lm:orderSup}, the function $f^I$ is ordered-supermodular whenever the principal's rewards are specified by an IR-supermodular function and the FOSD condition is satisfied.
	Hence, Corollary~\ref{cor:polyOrderSup} immediately provides a polynomial-time algorithm for finding a utility-maximizing contract.
\end{proof}

%% file: appendix_hard_dr_sub.tex
\section{Proof of Theorem~\ref{thm:hard_is}}\label{app:hard_dr_sub}

To prove the theorem, we employ a reduction from a {promise problem} related to the problem of finding large independent sets in graphs, whose definition follows.

\begin{definition}[$\textsf{GAP-IS}_{\alpha}$]
	For every $\alpha \in [0,1]$, we define \emph{$\textsf{GAP-IS}_{\alpha}$} as the following promise problem:
	\begin{itemize}
		\item \textnormal{\texttt{Input:}} An undirected graph $G=(V,E)$ such that either one of the following is true:
		\begin{itemize}
			\item there exists an independent set (i.e., a subset of vertices such that there is no edge connecting two of them) of size at least $|V|^{1-\alpha}$;
			\item all the independent sets have size at most $|V|^\alpha$.
		\end{itemize}
		\item \textnormal{\texttt{Output:}} Determine which of the above two cases hold.
	\end{itemize}
\end{definition}

$\textsf{GAP-IS}_{\alpha}$ is known to be \NPHard~for any $\alpha  > 0$~\citep{hastad1999clique,Zuckerman2007linear}.

Now, we are ready to prove of Theorem~\ref{thm:hard_is}.

\begin{proof}[Proof of Theorem~\ref{thm:hard_is}]
	Given a constant $\alpha>0$, we reduce from the problem $\textsf{GAP-IS}_{\alpha}$.
	
	Our construction is such that, if the $\textsf{GAP-IS}_{\alpha}$ instance $G=(V,E)$ admits an independent set of size at least $|V|^{1-\alpha}$, then the corresponding  contract design problem admits a solution providing the principal with an overall expected utility at least of $\delta |V|^{1-\alpha}$, where $\delta$ will be defined in the following.
	Otherwise, if all the independent sets have size at most $|V|^{\alpha}$, then  any contract provides the principal with an overall expected utility at most of $\delta |V|^{\alpha}$. Moreover, we will see that in the multi-agent principal-agent problem in our reduction  it holds $n =|V|$.   Since $\textsf{GAP-IS}_{\alpha}$ is \NPHard \ for each constant $\alpha>0$ this is sufficient to prove the statement.
	
	%

	\paragraph{Construction.}
	%
	%
	Given an instance of $\textsf{GAP-IS}_{\alpha}$ $G=(V,E)$, we build an instance of the multi-agent principal-agent problem as follows.
	For each vertex $v \in V$, there exists an agent $n_v$ with actions $a_1$ and $a_0$.
	The outcome space is given by $\mathbb{R}_+$.
	Then, for each agent $v \in V$ action $a_1$ induces deterministically outcome $\omega^1=1$, \emph{i.e.}, $F_{n_v,a_1,\omega^1}=1$.
	Moreover, action $a_1$ has cost $1-\delta$, \emph{i.e.}, $c_{n_v,a_1}=1-\delta$, where $\delta=\frac{1}{|V|^2}$.
	For each agent $v \in V$ action $a_0$ induces deterministically outcome $\omega^0=0$, \emph{i.e.}, $F_{n_v,a_0m\omega^0}=1$.
	Moreover, action $a_0$ has cost $0$, \emph{i.e.}, $c_{n_v,a_0}=0$.
	Given a node $v \in V$, let $k_v \le |V|$ be the degree of node $v$.
	Finally, the utility function $g$ is defined as \[g(\vomega)=\sum_{(u,v)\in E}  \max \{  \frac{1}{k_u} \omega_u,  \frac{1}{k_v} \omega_v \}.\]
	Notice that the function is DR-submodular since it is the sum of DR-submodular functions. Indeed, given and edge $e=(u,v)$, $ \max \{  \frac{1}{k_u} \omega_u,  \frac{1}{k_v} \omega_v \}$ is the maximum of two linear functions and hence is DR-submodular.
	
	\paragraph{Completeness.}
	Suppose that there exists an independent set $V^\ast \subseteq V$ of $G$ with size at least $ |V|^{1-\alpha}$.
	We can build a contracts $(p,a)$, $p \in \mathbb{R}^{n \times m}_+$, $\va \in A^n$ such that for each $v \in V^*$, it holds $p_{n_v,\omega^1}=1-\delta$, while all the other payments are set to $0$. 
	Finally, we recommend to all the agent $n_v$, $v \in V^*$, to play $a_1$, \emph{i.e.}, $a_{n_v}=a_1$, and to all the agents $n_v$, $v \in V\setminus V^*$ to play $a_0$, \emph{i.e.}, $a_{n_v}=a_0$.
	It is easy to see that the action profile $\va$ is such that $a_i$ is IC under $p$ for each agent $i \in N$.
	The total reward is 
	\begin{align*}
		\sum_{(u,v)\in E}  \max \{  \frac{1}{k_u} \omega_u,  \frac{1}{k_v} \omega_v \}&= \sum_{v\in V^*} \sum_{u \in V: (u,v)\in E} \max \{  \frac{1}{k_u} \omega_u,  \frac{1}{k_v} \omega_v\}\\
		&=  \sum_{v\in V^*} \sum_{u \in V: (u,v)\in E}   \frac{1}{k_v} 1\\
		&=|V^*| |k_v| \frac{1}{k_v} \\
		&=|V^*|,
	\end{align*}
	where the second inequality holds since for each $v \in V^*$ we have that $\omega_{v}=1$ and $\omega_{u}=0$ for each $u:(u,v)\in V^*$. 
	Moreover, the total payment is given by $\sum_{v \in V^*} (1-\delta)= |V^*|(1-\delta)$.
	Thus, the total principal's utility is given by $ |V^\ast|-  (1-\delta) |V^\ast|= \delta |V^\ast|$.
	
	\paragraph{Soundness.}
	We prove that, if all the independent sets of $G$ have size at most $ |V|^{\alpha}$, then the principal's expected utility is at most $\delta |V|^{\alpha}$ for any  contract $p \in \mathbb{R}_+{n\times m}$, $\va \in A^n$.
	First, we show that if the contract incentivizes two agents $n_u$ and $n_v$ with $(u,v) \in E$, \emph{i.e.}, $u$ and $v$ are adjacent vertexes, to play action $a_1$, then the principal's utility is negative.
	Let $\bar V$ be the set of nodes relative to agents incentivized to play $a_1$, \emph{i.e.}, the set of $i \in N$ such that $a_i=a_1$, and $\bar E$ be the set of edges connecting two nodes in $\bar V$.
	Then, the principal's reward is at most 
	\begin{align*}
	 \sum_{(u,v)\in E}  \max \{  \frac{1}{k_u} \omega_u,  \frac{1}{k_v} \omega_v \}&= \sum_{(u,v)\in E \setminus \bar E} \max \{  \frac{1}{k_u} \omega_u,  \frac{1}{k_v} \omega_v\} + \sum_{(u,v)\in \bar E } \max \{  \frac{1}{k_u} \omega_u,  \frac{1}{k_v} \omega_v\}\\
	 &=\sum_{v\in \bar V: (u,v)\in E \setminus \bar E}  \frac{1}{k_v} + \sum_{(u,v)\in \bar E } [ \frac{1}{k_u} \omega_u+  \frac{1}{k_v} \omega_v-1/|V| ]\\
	 &=\sum_{v\in \bar V: (u,v)\in E \setminus \bar E}  \frac{1}{k_v} + \sum_{(u,v)\in \bar E } [ \frac{1}{k_u} \omega_u+  \frac{1}{k_v} \omega_v]-1/|V| \\
	 &=\sum_{v\in \bar V: (u,v)\in E \setminus \bar E}  \frac{1}{k_v} + \sum_{v \in \bar V:(u,v)\in \bar E } \frac{1}{k_v}  -1/|V|\\
	 &  =\sum_{v\in \bar V} \sum_{u \in V: (u,v)\in E} \frac{1}{k_v} - 1/|V|\\
	 &\le |\bar V|-1/|V|.
	\end{align*}
	At the same time the payment is at least $(1-\delta)|\bar V|$ since for each agent $n_v$, $v \in \bar V$ it holds $p_{n_v,\omega^1}\ge 1-\delta$.  Hence, the principal's utility is at most $|\bar V|-1/|V|-(1-\delta)|\bar V|= \delta |\bar V|-1/|V|< 0$.
	
	Hence,  in any contract with positive utility there are not two agents $n_{v}$, $n_u$ relative to adjacent vertexes, \emph{i.e.}, such that $(v,u) \in E$ playing action $a_1$.
	Since all the independent sets has size at most $|V|^\alpha$, this implies that $|\bar V| \le  |V|^\alpha$.
	Then, the reward of the contract is given by $\sum_{v \in \bar V} \sum_{u:(u,v) \in E} \frac{1}{k_v} 1=|\bar V|$. Moreover, the payment is at least $(1-\delta)|\bar V|$ since for each agent $n_v$, $v \in \bar V$ it holds $p_{n_v,\omega^1}\ge 1-\delta$.   
	However, the principal's utility is at most $|\bar V|-(1-\delta)|\bar V|=\delta |\bar V|\le \delta |V|^{\alpha}$.
\end{proof}

%% file: appendix_dr.tex
\section{Proofs Omitted from Section~\ref{sec:dr_sub}}

\subdr*

\begin{proof}
	By letting $\mathsf{f}^I: 2^{\cG^I}\rightarrow \mathbb{R}_+$ and $\mathsf{l}^I:2^{\cG^I}\rightarrow \mathbb{R}$ be defined so that $\mathsf{f}^I(S) \coloneqq R_S$ and $\mathsf{l}^I(S) \coloneqq \sum_{(i,a)\in S} \widehat P_{i,a}$ for every $S \subseteq \cG^I$, in order to prove the statement it is sufficient to show that $\mathsf{f}^I$ is a monotone-increasing submodular function (notice that $\mathsf{l}^I$ is linear by definition).
	
	It is easy to check that $\mathsf{f}^I$ is monotone-increasing, since the reward function $g$ is increasing by assumption (see Assumption~\ref{ass:inc_rew}).
	Thus, we are left to show that $\mathsf{f}^I$ is also submodular.

	In the following, for ease of presentation, given an agent $i \in N$ and an outcome $\omega \in \Omega$, we let $\vomega^{i,\omega} \in \Omega^n$ be the tuple of agents' outcomes such that $\omega^{i,\omega}_i =\omega$ and $\omega^{i,\omega}_j = \vomega_\varnothing$ for all $j \in N : j \neq i$.
	Moreover, given two tuples $\vomega, \vomega' \in \Omega^n$, we let $\vomega +  \vomega'$ be the tuple whose $i$-th outcome is $\omega_i+\omega'_i$.

	In order to prove that $\mathsf{f}^I$ is submodular, we need to show that, for any two subsets $S\subset S'\subseteq \cG^I$ and element $(i,a)\in \cG^I$, it holds $\mathsf{f}^I(S\cup \{(i,a)\})-\mathsf{f}^I(S)\le \mathsf{f}^I(S' \cup \{(i,a)\})-\mathsf{f}^I(S')$:
	%
	%
	\begin{align*}
		\mathsf{f}^I(S'\cup\{(i,a)\}) & -\mathsf{f}^I(S')  = R_{S \cup \{(i,a)\} } - R_{S'}\\
		& =\sum_{\vomega \in \tilde\Omega^n} r_\vomega \prod_{j \in N} F_{j,S'\cup\{(i,a)\}, \omega_j}- \sum_{\tilde \vomega \in \Omega^n} r_\vomega \prod_{j \in N} F_{j,S',\omega_j}\\
		&=\sum_{\vomega \in \tilde \Omega^n} \sum_{\vomega' \in \tilde \Omega^n} \sum_{\omega \in \Omega} r_{\vomega+ \vomega' +\vomega^{i,\omega}} \, F_{i,a, \omega} \, \left( \prod_{j  \in N} F_{j,S,\omega_j} \right) \left( \prod_{j  \in N} F_{j,S'\setminus  S,\omega_j} \right)\\
		&\quad - \sum_{\vomega \in \tilde \Omega^n} \sum_{ \vomega' \in \tilde \Omega^n} r_{\vomega+ \vomega' }   \left( \prod_{j  \in N} F_{j,S,\omega_j} \right) \left( \prod_{j  \in N} F_{j,S'\setminus  S,\omega_j} \right)\\
		&=\sum_{\vomega \in \tilde \Omega^n} \sum_{ \vomega' \in \tilde \Omega^n} \sum_{ \omega \in \Omega} \, F_{i,a,\omega} \left( \prod_{j  \in N} F_{j,S,\omega_j} \right) \left( \prod_{j  \in N} F_{j,S'\setminus  S,\omega_j} \right) \left( r_{\vomega+ \vomega' +\vomega^{i,\omega}} - r_{\vomega+  \vomega'} \right) \\
		& \le \sum_{\vomega \in \tilde \Omega^n} \sum_{ \vomega' \in \tilde \Omega^n} \sum_{\omega \in \Omega}  F_{i,a,\omega} \, \left( \prod_{j  \in N} F_{j,S,\omega_j} \right) \left( \prod_{j  \in N} F_{j,S'\setminus  S,\omega_j} \right) \left( r_{\vomega +\vomega^{i,\omega}} - r_{\vomega } \right)  \\
		&=\sum_{\vomega \in \tilde \Omega^n} \sum_{ \omega \in \Omega}  F_{i,a, \omega} \left( \prod_{j  \in N} F_{j,S,\omega_j} \right) \left( r_{\vomega +1_i(\bar \omega)} - r_{\vomega } \right)   \\
		& =\sum_{\vomega \in \tilde \Omega^n} \sum_{ \omega \in \Omega}  F_{i,a, \omega} \left( \prod_{j  \in N} F_{j,S,\omega_j} \right) r_{\vomega +\vomega^{i,\omega}} - \sum_{\vomega \in \tilde \Omega^n} \left( \prod_{j  \in N} F_{j,S,\omega_j} \right)    r_{\vomega }   \\
		&= \sum_{\vomega \in \tilde \Omega^n} r_\vomega \prod_{j \in N} F_{j,S\cup\{(i,a), \omega_j \}}- \sum_{\vomega \in \tilde \Omega^n} r_\vomega \prod_{j \in N} F_{j,S, \omega_j}\\
		& =\mathsf{f}^I(S \cup\{(i,a)\})-\mathsf{f}^I(S),\\
	\end{align*}
	where the inequality hold by DR-submodularity.
	This concludes the proof.
	%
	%
\end{proof}

\drdue*

\begin{proof}
	The result easily follows by noticing that, thanks to Lemma~\ref{lem:sub_dr}, the problem is a specific case of the ones studied in~\citep{sviridenko2017optimal}.

	In particular, Theorem~3.1~in~\citep{sviridenko2017optimal} shows that there exists a polynomial-time algorithm that, given as input an $\epsilon>0$, a matroid $\M \coloneqq (\cG,\I)$, a monotone-increasing submodular function $\mathsf{f}: 2^{\cG} \rightarrow \mathbb{R}_+$, and a linear function $\mathsf{l}: 2^{\cG} \rightarrow \mathbb{R}$, outputs an independent set $S \in \I$ satisfying
	$\mathsf{f}( S ) + \mathsf{l} ( S ) \ge (1 -1/e)\mathsf{f} ( S' ) + \mathsf{l} (S')-\epsilon \hat v$ for every $S' \in \I$ with high probability, where, for ease of presentation, we let $\hat v \coloneqq \max\{ \max_{x \in \cG}(\mathsf{f}(\{ x \}),\max_{x \in \cG}|\mathsf{l}(\{ x \})| \} $.

	It is easy to see that, by definition of reward function, it holds $\max_{(i,a) \in \cG^I} \mathsf{f}^I( \{ (i,a) \} ) \le 1$.
	Moreover, it is always possible to build a matroid which is equivalent (for our purposes) to $\M^I$ and satisfies $\max_{(i,a) \in \cG^I}|\mathsf{l}^I( \{ (i,a) \}  )| \le 1$.
	To do that, let $\tilde \cG^I \subseteq \cG^I$ be the set of elements $(i,a)$ such that $\widehat P_{i,a}> 1$.
	Then, any independent set including an element of $\tilde \cG^I$ cannot be optimal, since it has negative value (recall that the values of $\textsf{f}^I$ are in $[0,1]$).
	Hence, we can optimize over the matroid that only includes the elements in $\cG^I \setminus \tilde \cG^I$, so that we get $\hat v\le 1$.
	This concludes the proof.
\end{proof}

%% file: appendix_bayesian.tex
\section{Proofs Omitted from Section~\ref{sec:bayesian}}

\smallSup*

\begin{proof}
	Consider a menu of randomized contract $\hat  \gamma^{\vtheta}$ for each $\theta \in \Theta^n$ such that given a $\hat \vtheta \in \Theta^n$ and $i \in N$, there exists two contracts $p,p' \in \textnormal{supp}(\gamma^{\hat \vtheta})$ such that $p_i \neq p'_i$ and  $ \{p_i , p'_i\} \subseteq \mP^{i, \hat \theta_i,a}$. Let $ \bar p$ be such that $\bar p_j= p_j$ for each $j \neq i$ and $\bar p_i = \hat  \gamma^{\hat \vtheta}_{p_i}p_i+\hat  \gamma^{\hat \vtheta}_{p'_i}p'_i $. Moreover,  let $\bar p'$ be such that $\bar p'_j= p'_j$ for each $j \neq i$ and $\bar p'_i = \hat  \gamma^{\hat \vtheta}_{p_i}p_i+\hat  \gamma^{\hat \vtheta}_{p'_i}p'_i $.
	We build a DSIC menu of randomized contracts $\gamma^{\vtheta}$ for each $\theta \in \Theta^n$ with at least the same principal's utility such that 
	\begin{itemize}
		\item $\gamma^{\vtheta}= \hat  \gamma^{\vtheta}$ for each $\theta \neq \hat\theta$,
		\item $\gamma^{\hat \vtheta}_{\bar p}=\hat \gamma^{\hat \vtheta}_{p}+\hat \gamma^{\hat \vtheta}_{\bar p}$,
		\item  $\gamma^{\hat \vtheta}_{\bar p'}= \hat\gamma^{\hat\vtheta}_{  p'}+\hat \gamma^{\hat\vtheta}_{\bar p'}$,
		\item $ \gamma^{\hat \vtheta}_{\hat p}=\hat \gamma^{\hat \vtheta}_{\hat p}$ for each $p \notin\{\bar p,\bar p',p\}$. 
		\end{itemize}
	It is easy to see that $\gamma^{\vtheta}$ satisfies $$\left| \left\{  p_i \, \mid \, p \in \textnormal{supp}(\gamma^{\vtheta}) \wedge p \in \mP^{i,\theta_i,a}  \right\} \right| < \left| \left\{  p_i \, \mid \, p \in \textnormal{supp}(\hat\gamma^{\vtheta}) \wedge p \in \mP^{i,\theta_i,a}  \right\} \right| -1.$$
	Moreover, $\gamma^{\vtheta}$ for each $\theta \in \Theta^n$ provides the same utility of $\hat \gamma^{\vtheta}$ for each $\theta \in \Theta^n$.
	To conclude the proof we need to proof that $\gamma^{\vtheta}$ for each $\theta \in \Theta^n$ is DSIC. Following an analysis similar to the one in \cite{castiglioni2022bayesian} we can show that replacing the marginal contracts $p_i$ and $p'_i$ with the weighted combination $\bar p_i$ the DSIC constraint continue to hold.
	Applying this operation until such two contracts does not exist is sufficient to prove the statement.
\end{proof}

\bounded*

\begin{proof}
	For every $I \coloneqq (N,\Theta,\Omega,A)$, agent $i \in N$, type $\theta\in \Theta$, and inducible action $a \in A_{i,\theta}$, the set $\mathcal{P}^{i,\theta,a}$ of contracts under which action $a$ is IC can be defined by means of a system of linear inequalities such that its number of variables, its number of inequalities, and the size of the binary representation of its coefficients can all be bounded by polynomials in $|I|$.
	Hence, given that $\mathcal{P}^{i,\theta,a}$ cannot be empty (otherwise $a \in A_{i,\theta}$ would be contradicted), there must exist a contract $p \in \mathcal{P}^{i,\theta,a}$ such that, for every outcome $\omega \in \Omega$, the payment $p_{i,\omega}$ is upper bounded by a $O(2^{\poly(|I|)})$ term, where $\poly(|I|)$ is a polynomial in the size $|I|$ (in terms of number of bits) of instance $I$ (this easily follows from standard LP arguments, see, \emph{e.g.},~\citep{bertsimas1997introduction}). 
	The result is readily proved by choosing a suitable function $\tau : \mathbb{N} \to \mathbb{R}$ so that such upper bound holds for every instance $I$, agent $i \in N$, type $\theta \in \Theta$, and inducible action $a \in A_{i,\theta}$.
	%
\end{proof}

\lpGeq*

\begin{proof}
	To prove the result, we show that, given any feasible solution to Problem~\eqref{pr:types}, it is possible to recover a feasible solution to LP~\eqref{lp:types} having the same objective function value.

	Let $(t_{\vtheta, \avec}$, $\xi_{i,\vtheta,a}$, $p_{i,\vtheta,a,\omega})$ be a feasible solution to Problem~\eqref{pr:types}.
	Then, we define a solution to LP~\eqref{lp:types} by letting $y_{i,\vtheta,a,\omega} = \xi_{i,\vtheta,a} \, p_{i,\vtheta,a,\omega}$ for every agent $i \in N$, tuple of agents' types $\vtheta \in \tilde \Theta^n$, inducible action $a \in A_{i,\theta_i}$, and outcome $\omega \in \Omega$.
	Additionally, all the variables that also appear in Problem~\eqref{pr:types} keep their values, while variables $\gamma_{i,\vtheta, \theta, ,a}$ are defined so that they are equal to their corresponding terms in the sums appearing in the right-had sides of Constraints~\eqref{eq:linearize}.
	It is immediate to see that the solution defined above is indeed feasible for LP~\eqref{lp:types}, after noticing that, for every $i \in N$, $\vtheta \in \tilde \Theta^n$, and action $a \in A$ which is \emph{not} inducible for an agent $i$ of type $\theta_i$, it holds that $\xi_{i,\vtheta,a} = 0$.
	Indeed, since Constraints~\eqref{eq:linearize} hold, if $\xi_{i,\vtheta,a}>0$ then there exists a contract $p \in \mathbb{R}_+^{n \times m}$ under which action $a$ is IC for an agent $i$ of type $\theta_i$, and, thus, $a\in A_{i,\theta_i}$.

	It is easy to check that the feasible solution to LP~\eqref{lp:types} defined above has exactly the same objective function value as its corresponding feasible solution to Problem~\eqref{pr:types}.
	Thus, the result is readily proved by observing that the objective functions of Problem~\eqref{pr:types} and LP~\eqref{lp:types} are continuous and, for any $\epsilon > 0$, there always exists a feasible solution to Problem~\eqref{pr:types} with value at least $\SUP - \epsilon$.
\end{proof}

\fromIRR*

\begin{proof}
	Let $(t_{\vtheta, \avec}, \xi_{i,\vtheta,a}, y_{i,\vtheta,a,\omega}, \gamma_{i,\vtheta, \theta, ,a})$ be a feasible solution to LP~\eqref{lp:types}.
	Moreover, let us define $W$ as the set of tuples $w=(i,\vtheta,a)$ such that $y_{i,\vtheta,a,\omega}>0$ and $\xi_{i,\vtheta,a}=0$ (\emph{i.e.}, the tuples of indexes identifying the pairs of variables that do \emph{not} meet regularity conditions).

	As a first step, we show that, for every tuple $w=(i,\vtheta,a) \in W$, it is possible to build a feasible solution to LP~\eqref{lp:types}, which we refer to as $\left( t_{\vtheta, \avec}^w, \xi_{i,\vtheta,a}^w, y_{i,\vtheta,a,\omega}^w, \gamma_{i,\vtheta, \theta, ,a}^w \right)$ for clarity of exposition, such that its corresponding DSIC menu of randomized contracts always recommends action $a$ with probability $1$ to an agent $i$ that truthfully reports their type to be $\theta_i$.
	Since $a \in A_{i, \theta_i}$ thanks to how LP~\eqref{lp:types} is constructed, Lemma~\ref{lem:bounded} says that there exists a contract $p^w \in \mathbb{R}_+^{n \times m}$ (depending on the tuple $w = (i,\vtheta,a)$) such that $a \in A^*_{i,\theta_i}(p^w)$ and $p^w_{i, \omega} \le \tau(|I|)$ for all $\omega\in \Omega$, where $\tau: \mathbb{N} \to \mathbb{R}$ is a suitably-defined function such that $\tau(x)$ is $O(2^{\poly(x)})$.
	Then, let us define $y^w_{i,\vtheta,a,\omega}=p^w_{i,\omega}$ for all $\omega \in \Omega$, while $\xi^w_{i,\vtheta,a}=1$.
	Additionally, for every $\vtheta' \in \tilde \Theta^n$ and $j \in N$ such that $(\vtheta',j) \neq (\vtheta,i)$, by letting $a' \in A_{j,\theta'_j}$ be any action that is inducible for an agent $j$ of type $\theta'_j$, we define $\xi^w_{i,\vtheta',a'}=1$.
	Finally, we let $t^w_{\vtheta,\va}=1$.
	It is easy to check that, by suitably defining all the unspecified variables, the solution $\left( t_{\vtheta, \avec}^w, \xi_{i,\vtheta,a}^w, y_{i,\vtheta,a,\omega}^w, \gamma_{i,\vtheta, \theta, ,a}^w \right)$ is feasible for LP~\eqref{lp:types} and it has value at least $-n \, \tau(|I|)$.
	%
	%

	In conclusion, for any $\epsilon > 0$, let us consider a solution $\left( t'_{\vtheta, \avec}, \xi'_{i,\vtheta,a}, y'_{i,\vtheta,a,\omega}, \gamma'_{i,\vtheta, \theta, ,a} \right)$ to LP~\eqref{lp:types} whose components are defined as follows (by applying operations component wise):
	\[
	(1-\epsilon) \, \left( t_{\vtheta, \avec}, \xi_{i,\vtheta,a}, y_{i,\vtheta,a,\omega}, \gamma_{i,\vtheta, \theta, ,a} \right) + \sum_{w \in W} \frac{\epsilon}{|W|} \, \left( t^w_{\vtheta, \avec}, \xi^w_{i,\vtheta,a}, y^w_{i,\vtheta,a,\omega}, \gamma^w_{i,\vtheta, \theta, ,a} \right).
	\]
	It is easy to see that such a solution is feasible for LP~\eqref{lp:types}, it is regular, and its objective function value is at least $\VAL- \epsilon (n \, \tau(|I|)+1)$, where $\VAL$ is the value of the original (irregular) solution.
	Moreover, it can be computed in time polynomial in $|I|$ and $\frac{1}{\epsilon}$, concluding the proof.
	%
	%
\end{proof}

\regularize*

\begin{proof}
	First, let us recall that, given any regular feasible solution $(t_{\vtheta, \avec}, \xi_{i,\vtheta,a}, y_{i,\vtheta,a,\omega}, \gamma_{i,\vtheta, \theta, ,a})$ to LP~\eqref{lp:types}, it is sufficient to set $p_{i,\vtheta,a,\omega}= y_{i,\vtheta,a,\omega}/\xi_{i,\vtheta,a}$ for every $i \in N$, $\vtheta \in \tilde \Theta^n$, $a \in A_{i , \theta_i}$, and $\omega \in \Omega$, in order to recover a feasible solution to Problem~\eqref{pr:types} having the same value.
	Thus, if the given optimal solution to LP~\eqref{lp:types} is regular, then the result immediately follows by Lemma~\ref{lem:lp_geq_sup}, since $\LP = \SUP$.
	Instead, if the given optimal solution is irregular, by applying Lemma~\ref{lem:from_irr_reg} we can recover a regular solution to LP~\eqref{pr:types} with value at least $\LP -\epsilon (n \, \tau(|I|)+1)$ in time polynomial in $|I|$ and $\frac{1}{\epsilon}$, and from that we can easily obtain a feasible solution to Problem~\eqref{pr:types} with value at least $\SUP -\epsilon (n \, \tau(|I|)+1)$ (using the bound in Lemma~\ref{lem:lp_geq_sup}), proving the result.
	%
	%
\end{proof}

\lpsEquiv*

\begin{proof}
	Since LP~\eqref{lp:typesRelaxed} is a relaxation of LP~\eqref{lp:types}, in order to prove the statement it is sufficient to show that, given a feasible solution to LP~\eqref{lp:typesRelaxed}, it is possible to build a feasible solution to LP~\eqref{lp:types} having at least the same value, in time polynomial in the size of the instance.

	Let $(t_{\vtheta, \avec}, \xi_{i,\vtheta,a}, y_{i,\vtheta,a,\omega}, \gamma_{i,\vtheta, \theta, ,a})$ be a feasible solution to LP~\eqref{lp:typesRelaxed}.

	As a first step, we show that, for every tuple of agents' types $\vtheta \in \tilde \Theta^n$, it is possible to compute new values for (some of the) variables $t_{\vtheta, \avec}$ so as to obtain a new set of variables $\hat t_{\vtheta,\va}$ for $\va \in A^{n,\vtheta}$ such that: (i) $\sum_{\va \in A^{n,\vtheta}:a_i=a} \hat t_{\vtheta,\va}=\xi_{i,\vtheta,a}-\sum_{\va:a_i=a} t_{\vtheta,a}$ for every $i \in  N$ and $a \in A_{i,\theta_i}$, (ii) $\hat t_{\vtheta,\va}\ge 0$ for all $\va\in A^{n,\vtheta}$, and (iii) the new values can be computed in polynomial time.
	For ease of presentation, for every tuple of agents' types $\vtheta \in \tilde \Theta^n$, agent $i \in N$, and action $a \in A_{i,\theta_i}$, let $\delta_{i,\vtheta, a} \coloneqq \xi_{i,\vtheta_i,a} - \sum_{\va \in A^{n,\vtheta}: a_i= a} t_{\vtheta,\va}$. Moreover, let $\delta_{\vtheta}\coloneqq 1-\sum_{\va \in A^{n,\vtheta}} t_{\vtheta,\va}$.
	Then, for every agent $i \in N$, it holds: 
	\begin{align*}
		\sum_{a \in A_{i,\theta_i}} \delta_{i,\vtheta,a}= \sum_{a \in A_{i,\theta_i}}\xi_{i,\theta_i,a} - \sum_{a \in A_{i,\theta_i}} \sum_{\va \in A^{n,\vtheta}: a_i= a} t_{\vtheta,\va}  
		=  1- \sum_{\va \in A^{n,\vtheta}} t_{\vtheta,\va}
		=  \delta_{\vtheta}.
	\end{align*}
	Now, let $\bar t_{\vtheta,\va}$ or $\va \in A^{n,\vtheta}$ be variable values identifying a probability distribution over action profiles in $A^{n,\vtheta}$ having marginal probabilities equal to $\delta_{i,\vtheta,a}/\delta_{\vtheta}$, \emph{i.e.}, for every $i \in N$ and $a \in A_{i,\theta_i}$, it holds $\sum_{\va\in A^{n,\vtheta}:a_i=a} \bar t_{\vtheta,\va} =\delta_{i,\vtheta,a}/\delta_{\vtheta}$.
	Notice that, since $\sum_{a \in A_{i,\theta_i}} \delta_{i,\vtheta,a}= \delta_{\vtheta}$ for every $i \in N$, the marginal probabilities are well defined and the (joint) probability distribution exists.
	Moreover, it is easy to see that such values $\bar t_{\vtheta,\va}$ can be computed in polynomial time, since there always exists a probability distribution as desired having a polynomially-sized support.
	%
	Then, let us define $\hat t_{\vtheta,\va}=\delta_{\vtheta} \bar t_{\vtheta,\va}$ for every $\va \in A^{n,\vtheta}$.
	Notice that such values satisfy all the conditions (i)--(iii), since
	\[
	\sum_{\va:a_i=a} \hat t_{\vtheta,\va}=\delta_{\vtheta} \sum_{\va:a_i=a}  \bar t_{\vtheta,\va} =\delta_{i,\vtheta,a}= \xi_{i,\vtheta,a}-\sum_{\va:a_i=a} t_{\vtheta,a},
	\]
	and $\hat t_{\vtheta,\va}\ge 0$ for all $\va\in A^{n,\vtheta}$, as the values $\bar t_{\vtheta,\va}$ identify a probability distribution. 

	Now, let us consider a new solution to LP~\eqref{lp:typesRelaxed}, namely $(t'_{\vtheta, \avec}, \xi_{i,\vtheta,a}, y_{i,\vtheta,a,\omega}, \gamma_{i,\vtheta, \theta, ,a})$, which is such that, for every $\vtheta \in \tilde \Theta^n$ and $ \va \in A^{n,\vtheta}$, it holds $t'_{\vtheta,\va}= t_{\vtheta,\va} +\hat t_{\vtheta,\va}$.
	In the rest of the proof, we show that the solution defined above is feasible for LP~\eqref{lp:types} and it has at least the same objective function value as the original feasible solution to LP~\eqref{lp:typesRelaxed}.

	First, it is easy to check that the objective value does \emph{not} decrease, since each $t'_{\vtheta,\va}$ increases its value with respect to $t_{\vtheta,\va}$ and such variables appear with non-negative coefficients in the objective function.
	Second, for every agent $i\in N$, tuple of agents' types $\vtheta \in \tilde \Theta^n$, and action $ a \in A_{i \theta_i}$, it holds:
	\[
	\sum_{\va \in A^{n,\vtheta}:a_i= a} t'_{\vtheta,\va}= \sum_{\va \in A^{n,\vtheta}:a_i= a} \left( t_{\vtheta,\va}+\hat t_{\vtheta,\va}\right)=\sum_{\va \in A^{n,\vtheta}:a_i= a} t_{\vtheta,\va}+ \xi_{i,\vtheta, a} - \sum_{\va \in A^{n,\vtheta}:a_i= a} t_{\vtheta,\va} =\xi_{i,\vtheta, a} ,
	\]
	where the second-to-last equality comes from condition (i) on $\hat t_{\vtheta,\va}$.
	Thus, such a solution is also feasible for LP~\eqref{lp:types}.
	Moreover, it is easy to see that it can be computed in polynomial time.
\end{proof}

\bayesian*

\begin{proof}
	We start by providing the general procedure underlining the approximation algorithm (see Algorithm~\ref{alg:bisection}).
	The algorithm implements a binary search scheme to find a value $\eta^\star\in [0,1]$ such that a feasibility-version of LP~\eqref{lp:typesDual} with the objective constrained to be at most $\eta^\star$ is ``approximately" feasible, while the same problem with the objective constrained to be at most $\eta^\star-\beta$ is infeasible.
	The constant $\beta \geq 0$ will be specified later in the proof.
	
	Algorithm~\ref{alg:bisection} requires $\log(\beta)$ steps and, at each step, it works by determining, for a given value $\eta \in [0,1]$, whether there exists an ``approximately" feasible solution to the following feasibility-version of LP~\eqref{lp:typesDual}---called \circled{F} for ease of presentation---, which is obtained by dropping the objective function from LP~\eqref{lp:typesDual} and adding a constraint enforcing that the value of the objective is at most $\eta$.
	\begin{equation*}
		\circled{F} \, \, \left\{ \hspace{-1.25mm}\begin{array}{l}
			\displaystyle\sum_{i \in N} \sum_{\theta \in \Theta}  \sum_{\vtheta_{-i} \in \tilde \Theta^n_{-i}} x_{i,\theta, \vtheta_{-i}} \le \eta \\
			\textnormal{Constraints~\eqref{eq:conFirst}---\eqref{eq:conLast}} .
		\end{array} \right.
	\end{equation*}

	The algorithm is initialized with $l=0$, $h=1$.
	At each iteration of the binary search scheme, the feasibility problem \circled{F} with objective $\leq \eta = \frac{l+h}{2}$ is solved via an \emph{ad hoc} implementation of the ellipsoid method (see the following for more details).
	If \circled{F} is found to be infeasible by the ellipsoid method, the algorithm sets $l \gets \eta$.
	Otherwise, if \circled{F} is found to be ``approximately'' feasible, the algorithm sets $h\gets \eta$.
	Then, the procedure is repeated with the updated values of $l$ and $h$, and it terminates when it determines a value $\eta^\star = h$ such that \circled{F} with objective $\leq \eta^\star$ is ``approximately'' feasible and \circled{F} with objective $\leq \eta^\star-\beta$ is infeasible.\footnote{Notice that, in the case in which the \emph{ad hoc} implementation of the ellipsoid method concludes that \circled{F} is ``approximately'' feasible at every iteration, or, similarly, when it always returns infeasible, we can perform a similar analysis by observing that there always exists a feasible solution with objective $1$, while all the feasible solutions have objective at least $0$.}
	
	In the following, we first describe in details the \emph{ad hoc} implementation of the ellipsoid method employed by Algorithm~\ref{alg:bisection}.
	%
	Then, we provide a bound on the principal's expected utility in an optimal DISC menu of randomized contracts in terms of the value $\eta^\star$ found by Algorithm~\ref{alg:bisection}, as well as a $\eta^\star$-depending bound on the value of the solution returned by Algorithm~\ref{alg:bisection}.
	Finally, we put all the bounds together in order to prove the statement of the theorem.

	\begin{algorithm}[!htp]
		\caption{Approximation algorithm introduced in the proof of Theorem~\ref{th:bayesian}}
		\begin{algorithmic}[1]
			\Statex \textbf{Input}: Bayesian principal-multi-agent problem instance $I \coloneqq (N,\Theta,\Omega,A)$; Multiplicative approximation factor $\alpha \in (0,1]$; Additive approximation error $\rho > 0$; Approximate separation oracle $\Ocal_{\alpha}(\cdot,\cdot,\cdot,\cdot)$ for Constraints~\eqref{eq:useOracle}
			\State \textbf{Initialization}: $l\gets0$; $h\gets1$;  $\mathcal{H} \gets\varnothing$; $\mathcal{H}^\star \gets\varnothing$; $\beta \gets \frac{\rho}{4}$; $\epsilon \gets \frac{\rho}{4 |\tilde \Theta^n|}$
			\While{ $h-l>\beta$}
			\State $\eta\gets \frac{h+l}{2}$
			\State Run \emph{ad hoc} ellipsoid method on \circled{F} with objective $\leq \eta$, using additive error $\epsilon$ as input in the calls to the approximate separation oracle $\Ocal_{\alpha}(\cdot,\cdot,\cdot,\cdot)$ for Constraints~\eqref{eq:useOracle}
			\State $\mathcal{H} \gets\{\text{\normalfont Constraints~\eqref{eq:useOracle} found to be violated during the ellipsoid method} \}$
			\If  {ellipsoid method returned infeasible}
			\State $l \gets \eta$; $\mathcal{H}^\star \gets \mathcal{H}$
			\Else{
				\State $h \gets \eta$}
			\EndIf
			\State\Return $\eta^\star \gets h$; Optimal solution to LP~\eqref{lp:typesRelaxed} in which only the variables $t_{\vtheta,\va}$ corresponding to the dual constraints in $\mathcal{H}^\star$ are specified
			\EndWhile
		\end{algorithmic}\label{alg:bisection}
	\end{algorithm}
	
	\paragraph{Implementation of the ellipsoid method}
	Given a point $(x_{i,\theta,\vtheta_{-i}},y_{i,\vtheta,a}, z_{i, \vtheta,\theta,a, a'}, d_{i,\theta,\vtheta_{-i}})$ in the variable domain of LP~\eqref{lp:typesDual} and $\eta \in[0,1]$, our implementation of the ellipsoid method employs an \emph{ad hoc} separation oracle to determine whether the point $(x_{i,\theta,\vtheta_{-i}},y_{i,\vtheta,a}, z_{i, \vtheta,\theta,a, a'}, d_{i,\theta,\vtheta_{-i}})$ is ``approximately'' feasible for problem \circled{F} with objective $\leq \eta$ or there exists a constraint of \circled{F} that is violated in such a point.
	In the latter case, the separation oracle returns the violated constraint.
	
	First, the oracle checks if one among Constraints~\eqref{eq:conFirst}---\eqref{eq:conMid1}~and~Constraints~\eqref{eq:conMid2}---\eqref{eq:conLast} is violated, which can be done in polynomial time by checking them one by one, since such constraints are polynomially many.
	If a violated constraint is found, the oracle returns it.

	If all the Constraints~\eqref{eq:conFirst}---\eqref{eq:conMid1}~and~the~Constraints~\eqref{eq:conMid2}---\eqref{eq:conLast} are \emph{not} violated, the oracle has to check the exponentially-many Constraints~\eqref{eq:useOracle}. 
	In order to do so, for every tuple of agents' types $\vtheta \in \tilde \Theta^n$, the oracle runs the procedure $\Ocal_{\alpha}(\cdot,\cdot,\cdot,\cdot)$, feeding it with the following inputs: the instance $I$, the tuple $\vtheta$, weights $w \in \mathbb{R}^{n \ell}$ such that $w_{i,a} = \min\{y_{i,\vtheta,a},2\}$ for all $i \in N$ and $a \in A_{i,\theta_i}$, and an additive error $\epsilon$.
	If the call $\Ocal_{\alpha}(I,w,\vtheta,\epsilon)$ returns a tuple of agents' actions $\va \in A^{n,\vtheta}$ such that $\lambda_{\vtheta} R_{\vtheta, \va} - \sum_{i \in N} w_{i,a_i} \leq 0$, then it also holds that
	$
	\alpha \lambda_{\vtheta} R_{\vtheta,\va} -\sum_{i \in N} y_{i,\vtheta,a_i} \le \epsilon
	$
	for every $\va \in A^{n,\vtheta}$, since $w_{i,a_i} \leq y_{i,\vtheta,a_i}$ for all $i \in N$ by definition.
	If this happens for every tuple of agents' types $\vtheta \in \tilde \Theta^n$, the separation oracle then concludes that \circled{F} is ``approximately'' feasible, meaning that Constraints~\eqref{eq:useOracle} are satisfied up to a reward-multiplying approximation factor $\alpha$ and an additive error $\epsilon$.
	Instead, if for some $\vtheta \in \tilde \Theta^n$ the call $\Ocal_{\alpha}(I,w,\vtheta,\epsilon)$ returns an action profile $\va \in A^{n,\vtheta}$ such that $\lambda_{\vtheta} R_{\vtheta, \va} - \sum_{i \in N} w_{i,a_i} > 0$, then it must be $w_{i,a_i} \le 1 $ for all $i \in N$ (as rewards belong to $[0,1]$). 
	Hence, it must be $ w_{i,a_i}=y_{i,\vtheta,a_i}$ for all $i \in N$, and, thus, $\lambda_{\vtheta} R_{\vtheta,\va} -\sum_{i \in N} y_{i,\vtheta,a_i}= \lambda_{\vtheta} R_{\vtheta,\va} -\sum_{i \in N} w_{i,a_i} > 0$.
	Then, the separation oracle concludes that the feasibility problem \circled{F} is infeasible and outputs the constraint in Constraints~\eqref{eq:useOracle} related to $\vtheta \in \tilde \Theta^n$ and $\avec \in A^{n,\vtheta}$.

	\paragraph{Bounding the value of an optimal solution}
	Next, prove that $\alpha R_\OPT- P_\OPT-|\tilde \Theta^n| \epsilon \le \eta^\star$ for any optimal DSIC menu of randomized contracts.

	First, let us recall that the binary search scheme in Algorithm~\ref{alg:bisection} terminates with an $\eta^\star\in [0,1]$ such that the \emph{ad hoc} implementation of the ellipsoid method applied to \circled{F} with objective $\leq \eta^\star$ concludes that the problem is ``approximately'' feasible.
	This implies that there exists a point $(x_{i,\theta,\vtheta_{-i}},y_{i,\vtheta,a}, z_{i, \vtheta,\theta,a, a'}, d_{i,\theta,\vtheta_{-i}})$ in the variable domain of LP~\eqref{lp:typesDual} such that Constraints~\eqref{eq:conFirst}---\eqref{eq:conMid1}~and~Constraints~\eqref{eq:conMid2}---\eqref{eq:conLast} are satisfied and, additionally, $\alpha \lambda_{\vtheta} R_{\vtheta,\va} -\sum_{i \in N} y_{i,\vtheta,a_i} \le \epsilon$ for every tuple of agents' types $\vtheta \in \tilde \Theta^n$ and tuple of agents' actions $\va \in A^{n,\vtheta}$.

	In the following, we define a modified version of LP~\eqref{lp:typesDual} (see LP~\eqref{lp:dualviolated} below) and we show that $(x_{i,\theta,\vtheta_{-i}},y_{i,\vtheta,a}, z_{i, \vtheta,\theta,a, a'}, d_{i,\theta,\vtheta_{-i}})$ is a feasible solution to such a problem having value at most $\eta^\star$.
	\begin{subequations}\label{lp:dualviolated}
		\begin{align}
			\min & \,\, \sum_{i \in N} \sum_{\theta \in \Theta} \sum_{\vtheta_{-i} \in \tilde \Theta^n_{-i}} x_{i,\theta, \vtheta_{-i}} \quad \text{s.t.} \quad\\
			&  \alpha \, \lambda_{\vtheta} \, R_{\vtheta,\va}- \sum_{i \in N} y_{i,\vtheta,a_i} \le  \epsilon & \forall  \vtheta \in \tilde \Theta^n, \forall \va \in A^{n,\vtheta} \label{eq:OracleAPX}\\
			&\textnormal{Constraints~\eqref{eq:conFirst}---\eqref{eq:conMid1} and ~\eqref{eq:conMid2}---\eqref{eq:conLast}.} \nonumber
		\end{align}
	\end{subequations}
	Since Constraints~\eqref{eq:conFirst}---\eqref{eq:conMid1} and ~\eqref{eq:conMid2}---\eqref{eq:conLast} are satisfied by $(x_{i,\theta,\vtheta_{-i}},y_{i,\vtheta,a}, z_{i, \vtheta,\theta,a, a'}, d_{i,\theta,\vtheta_{-i}})$, we only need to show that also Constraints~\eqref{eq:OracleAPX} are satisfied by such a point.
	By contradiction, suppose that Constraint~\eqref{eq:OracleAPX} relative to $\vtheta \in \tilde \Theta^n$ and $\va \in A^{n,\vtheta}$ is violated by $(x_{i,\theta,\vtheta_{-i}},y_{i,\vtheta,a}, z_{i, \vtheta,\theta,a, a'}, d_{i,\theta,\vtheta_{-i}})$.
	Then, it must the case that $\alpha \lambda_{\vtheta} R_{\vtheta, \va} -\sum_{i \in N} y_{i,\vtheta,a_i}-\epsilon > 0$, contradicting the fact the ellipsoid method classified problem \circled{F} with objective $\leq \eta^\star$ as ``approximately'' feasible.
	%
	This shows that $(x_{i,\theta,\vtheta_{-i}},y_{i,\vtheta,a}, z_{i, \vtheta,\theta,a, a'}, d_{i,\theta,\vtheta_{-i}})$ is feasible for LP~\eqref{lp:dualviolated}.
	%
	%

	The dual formulation of LP~\eqref{lp:dualviolated} reads as follows:
	\begin{subequations}\label{lp:types_dual}
		\begin{align}
			\max & \,\, \sum_{\vtheta \in \tilde \Theta^n} \sum_{\avec \in A^{n,\vtheta}} t_{\vtheta,\avec} \, (\alpha   \lambda_{\vtheta}R_{\vtheta, \va} - \epsilon)- \sum_{i \in N} \sum_{\vtheta \in \tilde \Theta^n} \lambda_{\vtheta} \sum_{a\in A_{i,\theta_i}} \sum_{\omega \in \Omega} F_{i,\theta_i,a,\omega} \, y_{i,\vtheta,a,\omega}  \quad \text{s.t.}\\
			& \specialcell{\sum_{\va \in A^{n,\vtheta}:a_i= a} t_{\vtheta,\va} \leq \xi_{i,\vtheta, a}    \hfill \forall i \in N, \forall \vtheta \in \tilde \Theta^n, \forall a \in A_{i,\theta_i}} \\
			& \textnormal{Constraints~\eqref{eq:lp_cons_1}---\eqref{eq:lp_cons_3}~and~\eqref{eq:lp_cons_5}---\eqref{eq:lp_cons_8},} \nonumber
		\end{align}
	\end{subequations}
	where, for ease of presentation, we used the same variable names as in LP~\eqref{lp:typesRelaxed}.

	By strong duality, we have that the optimal value of LP~\eqref{lp:types_dual} is at most $\eta^\star$.
	Then, since an optimal DSIC menu of randomized contracts identifies an optimal solution to LP~\eqref{lp:typesRelaxed} and such a solution is clearly feasible for LP~\eqref{lp:types_dual}, we have that the optimal value of LP~\eqref{lp:types_dual} is at least
	\begin{equation}\label{alfaopt}
	\alpha \, R_\OPT- P_\OPT-|\tilde \Theta^n| \epsilon ,
	\end{equation}
	where we used the fact that, in any feasible solution, it holds $ \sum_{\va \in A^{n,\theta}}  t_{\vtheta,\va}\le 1$ for every $\vtheta \in \tilde \Theta^n$.
	This proves that $\alpha R_\OPT- P_\OPT-|\tilde \Theta^n| \epsilon \le \eta^\star$.
	%
	%
	%

	\paragraph{Bounding the value of the solution returned by Algorithm~\ref{alg:bisection}}
	Next, we show that Algorithm~\ref{alg:bisection} gives as output a solution with value at least $\eta^\star-\beta$.

	Let $\Hcal^\star \subset \tilde \Theta^n \times A^{n}$ be the set of tuples of agents' types and tuples of actions corresponding to Constraints~\eqref{eq:useOracle} which are identified as violated by the \emph{ad hoc} ellipsoid method during the last iteration of the binary search scheme in which it returned infeasible.
	It is immediate to see that, during such an iteration, the ellipsoid method is applied to the feasibility problem \circled{F} with objective $\leq l$ with $l \leq \eta^\star-\beta$, by definition of $\eta^\star$ and given how the binary search scheme terminates.

	LP~\eqref{lp:typesDual} with only the Constraints~\eqref{eq:useOracle} corresponding to elements in $\Hcal^\star$ (and all the other Constraints~\eqref{eq:conFirst}---\eqref{eq:conMid1} and ~\eqref{eq:conMid2}---\eqref{eq:conLast}) is infeasible, and the ellipsoid method guarantees that the elements in $\Hcal^\star$ are polynomially many.
	Moreover, the dual of such an LP is LP~\eqref{lp:typesRelaxed} in which only the variables $t_{\vtheta,\va}$ corresponding to the elements in $\mathcal{H}^\star$ are specified.
	Formally, it can be written as:
	\begin{subequations}\label{lp:newPrimal}
		\begin{align}
			\max & \,\, \sum_{(\vtheta, \va) \in \Hcal^\star} \lambda_{\vtheta}  t_{\vtheta,\avec} \, R_{\vtheta, \va} - \sum_{i \in N} \sum_{\vtheta \in \tilde \Theta^n} \lambda_{\vtheta} \sum_{a\in A_{i,\theta_i}} \sum_{\omega \in \Omega} F_{i,\theta_i,a,\omega} \, y_{i,\vtheta,a,\omega}  \quad \text{s.t.} \\
			& \specialcell{\sum_{\va \in A^{n,\vtheta}:a_i= a\land (\vtheta,\va)\in \Hcal} t_{\vtheta,\va} \leq \xi_{i,\vtheta, a}    \hfill \forall i \in N, \forall \vtheta \in \tilde \Theta^n, \forall a \in A_{i,\theta_i}} \\
			& \textnormal{Constraints~\eqref{eq:lp_cons_1}---\eqref{eq:lp_cons_3}~and~\eqref{eq:lp_cons_5}---\eqref{eq:lp_cons_8}.} \nonumber
		\end{align}
	\end{subequations}

	By strong duality, LP~\eqref{lp:newPrimal} has optimal value at least $\eta^\star-\beta$.
	Moreover, an optimal solution can be computed in polynomial time since there are polynomially-many constraints in $\Hcal^\star$ and, thus, LP~\eqref{lp:newPrimal} has polynomially-many variables and constraints.
	%
	
	\paragraph{Putting all together}
	We conclude the proof by providing the desired approximation guarantees for an optimal solution to LP~\eqref{lp:newPrimal} returned by Algorithm~\ref{alg:bisection}.
	Let $\APX$ be the value of an optimal solution to LP~\eqref{lp:newPrimal}.
	Moreover, set $\beta \coloneqq \frac{\rho}{4}$ and $\epsilon \coloneqq \frac{\rho}{4|\tilde \Theta^n|}$.
	Then,
	\begin{align*}
		\APX & \ge \eta^\star - \beta \\
		& \ge \alpha R_\OPT- P_\OPT-|\tilde \Theta^n| \epsilon-\beta\\
		& \ge \alpha R_\OPT- P_\OPT-\frac{\rho}{2}.
	\end{align*}
	
	Finally, given a feasible solution to LP~\eqref{lp:newPrimal}, by applying Lemma~\ref{lem:lps_equiv}, we can recover in polynomial time a feasible solution to LP~\eqref{lp:types} with the same objective function value (notice that any solution that is feasible for LP~\eqref{lp:newPrimal} is also feasible for LP~\eqref{lp:typesRelaxed}).
	Then, by applying Theorem~\ref{thm:regularize} with $\epsilon=\frac{\rho}{n \tau(|i|)+1}$ to the just computed solution, we can recover in polynomial time a feasible solution to Problem~\eqref{pr:types}, which corresponds to a DISC menu of randomized contracts with principal's expected utility at least $\APX- \frac{\rho}{2}- \epsilon (n \, \tau(|I|)+1) = \alpha R_\OPT- P_\OPT - \rho$, concluding the proof.
\end{proof}

\corIR*

\begin{proof}
	We show that the problem admits a polynomial-time approximate separation oracle $\Ocal_{1}(\cdot,\cdot,\cdot,\cdot)$.
	Then, the result directly follows from Theorem~\ref{th:bayesian}.

	A call $\Ocal_{1}(I,w,\theta,\epsilon)$ to the approximation oracle can simply implement the polynomial-time algorithm for non-Bayesian problem (see Theorem~\ref{thm:IR}).
	Indeed, it is sufficient to rescale the function $g$ (and hence the rewards) by a factor $\lambda_{\vtheta}$, while replacing each value $\widehat P_{i,a}$ with the weight $w_{i,a}$.

	It is easy to see that the arguments proving Theorem~\ref{thm:IR} continue to hold. 
\end{proof}

\corDR*

\begin{proof}
	We show that the problem admits a polynomial-time approximate separation oracle $\Ocal_{1-1/e}(\cdot,\cdot,\cdot,\cdot)$.
	Then, the result readily follows from Theorem~\ref{th:bayesian}.
	
	In particular, a call $\Ocal_{1-1/e}(I,w,\theta,\epsilon)$ to the oracle oracle can be implemented by means of the polynomial-time approximation algorithm introduced for non-Bayesian instances (see Theorem~\ref{thm:DR}).
	Indeed, we can rescale the reward function $g$ (and hence the rewards) by a factor $\lambda_{\vtheta}$, while we can replace the value $\widehat P_{i,a}$ with the weights $w_{i,a}$.
	It is easy to see that Theorem~\ref{thm:DR} continues to hold.
	
	Finally, by Theorem~\ref{thm:DR}, it is easy to show that the approximation guarantees of the oracle hold with high probability
	Indeed, it is sufficient to apply a union bound over the polynomially-many calls to the oracle in order to get that the approximation guarantees of all the calls hold simultaneously with high probability, proving the result.
\end{proof}

%% file: ec23.multiagent.contracts.bbl

\begin{thebibliography}{35}


\ifx \showCODEN    \undefined \def \showCODEN     #1{\unskip}     \fi
\ifx \showDOI      \undefined \def \showDOI       #1{#1}\fi
\ifx \showISBNx    \undefined \def \showISBNx     #1{\unskip}     \fi
\ifx \showISBNxiii \undefined \def \showISBNxiii  #1{\unskip}     \fi
\ifx \showISSN     \undefined \def \showISSN      #1{\unskip}     \fi
\ifx \showLCCN     \undefined \def \showLCCN      #1{\unskip}     \fi
\ifx \shownote     \undefined \def \shownote      #1{#1}          \fi
\ifx \showarticletitle \undefined \def \showarticletitle #1{#1}   \fi
\ifx \showURL      \undefined \def \showURL       {\relax}        \fi
\providecommand\bibfield[2]{#2}
\providecommand\bibinfo[2]{#2}
\providecommand\natexlab[1]{#1}
\providecommand\showeprint[2][]{arXiv:#2}

\bibitem[\protect\citeauthoryear{Alon, D{\"u}tting, and Talgam-Cohen}{Alon
  et~al\mbox{.}}{2021}]%
        {alon2021contracts}
\bibfield{author}{\bibinfo{person}{Tal Alon}, \bibinfo{person}{Paul
  D{\"u}tting}, {and} \bibinfo{person}{Inbal Talgam-Cohen}.}
  \bibinfo{year}{2021}\natexlab{}.
\newblock \showarticletitle{Contracts with Private Cost per Unit-of-Effort}. In
  \bibinfo{booktitle}{\emph{Proceedings of the 22nd ACM Conference on Economics
  and Computation}}. \bibinfo{pages}{52--69}.
\newblock


\bibitem[\protect\citeauthoryear{Arora, Lund, Motwani, Sudan, and
  Szegedy}{Arora et~al\mbox{.}}{1998}]%
        {arora1998proof}
\bibfield{author}{\bibinfo{person}{Sanjeev Arora}, \bibinfo{person}{Carsten
  Lund}, \bibinfo{person}{Rajeev Motwani}, \bibinfo{person}{Madhu Sudan}, {and}
  \bibinfo{person}{Mario Szegedy}.} \bibinfo{year}{1998}\natexlab{}.
\newblock \showarticletitle{Proof verification and the hardness of
  approximation problems}.
\newblock \bibinfo{journal}{\emph{Journal of the ACM (JACM)}}
  \bibinfo{volume}{45}, \bibinfo{number}{3} (\bibinfo{year}{1998}),
  \bibinfo{pages}{501--555}.
\newblock


\bibitem[\protect\citeauthoryear{Babaioff, Feldman, and Nisan}{Babaioff
  et~al\mbox{.}}{2006}]%
        {babaioff2006combinatorial}
\bibfield{author}{\bibinfo{person}{Moshe Babaioff}, \bibinfo{person}{Michal
  Feldman}, {and} \bibinfo{person}{Noam Nisan}.}
  \bibinfo{year}{2006}\natexlab{}.
\newblock \showarticletitle{Combinatorial agency}. In
  \bibinfo{booktitle}{\emph{Proceedings of the 7th ACM Conference on Electronic
  Commerce}}. \bibinfo{pages}{18--28}.
\newblock


\bibitem[\protect\citeauthoryear{Babaioff, Feldman, and Nisan}{Babaioff
  et~al\mbox{.}}{2009}]%
        {babaioff2009free}
\bibfield{author}{\bibinfo{person}{Moshe Babaioff}, \bibinfo{person}{Michal
  Feldman}, {and} \bibinfo{person}{Noam Nisan}.}
  \bibinfo{year}{2009}\natexlab{}.
\newblock \showarticletitle{Free-riding and free-labor in combinatorial
  agency}. In \bibinfo{booktitle}{\emph{International Symposium on Algorithmic
  Game Theory}}. Springer, \bibinfo{pages}{109--121}.
\newblock


\bibitem[\protect\citeauthoryear{Babaioff, Feldman, and Nisan}{Babaioff
  et~al\mbox{.}}{2010}]%
        {babaioff2010mixed}
\bibfield{author}{\bibinfo{person}{Moshe Babaioff}, \bibinfo{person}{Michal
  Feldman}, {and} \bibinfo{person}{Noam Nisan}.}
  \bibinfo{year}{2010}\natexlab{}.
\newblock \showarticletitle{Mixed strategies in combinatorial agency}.
\newblock \bibinfo{journal}{\emph{Journal of Artificial Intelligence Research}}
   \bibinfo{volume}{38} (\bibinfo{year}{2010}), \bibinfo{pages}{339--369}.
\newblock


\bibitem[\protect\citeauthoryear{Babaioff, Feldman, Nisan, and Winter}{Babaioff
  et~al\mbox{.}}{2012}]%
        {babaioff2012combinatorial}
\bibfield{author}{\bibinfo{person}{Moshe Babaioff}, \bibinfo{person}{Michal
  Feldman}, \bibinfo{person}{Noam Nisan}, {and} \bibinfo{person}{Eyal Winter}.}
  \bibinfo{year}{2012}\natexlab{}.
\newblock \showarticletitle{Combinatorial agency}.
\newblock \bibinfo{journal}{\emph{Journal of Economic Theory}}
  \bibinfo{volume}{147}, \bibinfo{number}{3} (\bibinfo{year}{2012}),
  \bibinfo{pages}{999--1034}.
\newblock


\bibitem[\protect\citeauthoryear{Babaioff and Winter}{Babaioff and
  Winter}{2014}]%
        {babaioff2014contract}
\bibfield{author}{\bibinfo{person}{Moshe Babaioff} {and} \bibinfo{person}{Eyal
  Winter}.} \bibinfo{year}{2014}\natexlab{}.
\newblock \showarticletitle{Contract complexity.}
\newblock \bibinfo{journal}{\emph{EC}}  \bibinfo{volume}{14}
  (\bibinfo{year}{2014}), \bibinfo{pages}{911}.
\newblock


\bibitem[\protect\citeauthoryear{Bach}{Bach}{2019}]%
        {bach2019submodular}
\bibfield{author}{\bibinfo{person}{Francis Bach}.}
  \bibinfo{year}{2019}\natexlab{}.
\newblock \showarticletitle{Submodular functions: from discrete to continuous
  domains}.
\newblock \bibinfo{journal}{\emph{Mathematical Programming}}
  \bibinfo{volume}{175}, \bibinfo{number}{1} (\bibinfo{year}{2019}),
  \bibinfo{pages}{419--459}.
\newblock


\bibitem[\protect\citeauthoryear{Bastani, Bayati, Braverman, Gummadi, and
  Johari}{Bastani et~al\mbox{.}}{2016}]%
        {bastani2016analysis}
\bibfield{author}{\bibinfo{person}{Hamsa Bastani}, \bibinfo{person}{Mohsen
  Bayati}, \bibinfo{person}{Mark Braverman}, \bibinfo{person}{Ramki Gummadi},
  {and} \bibinfo{person}{Ramesh Johari}.} \bibinfo{year}{2016}\natexlab{}.
\newblock \showarticletitle{Analysis of medicare pay-for-performance
  contracts}.
\newblock \bibinfo{journal}{\emph{Available at SSRN 2839143}}
  (\bibinfo{year}{2016}).
\newblock


\bibitem[\protect\citeauthoryear{Bertsimas and Tsitsiklis}{Bertsimas and
  Tsitsiklis}{1997}]%
        {bertsimas1997introduction}
\bibfield{author}{\bibinfo{person}{Dimitris Bertsimas} {and}
  \bibinfo{person}{John~N Tsitsiklis}.} \bibinfo{year}{1997}\natexlab{}.
\newblock \bibinfo{booktitle}{\emph{Introduction to linear optimization}}.
  Vol.~\bibinfo{volume}{6}.
\newblock \bibinfo{publisher}{Athena scientific Belmont, MA}.
\newblock


\bibitem[\protect\citeauthoryear{Bian, Mirzasoleiman, Buhmann, and Krause}{Bian
  et~al\mbox{.}}{2017}]%
        {Bian2017Guaranteed}
\bibfield{author}{\bibinfo{person}{Andrew~An Bian}, \bibinfo{person}{Baharan
  Mirzasoleiman}, \bibinfo{person}{Joachim Buhmann}, {and}
  \bibinfo{person}{Andreas Krause}.} \bibinfo{year}{2017}\natexlab{}.
\newblock \showarticletitle{{Guaranteed Non-convex Optimization: Submodular
  Maximization over Continuous Domains}}. In
  \bibinfo{booktitle}{\emph{Proceedings of the 20th International Conference on
  Artificial Intelligence and Statistics}} \emph{(\bibinfo{series}{Proceedings
  of Machine Learning Research})}, \bibfield{editor}{\bibinfo{person}{Aarti
  Singh} {and} \bibinfo{person}{Jerry Zhu}} (Eds.), Vol.~\bibinfo{volume}{54}.
  \bibinfo{publisher}{PMLR}, \bibinfo{pages}{111--120}.
\newblock
\urldef\tempurl%
\url{https://proceedings.mlr.press/v54/bian17a.html}
\showURL{%
\tempurl}


\bibitem[\protect\citeauthoryear{Birkhoff}{Birkhoff}{1937}]%
        {birkhoff1937rings}
\bibfield{author}{\bibinfo{person}{Garrett Birkhoff}.}
  \bibinfo{year}{1937}\natexlab{}.
\newblock \showarticletitle{Rings of sets}.
\newblock \bibinfo{journal}{\emph{Duke Mathematical Journal}}
  \bibinfo{volume}{3}, \bibinfo{number}{3} (\bibinfo{year}{1937}),
  \bibinfo{pages}{443--454}.
\newblock


\bibitem[\protect\citeauthoryear{Carroll}{Carroll}{2015}]%
        {carroll2015robustness}
\bibfield{author}{\bibinfo{person}{Gabriel Carroll}.}
  \bibinfo{year}{2015}\natexlab{}.
\newblock \showarticletitle{Robustness and linear contracts}.
\newblock \bibinfo{journal}{\emph{American Economic Review}}
  \bibinfo{volume}{105}, \bibinfo{number}{2} (\bibinfo{year}{2015}),
  \bibinfo{pages}{536--63}.
\newblock


\bibitem[\protect\citeauthoryear{Castiglioni, Marchesi, and Gatti}{Castiglioni
  et~al\mbox{.}}{2022a}]%
        {castiglioni2022bayesian}
\bibfield{author}{\bibinfo{person}{Matteo Castiglioni},
  \bibinfo{person}{Alberto Marchesi}, {and} \bibinfo{person}{Nicola Gatti}.}
  \bibinfo{year}{2022}\natexlab{a}.
\newblock \showarticletitle{Bayesian agency: Linear versus tractable
  contracts}.
\newblock \bibinfo{journal}{\emph{Artificial Intelligence}}
  \bibinfo{volume}{307} (\bibinfo{year}{2022}).
\newblock


\bibitem[\protect\citeauthoryear{Castiglioni, Marchesi, and Gatti}{Castiglioni
  et~al\mbox{.}}{2022b}]%
        {Castiglioni2022Designing}
\bibfield{author}{\bibinfo{person}{Matteo Castiglioni},
  \bibinfo{person}{Alberto Marchesi}, {and} \bibinfo{person}{Nicola Gatti}.}
  \bibinfo{year}{2022}\natexlab{b}.
\newblock \bibinfo{title}{Designing Menus of Contracts Efficiently: The Power
  of Randomization}.
\newblock
\newblock
\urldef\tempurl%
\url{https://doi.org/10.48550/ARXIV.2202.10966}
\showDOI{\tempurl}


\bibitem[\protect\citeauthoryear{Castiglioni, Marchesi, and Gatti}{Castiglioni
  et~al\mbox{.}}{2022c}]%
        {DBLP:conf/sigecom/CastiglioniM022}
\bibfield{author}{\bibinfo{person}{Matteo Castiglioni},
  \bibinfo{person}{Alberto Marchesi}, {and} \bibinfo{person}{Nicola Gatti}.}
  \bibinfo{year}{2022}\natexlab{c}.
\newblock \showarticletitle{Designing Menus of Contracts Efficiently: The Power
  of Randomization}. In \bibinfo{booktitle}{\emph{{EC} '22: The 23rd {ACM}
  Conference on Economics and Computation}}. \bibinfo{pages}{705--735}.
\newblock


\bibitem[\protect\citeauthoryear{Cong and He}{Cong and He}{2019}]%
        {cong2019blockchain}
\bibfield{author}{\bibinfo{person}{Lin~William Cong} {and}
  \bibinfo{person}{Zhiguo He}.} \bibinfo{year}{2019}\natexlab{}.
\newblock \showarticletitle{Blockchain disruption and smart contracts}.
\newblock \bibinfo{journal}{\emph{The Review of Financial Studies}}
  \bibinfo{volume}{32}, \bibinfo{number}{5} (\bibinfo{year}{2019}),
  \bibinfo{pages}{1754--1797}.
\newblock


\bibitem[\protect\citeauthoryear{Duetting, Ezra, Feldman, and
  Kesselheim}{Duetting et~al\mbox{.}}{2022}]%
        {duetting2022multi}
\bibfield{author}{\bibinfo{person}{Paul Duetting}, \bibinfo{person}{Tomer
  Ezra}, \bibinfo{person}{Michal Feldman}, {and} \bibinfo{person}{Thomas
  Kesselheim}.} \bibinfo{year}{2022}\natexlab{}.
\newblock \showarticletitle{Multi-Agent Contracts}.
\newblock \bibinfo{journal}{\emph{arXiv preprint arXiv:2211.05434}}
  (\bibinfo{year}{2022}).
\newblock


\bibitem[\protect\citeauthoryear{D{\"u}tting, Ezra, Feldman, and
  Kesselheim}{D{\"u}tting et~al\mbox{.}}{2022}]%
        {dutting2022combinatorial}
\bibfield{author}{\bibinfo{person}{Paul D{\"u}tting}, \bibinfo{person}{Tomer
  Ezra}, \bibinfo{person}{Michal Feldman}, {and} \bibinfo{person}{Thomas
  Kesselheim}.} \bibinfo{year}{2022}\natexlab{}.
\newblock \showarticletitle{Combinatorial contracts}. In
  \bibinfo{booktitle}{\emph{2021 IEEE 62nd Annual Symposium on Foundations of
  Computer Science (FOCS)}}. IEEE, \bibinfo{pages}{815--826}.
\newblock


\bibitem[\protect\citeauthoryear{D{\"u}tting, Roughgarden, and
  Talgam-Cohen}{D{\"u}tting et~al\mbox{.}}{2019}]%
        {dutting2019simple}
\bibfield{author}{\bibinfo{person}{Paul D{\"u}tting}, \bibinfo{person}{Tim
  Roughgarden}, {and} \bibinfo{person}{Inbal Talgam-Cohen}.}
  \bibinfo{year}{2019}\natexlab{}.
\newblock \showarticletitle{Simple versus optimal contracts}. In
  \bibinfo{booktitle}{\emph{Proceedings of the 2019 ACM Conference on Economics
  and Computation}}. \bibinfo{pages}{369--387}.
\newblock


\bibitem[\protect\citeauthoryear{Dutting, Roughgarden, and
  Talgam-Cohen}{Dutting et~al\mbox{.}}{2021}]%
        {dutting2021complexity}
\bibfield{author}{\bibinfo{person}{Paul Dutting}, \bibinfo{person}{Tim
  Roughgarden}, {and} \bibinfo{person}{Inbal Talgam-Cohen}.}
  \bibinfo{year}{2021}\natexlab{}.
\newblock \showarticletitle{The complexity of contracts}.
\newblock \bibinfo{journal}{\emph{SIAM J. Comput.}} \bibinfo{volume}{50},
  \bibinfo{number}{1} (\bibinfo{year}{2021}), \bibinfo{pages}{211--254}.
\newblock


\bibitem[\protect\citeauthoryear{Emek and Feldman}{Emek and Feldman}{2012}]%
        {emek2012computing}
\bibfield{author}{\bibinfo{person}{Yuval Emek} {and} \bibinfo{person}{Michal
  Feldman}.} \bibinfo{year}{2012}\natexlab{}.
\newblock \showarticletitle{Computing optimal contracts in combinatorial
  agencies}.
\newblock \bibinfo{journal}{\emph{Theoretical Computer Science}}
  \bibinfo{volume}{452} (\bibinfo{year}{2012}), \bibinfo{pages}{56--74}.
\newblock


\bibitem[\protect\citeauthoryear{Gan, Han, Wu, and Xu}{Gan
  et~al\mbox{.}}{2022}]%
        {gan2022optimal}
\bibfield{author}{\bibinfo{person}{Jiarui Gan}, \bibinfo{person}{Minbiao Han},
  \bibinfo{person}{Jibang Wu}, {and} \bibinfo{person}{Haifeng Xu}.}
  \bibinfo{year}{2022}\natexlab{}.
\newblock \showarticletitle{Optimal Coordination in Generalized Principal-Agent
  Problems: A Revisit and Extensions}.
\newblock \bibinfo{journal}{\emph{arXiv preprint arXiv:2209.01146}}
  (\bibinfo{year}{2022}).
\newblock


\bibitem[\protect\citeauthoryear{Gr{\"o}tschel, Lov{\'a}sz, and
  Schrijver}{Gr{\"o}tschel et~al\mbox{.}}{2012}]%
        {grotschel2012geometric}
\bibfield{author}{\bibinfo{person}{Martin Gr{\"o}tschel},
  \bibinfo{person}{L{\'a}szl{\'o} Lov{\'a}sz}, {and} \bibinfo{person}{Alexander
  Schrijver}.} \bibinfo{year}{2012}\natexlab{}.
\newblock \bibinfo{booktitle}{\emph{Geometric algorithms and combinatorial
  optimization}}. Vol.~\bibinfo{volume}{2}.
\newblock \bibinfo{publisher}{Springer Science \& Business Media}.
\newblock


\bibitem[\protect\citeauthoryear{Guruganesh, Schneider, and Wang}{Guruganesh
  et~al\mbox{.}}{2021}]%
        {guruganesh2021contracts}
\bibfield{author}{\bibinfo{person}{Guru Guruganesh}, \bibinfo{person}{Jon
  Schneider}, {and} \bibinfo{person}{Joshua~R Wang}.}
  \bibinfo{year}{2021}\natexlab{}.
\newblock \showarticletitle{Contracts under moral hazard and adverse
  selection}. In \bibinfo{booktitle}{\emph{{EC} '21: The 22nd {ACM} Conference
  on Economics and Computation}}. \bibinfo{pages}{563--582}.
\newblock


\bibitem[\protect\citeauthoryear{H{\aa}stad}{H{\aa}stad}{1999}]%
        {hastad1999clique}
\bibfield{author}{\bibinfo{person}{Johan H{\aa}stad}.}
  \bibinfo{year}{1999}\natexlab{}.
\newblock \showarticletitle{Clique is hard to approximate within
  $n^{1-\epsilon}$}.
\newblock \bibinfo{journal}{\emph{Acta Mathematica}} \bibinfo{volume}{182},
  \bibinfo{number}{1} (\bibinfo{year}{1999}), \bibinfo{pages}{105--142}.
\newblock
\showISBNx{1871-2509}


\bibitem[\protect\citeauthoryear{Ho, Slivkins, and Vaughan}{Ho
  et~al\mbox{.}}{2016}]%
        {ho2016adaptive}
\bibfield{author}{\bibinfo{person}{Chien-Ju Ho}, \bibinfo{person}{Aleksandrs
  Slivkins}, {and} \bibinfo{person}{Jennifer~Wortman Vaughan}.}
  \bibinfo{year}{2016}\natexlab{}.
\newblock \showarticletitle{Adaptive contract design for crowdsourcing markets:
  Bandit algorithms for repeated principal-agent problems}.
\newblock \bibinfo{journal}{\emph{Journal of Artificial Intelligence Research}}
   \bibinfo{volume}{55} (\bibinfo{year}{2016}), \bibinfo{pages}{317--359}.
\newblock


\bibitem[\protect\citeauthoryear{Raz}{Raz}{1998}]%
        {raz1998parallel}
\bibfield{author}{\bibinfo{person}{Ran Raz}.} \bibinfo{year}{1998}\natexlab{}.
\newblock \showarticletitle{A parallel repetition theorem}.
\newblock \bibinfo{journal}{\emph{SIAM J. Comput.}} \bibinfo{volume}{27},
  \bibinfo{number}{3} (\bibinfo{year}{1998}), \bibinfo{pages}{763--803}.
\newblock


\bibitem[\protect\citeauthoryear{Schrijver}{Schrijver}{2000}]%
        {schrijver2000combinatorial}
\bibfield{author}{\bibinfo{person}{Alexander Schrijver}.}
  \bibinfo{year}{2000}\natexlab{}.
\newblock \showarticletitle{A combinatorial algorithm minimizing submodular
  functions in strongly polynomial time}.
\newblock \bibinfo{journal}{\emph{Journal of Combinatorial Theory, Series B}}
  \bibinfo{volume}{80}, \bibinfo{number}{2} (\bibinfo{year}{2000}),
  \bibinfo{pages}{346--355}.
\newblock


\bibitem[\protect\citeauthoryear{Schrijver et~al\mbox{.}}{Schrijver
  et~al\mbox{.}}{2003}]%
        {schrijver2003combinatorial}
\bibfield{author}{\bibinfo{person}{Alexander Schrijver} {et~al\mbox{.}}}
  \bibinfo{year}{2003}\natexlab{}.
\newblock \bibinfo{booktitle}{\emph{Combinatorial optimization: polyhedra and
  efficiency}}. Vol.~\bibinfo{volume}{24}.
\newblock \bibinfo{publisher}{Springer}.
\newblock


\bibitem[\protect\citeauthoryear{Shoham and Leyton-Brown}{Shoham and
  Leyton-Brown}{2008}]%
        {shoham2008multiagent}
\bibfield{author}{\bibinfo{person}{Yoav Shoham} {and} \bibinfo{person}{Kevin
  Leyton-Brown}.} \bibinfo{year}{2008}\natexlab{}.
\newblock \bibinfo{booktitle}{\emph{Multiagent systems: Algorithmic,
  game-theoretic, and logical foundations}}.
\newblock \bibinfo{publisher}{Cambridge University Press}.
\newblock


\bibitem[\protect\citeauthoryear{Sviridenko, Vondr{\'a}k, and Ward}{Sviridenko
  et~al\mbox{.}}{2017}]%
        {sviridenko2017optimal}
\bibfield{author}{\bibinfo{person}{Maxim Sviridenko}, \bibinfo{person}{Jan
  Vondr{\'a}k}, {and} \bibinfo{person}{Justin Ward}.}
  \bibinfo{year}{2017}\natexlab{}.
\newblock \showarticletitle{Optimal approximation for submodular and
  supermodular optimization with bounded curvature}.
\newblock \bibinfo{journal}{\emph{Mathematics of Operations Research}}
  \bibinfo{volume}{42}, \bibinfo{number}{4} (\bibinfo{year}{2017}),
  \bibinfo{pages}{1197--1218}.
\newblock


\bibitem[\protect\citeauthoryear{Tadelis and Segal}{Tadelis and Segal}{2005}]%
        {tadelis2005lectures}
\bibfield{author}{\bibinfo{person}{Steve Tadelis} {and} \bibinfo{person}{Ilya
  Segal}.} \bibinfo{year}{2005}\natexlab{}.
\newblock \showarticletitle{Lectures in contract theory}.
\newblock \bibinfo{journal}{\emph{Lecture notes for UC Berkeley and Stanford
  University}} (\bibinfo{year}{2005}).
\newblock


\bibitem[\protect\citeauthoryear{Zuckerman}{Zuckerman}{2007}]%
        {Zuckerman2007linear}
\bibfield{author}{\bibinfo{person}{David Zuckerman}.}
  \bibinfo{year}{2007}\natexlab{}.
\newblock \showarticletitle{Linear Degree Extractors and the Inapproximability
  of Max Clique and Chromatic Number}.
\newblock \bibinfo{journal}{\emph{Theory of Computing}} \bibinfo{volume}{3},
  \bibinfo{number}{6} (\bibinfo{year}{2007}), \bibinfo{pages}{103--128}.
\newblock


\bibitem[\protect\citeauthoryear{Østerdal}{Østerdal}{2010}]%
        {OSTERDAL20101222}
\bibfield{author}{\bibinfo{person}{Lars~Peter Østerdal}.}
  \bibinfo{year}{2010}\natexlab{}.
\newblock \showarticletitle{The mass transfer approach to multivariate discrete
  first order stochastic dominance: Direct proof and implications}.
\newblock \bibinfo{journal}{\emph{Journal of Mathematical Economics}}
  \bibinfo{volume}{46}, \bibinfo{number}{6} (\bibinfo{year}{2010}),
  \bibinfo{pages}{1222--1228}.
\newblock
\showISSN{0304-4068}
\urldef\tempurl%
\url{https://doi.org/10.1016/j.jmateco.2010.08.018}
\showDOI{\tempurl}
\newblock
\shownote{The Conferences at Barcelona, Milan, New Haven, San Diego and Tokyo.}


\end{thebibliography}
